\documentclass[hyper]{JHEP3} 
\usepackage{graphicx}
\usepackage{epsfig}
\usepackage{cite}
\usepackage{axodraw}
\usepackage{amsmath}

\makeatletter

\newcommand{\fmslash}[2][0mu]{%
  \mathchoice
    {\fmsl@sh\displaystyle{#1}{#2}}%
    {\fmsl@sh\textstyle{#1}{#2}}%
    {\fmsl@sh\scriptstyle{#1}{#2}}%
    {\fmsl@sh\scriptscriptstyle{#1}{#2}}}
\newcommand{\fmsl@sh}[3]{%
  \m@th\ooalign{$\hfil#1\mkern#2/\hfil$\crcr$#1#3$}}
\makeatother

\newcommand{\beq}{\begin{equation}}
\newcommand{\eeq}{\end{equation}}
\newcommand{\bea}{\begin{eqnarray}}
\newcommand{\eea}{\end{eqnarray}}

\newcommand{\met}{\not \!\! E_T}

\newcommand{\mht}{\not \!\! H_T}
\newcommand{\mpt}{\not \!\! P_T}
\newcommand{\mmis}{\not \!\! M}
\newcommand{\mptvec}{\not \!\! \vec{P}_T}
\newcommand{\mhtvec}{\not \!\! \vec{H}_T}

\title{RECO level $\sqrt{s}_{min}$
and subsystem $\sqrt{s}_{min}$:
improved global inclusive variables for measuring the new physics
mass scale in $\met$ events at hadron colliders}

\author{Partha Konar \\
        Physics Department, University of Florida,
        Gainesville, FL 32611, USA \\
        E-mail: \email{konar@phys.ufl.edu}
        }

\author{Kyoungchul Kong\\
        Theoretical Physics Department, SLAC, 
        Menlo Park, CA 94025, USA \\
        E-mail: \email{kckong@slac.stanford.edu}
        }

\author{Konstantin T.~Matchev \\ 
        Physics Department, University of Florida,
        Gainesville, FL 32611, USA \\
        E-mail: \email{matchev@phys.ufl.edu}
        }

\author{Myeonghun Park\\
        Physics Department, University of Florida,
        Gainesville, FL 32611, USA\\
        E-mail: \email{ishaed@phys.ufl.edu}
        }

\received{\today}               
\accepted{\today}               

\preprint{SLAC-PUB-14136 \\
          June 3, 2010
          } 

\abstract{The variable $\sqrt{s}_{min}$ was originally proposed 
in \cite{Konar:2008ei} as a model-independent, global and fully 
inclusive measure of the new physics mass scale 
in missing energy events at hadron colliders. In the original
incarnation of $\sqrt{s}_{min}$, however, the connection
to the new physics mass scale was blurred by the effects of
the underlying event, most notably initial state radiation 
and multiple parton interactions.
In this paper we advertize two improved variants of the 
$\sqrt{s}_{min}$ variable, which overcome this problem.
First we show that by evaluating the $\sqrt{s}_{min}$
variable at the RECO level, in terms of the reconstructed 
objects in the event, the effects from the underlying event 
are significantly diminished and the nice correlation between 
the peak in the $\sqrt{s}_{min}^{(reco)}$ distribution and the 
new physics mass scale is restored. Secondly, the underlying 
event problem can be avoided altogether when the $\sqrt{s}_{min}$
concept is applied to a subsystem of the event which does
not involve any QCD jets. We supply an analytic formula
for the resulting subsystem $\sqrt{s}_{min}^{(sub)}$ variable
and show that its peak exhibits the usual correlation with the 
mass scale of the particles produced in the subsystem.
Finally, we contrast $\sqrt{s}_{min}$ to other popular inclusive 
variables such as $H_T$, $M_{Tgen}$ and $M_{TTgen}$.
We illustrate our discussion with several examples from 
supersymmetry, and with dilepton events from top quark 
pair production.
}


\begin{document} 

\section{Introduction and motivation}
\label{sec:intro}

\subsection{The need for a universal, global and inclusive mass variable}

It is generally believed that missing energy signatures
offer the best bet for discovering new physics 
Beyond the Standard Model (BSM) at colliders. 
This belief is reinforced by the dark matter puzzle - 
the Standard Model (SM) does not contain a suitable dark matter 
candidate. If dark matter particles are produced at colliders,
they will be invisible in the detector, and will in principle
lead to missing energy and missing momentum. 
However, at hadron colliders the total energy 
and longitudinal momentum of the event are unknown.
Therefore, the production of any invisible particles can only be inferred
from an imbalance in the total {\em transverse} momentum. 
The measured {\em missing} transverse momentum $\mptvec$
then gives the sum of the transverse momenta of
all invisible particles in the event.

Unfortunately, $\mptvec$ is the only measured quantity directly
related to the invisible particles. Without any further 
model-dependent assumptions, it is in general very difficult 
if not impossible to make any definitive statements 
about the nature and properties of the missing particles.
For example, leaving all theoretical prejudice aside, 
one would not be able to answer such basic and fundamental questions like
\cite{Konar:2008ei,Chang:2009dh,Barr:2009jv,Konar:2009qr,Agashe:2010gt}:
How many invisible particles were produced in the event?
Are all invisible particles SM neutrinos, or are there 
any new neutral, stable, weakly-interacting massive particles 
(WIMPs) among them? What are the masses of the new invisible particles?
What are their spins? What are the masses of any (parent) 
particles which may have decayed to invisible particles?

The recent literature is abundant with numerous proposals\footnote{See 
Ref.~\cite{Barr:2010zj} for a recent review.}
on how {\em under particular circumstances}
one might be able to measure the masses 
of the invisible particles.
Unfortunately, all of the proposed methods suffer 
from varying degrees of model-dependence\footnote{Worse still,
there are even fewer ideas for measuring the {\em spins} of the new
particles in a truly model-independent fashion 
\cite{Buckley:2007th,Burns:2008cp,Boudjema:2009fz}.}:
\begin{itemize}
\item {\em Limited applicability topology-wise.}
Most methods are model-dependent in the sense that each method
crucially relies on the assumption of a very specific event topology.
One common flaw of all methods on the market is that 
they usually do not allow any SM neutrinos to enter the 
targeted event topology, and the missing energy is 
typically assumed to arise only as a result of the 
production of (two) new dark matter particles. 
Furthermore, each method has its own limitations.
For example, the traditional invariant mass endpoint methods 
\cite{Hinchliffe:1996iu,Bachacou:1999zb,Hinchliffe:1999zc,Allanach:2000kt,Gjelsten:2004ki,%
Gjelsten:2005aw,Birkedal:2005cm,Miller:2005zp,Costanzo:2009mq,Burns:2009zi,Matchev:2009iw} 
require the identification of a sufficiently long cascade decay chain, with at least
three successive two-body decays \cite{Burns:2008va}. The polynomial methods of 
Refs.~\cite{Nojiri:2003tu,Kawagoe:2004rz,Cheng:2007xv,Nojiri:2007pq,%
Cheng:2008mg,Cheng:2009fw,Webber:2009vm,Nojiri:2010dk}
also require such long decay chains and furthermore, the events 
must be symmetric, i.e. must have two {\em identical} decay chains per event,
or else the decay chain must be even longer \cite{Burns:2008va}.
The recently popular $M_{T2}$ methods  \cite{Lester:1999tx,Barr:2003rg,Lester:2007fq,%
Cho:2007qv,Gripaios:2007is,Barr:2007hy,Cho:2007dh,Cheng:2008hk,%
Matchev:2009fh,Konar:2009wn} 
do not require long decay chains \cite{Burns:2008va}, but typically
assume that the parent particles are the same
and decay to two identical invisible particles\footnote{See
\cite{Barr:2009jv,Konar:2009qr} for a more general approach which avoids this assumption.}.
The limitations of the $M_{CT}$ methods 
\cite{Tovey:2008ui,Polesello:2009rn,Matchev:2009ad} are rather similar.
The kinematic cusp method \cite{Han:2009ss} is limited to the so called
``antler'' event topology, which contains two symmetric one-step decay chains 
originating from a single $s$-channel resonance.
In light of all these various assumptions, it is certainly
desirable to have a {\em universal} method which can be applied to {\em any} 
event topology. To the best of our knowledge, the only
such method in the literature is the one proposed in 
Ref.~\cite{Konar:2008ei}, where the $\sqrt{s}_{min}$
variable was first introduced. The $\sqrt{s}_{min}$ variable
is defined in terms of the {\em total} energy $E$ and 3-momentum $\vec{P}$ 
observed in the event, and thus does not make any reference to 
the actual event topology. In this sense $\sqrt{s}_{min}$ 
is a {\em universal} variable which can be applied under any circumstances.
\item {\em Limited applicability signature-wise.} As a rule,
most of the proposed methods work well only if the corresponding
signature contains some minimum number of high $p_T$ isolated leptons.
Leptonic signatures have the twofold advantage of lower SM backgrounds
and good lepton momentum measurement. The performance of the methods 
typically deteriorates as we lower the number of leptons in the 
signature. The most challenging signature of multijets plus $\met$
has rarely been studied in relation to mass and spin measurements
(see, however \cite{Alves:2006df,Butterworth:2007ke,%
Csaki:2007xm,Cho:2007qv,Nojiri:2008vq}).
Unfortunately, at hadron colliders like the Tevatron and LHC, 
one typically expects strong production to dominate the new physics
cross-sections, and this in turn guarantees the presence of 
some minimum number of jets in the signature. At the same time, a priori 
there are no theoretical arguments which would similarly 
guarantee the presence of any hard isolated leptons.
Therefore, one would like to have a general, sufficiently inclusive
method, which treats jets and leptons on an equal footing. The $\sqrt{s}_{min}$
method of Ref.~\cite{Konar:2008ei} satisfies this requirement as well,
since it does not differentiate between the type of reconstructed 
objects. In fact, the original proposal of Ref.~\cite{Konar:2008ei}
did not require any object reconstruction at all, and used 
(muon-corrected) calorimeter energy measurements to define 
the observed $E$ and $\vec{P}$ in the event.
\item {\em Combinatorics problem.} Even if one correctly guesses 
the new physics event topology, and the signature happens to be 
abundant in hard isolated leptons, one still has to face 
the usual combinatorics problem of how to properly associate the 
reconstructed objects with the individual particles in 
the assumed event topology. Here we shall be careful to
make the distinction between two different aspects of the 
combinatorics problem:
\begin{itemize}
\item {\em Partitioning ambiguity.} As a prototypical example, consider
a model of supersymmetry (SUSY) in which $R$-parity is conserved and
the lightest supersymmetric particle (LSP) is neutral and stable. 
Each SUSY event contains two 
independent cascade decay chains, so first one must decide which 
reconstructed objects belong to the first decay chain and which 
belong to the second \cite{Lester:2007fq,Nojiri:2008hy}. 
However, a priori there are no guiding principles on how to
do this partitioning into subsets. 
The decision is further complicated by
the inevitable presence of jets from initial state radiation,
which have nothing to do with the SUSY cascades \cite{Alwall:2009zu};
by final state radiation, which modifies the assumed 
event topology; and by the occasional overlapping of 
jets \cite{Nojiri:2008ir}.
\item {\em Ordering ambiguity.}
Having separated the objects into two groups, 
one must still decide on the sequential ordering of the 
reconstructed objects along each decay chain. One well-known 
example of this problem is the ambiguity between the ``near'' 
and ``far'' lepton in the standard jet-lepton-lepton 
squark decay chain \cite{Matchev:2009iw}. 
\end{itemize}
The severity of either one of these two combinatorics problems
depends on the type of signature --- simple signatures 
resulting from short decay chains suffer from less 
combinatorics but tend to have larger SM backgrounds.
By the same token, more complex signatures, which result
from longer decay chains, are easier to see over the 
SM backgrounds, but very quickly run into severe 
combinatorial problems. Thus ideally one would
like to have a method which treats {\em all} objects
in the event in a fully inclusive manner, so that 
{\em neither of these two} combinatorial issues can ever arise at all.
The $\sqrt{s}_{min}$ variable of Ref.~\cite{Konar:2008ei}
was proposed for exactly this reason, and is free of
the partitioning {\em and} ordering combinatorial ambiguities.
\item {\em Limited use of the available experimental information.}
At hadron colliders, events with invisible particles 
in the final state present an additional challenge: 
the total energy and longitudinal momentum of the 
initial state in the event are unknown. On the other hand,
the {\em transverse} momentum of the initial state 
is known, which has greatly motivated the use of {\em transverse} 
variables like the missing ``transverse energy'' $\met$, 
the scalar sum of transverse momenta $H_T$, the transverse 
mass $M_T$, the stransverse mass $M_{T2}$ \cite{Lester:1999tx}, 
the contransverse mass $M_{CT}$ \cite{Tovey:2008ui}, etc.
An unsettling feature of a purely transverse 
kinematical approach is that it completely
ignores the measured longitudinal momentum components 
of the visible particles. In principle, the longitudinal 
momenta also carry a certain amount of information about 
the underlying physics, although it is difficult to see 
immediately how this information can be utilized. 
(For example, one cannot take advantage of 
longitudinal momentum conservation, because the 
longitudinal momentum of the initial state is unknown.) 
By defining the $\sqrt{s}_{min}$ variable in a manifestly 
1+3 Lorentz invariant way, Ref.~\cite{Konar:2008ei}
proposed one possible way to utilize the additional information
encoded in the measured longitudinal momenta.
\end{itemize}

The above discussion makes it clear that the 
method of the $\sqrt{s}_{min}$ variable has several
unique advantages over all other known methods:
it is completely general and universal, is fully inclusive,
and to the fullest extent makes use of the available experimental information.
In spite of these advantages, the $\sqrt{s}_{min}$ 
variable has not yet found wide application. 
The one major perceived drawback of $\sqrt{s}_{min}$ 
is its sensitivity to initial state radiation (ISR)
and/or multiple parton interactions (MPI)
\cite{Konar:2008ei,Papaefstathiou:2009hp,%
Papaefstathiou:2010ru,Barr:2010zj,Brooijmans:2010tn}.
To see how this comes about, let us first review the formal 
definition of $\sqrt{s}_{min}$. 

\subsection{Definition of $\sqrt{s}_{min}$}

Consider the most generic missing energy event 
topology shown in Fig.~\ref{fig:metevent}.
\FIGURE[ht]{
{
\unitlength=1.5 pt
\SetScale{1.5}
\SetWidth{1.0}      
\normalsize    
{} \qquad\allowbreak
\begin{picture}(250,200)(0,0)
\SetColor{Gray}
\Line( 13,185)(130,185)
\Line( 50,150)(130,150)
\Line( 50,140)(130,140)
\Line( 50,130)(130,130)
\Line( 50,120)(130,120)
\Text(135,120)[l]{\Black{$X_1$}}
\Text(135,130)[l]{\Black{$X_2$}}
\Text(135,140)[l]{\Black{$X_3$}}
\Text(135,150)[l]{\Black{$X_4$}}
\Text(135,185)[l]{\Black{$X_{n_{vis}}$}}
\DashLine(50,110)(130,110){2}
\DashLine(50, 85)(130, 85){2}
\DashLine(50, 75)(130, 75){2}
\Text(135,110)[l]{$\chi_{n_{inv}}$}
\Text(135, 85)[l]{$\chi_{n_{\chi}+2}$}
\Text(135, 75)[l]{$\chi_{n_{\chi}+1}$}
\Line( 10,190)(55,120)
\Line( 10, 10)(55, 80)
\Text(20,155)[c]{\Black{$p(\bar{p})$}}
\Text(20, 45)[c]{\Black{$p(\bar{p})$}}
\DashLine(95,180)(95,155){1}
\DashLine(95,105)(95, 90){1}
\CArc(10,152.5)(150,-14,14)
\LongArrow(167,152.5)(197,152.5)
\SetColor{Red}
\DashLine(95, 60)(95, 45){1}
\CArc(15,   70)(144,-17,17)
\LongArrow(165,70)(195,70)
\SetWidth{1.2}      
\DashLine(50,65)(130,65){2}
\DashLine(50,40)(130,40){2}
\DashLine(50,30)(130,30){2}
\Text(135,65)[l]{\Red{$\chi_{n_\chi}$}}
\Text(135,40)[l]{\Red{$\chi_2$}}
\Text(135,30)[l]{\Red{$\chi_1$}}
\CBoxc(230,152.5)(50,20){Black}{Yellow}
\Text(230,152.5)[c]{$E,P_x,P_y,P_z$}
\CBoxc(215, 70)(25,20){Red}{Yellow}
\Text(215,70)[c]{\Red{$\mptvec$}}
\COval(50,100)(75,15)(0){Blue}{Green}
\end{picture}
}
\caption{The generic event topology used to define the 
$\sqrt{s}_{min}$ variable in Ref.~\cite{Konar:2008ei}.
Black (red) lines correspond to SM (BSM) particles.
The solid lines denote SM particles $X_i$, $i=1,2,\ldots, n_{vis}$,
which are visible in the detector, e.g.~jets, electrons, muons and photons.
The SM particles may originate either from initial 
state radiation (ISR), or from the hard scattering and subsequent 
cascade decays (indicated with the green-shaded ellipse).
The dashed lines denote neutral stable particles 
$\chi_i$, $i=1,2,\ldots, n_{inv}$, which are invisible in the detector.
In general, the set of invisible particles consists of some number 
$n_{\chi}$ of BSM particles (indicated with the red dashed lines),
as well as some number $n_{\nu}=n_{inv}-n_{\chi}$ of SM neutrinos 
(denoted with the black dashed lines). The identities and the masses
$m_i$ of the BSM invisible particles $\chi_i$, ($i=1,2,\ldots,n_{\chi}$)
do not necessarily have to be all the same, i.e.~we allow for the
simultaneous production of several {\em different} species of dark 
matter particles. The global event variables describing the visible
particles are: the total energy $E$, the transverse components 
$P_x$ and $P_y$ and the longitudinal component $P_z$ 
of the total visible momentum $\vec{P}$. The only experimentally available
information regarding the invisible particles is the missing transverse
momentum $\mptvec$.
}
\label{fig:metevent} 
}
As seen from the figure, in defining $\sqrt{s}_{min}$, one 
imagines a completely general setup -- each event contains
some number $n_{vis}$ of Standard Model (SM) particles 
$X_i$, $i=1,2,\ldots, n_{vis}$, which are {\em visible} in the detector,
i.e. their energies and momenta are in principle measured.
Examples of such visible SM particles 
are the basic reconstructed objects, e.g.~jets, photons, electrons and muons.
The visible particles $X_i$ are denoted in Fig.~\ref{fig:metevent} 
with solid black lines and may originate either from ISR, 
or from the hard scattering and subsequent 
cascade decays (indicated with the green-shaded ellipse).
In turn, the missing transverse momentum $\mptvec$ 
arises from a certain number $n_{inv}$
of stable neutral particles $\chi_i$, $i=1,2,\ldots, n_{inv}$, 
which are {\em invisible} in the detector. 
In general, the set of invisible particles consists of some number
$n_{\chi}$ of BSM particles (indicated with the red dashed lines),
as well as some number $n_{\nu}=n_{inv}-n_{\chi}$ 
of SM neutrinos (denoted with the black dashed lines).
As already mentioned earlier, the $\mptvec$ measurement alone does 
not reveal the number $n_{inv}$ of missing particles, nor how 
many of them are neutrinos and how many are BSM (dark matter) particles.
This general setup also allows the identities and the masses
$m_i$ of the BSM invisible particles $\chi_i$, ($i=1,2,\ldots,n_{\chi}$)
in principle to be different, as in models with several {\em different} species of dark 
matter particles \cite{Hur:2007ur,Cao:2007fy,SungCheon:2008ts,Hur:2008sy}. 
Of course, the neutrino masses can be safely taken to be zero
\beq
m_i=0, \quad{\rm for}\ i=n_\chi+1,n_\chi+2,\ldots,n_{inv}\ .
\label{zeromnu}
\eeq

Given this very general setup, Ref.~\cite{Konar:2008ei}
asked the following question: What is the {\em minimum}
value $\sqrt{s}_{min}$ of the parton-level
Mandelstam invariant mass variable $\sqrt{s}$ which is consistent with the 
observed visible 4-momentum vector $P^\mu \equiv (E, \vec{P})$?
As it turned out, the answer to this question 
is given by the universal formula \cite{Konar:2008ei}
\beq
\sqrt{s}_{min}(\mmis) \equiv \sqrt{E^2-P_z^2}+\sqrt{\mmis^2+\mpt^2}\ ,
\label{smin_def}
\eeq
where the mass parameter $\mmis$ is nothing but 
the total mass of all invisible particles in the event:
\beq
\mmis \equiv \sum_{i=1}^{n_{inv}} m_i = \sum_{i=1}^{n_{\chi}} m_i\ ,
\label{minv}
\eeq
and the second equality follows from the assumption 
of vanishing neutrino masses (\ref{zeromnu}).
The result (\ref{smin_def}) can be equivalently rewritten 
in a more symmetric form
\begin{equation}
\sqrt{s}_{min}(\mmis) 
= \sqrt{M^2+P_T^2}+\sqrt{\mmis^2+\mpt^2}
\label{smin_def_mvis}
\end{equation}
in terms of the total visible invariant mass $M$ defined as
\beq
M^2\equiv E^2 - P_x^2- P_y^2 - P_z^2 \equiv E^2 - P_T^2 - P_z^2\, .
\eeq
Notice that in spite of the complete arbitrariness of the 
invisible particle sector at this point, the definition of
$\sqrt{s}_{min}$ depends on a single unknown parameter $\mmis$ - 
the {\em sum} of all the masses of the invisible particles 
in the event. For future reference, one should keep in mind that 
transverse momentum conservation at this point implies that 
\beq
\vec{P}_T + \mptvec = 0.
\label{PTcons}
\eeq

The main result from Ref.~\cite{Konar:2008ei} was that
in the absence of ISR and MPI, the peak in the $\sqrt{s}_{min}$
distribution nicely correlates with the
mass threshold of the newly produced particles.
This observation provides one generic relation between the 
total mass of the produced particles and the 
total mass $\mmis$ of the invisible particles.
Based on several SUSY examples involving fully 
hadronic signatures in symmetric as well as asymmetric topologies, 
Ref.~\cite{Konar:2008ei} showed that the accuracy of this measurement 
rivals the one achieved with the more 
traditional $M_{T2}$ methods.

\subsection{$\sqrt{s}_{min}$ and the underlying event problem}

At the same time, it was also recognized that effects
from the underlying event (UE), most notably ISR and MPI,
severely jeopardize this measurement.
The problem is that in the presence of the UE, 
the $\sqrt{s}_{min}$ variable would be measuring
{\em the total} energy of the full system shown in
Fig.~\ref{fig:metevent}, while for studying any new physics
we are mostly interested in the energy of the hard scattering,
as represented by the green-shaded ellipse in Fig.~\ref{fig:metevent}.
The inclusion of the UE causes a drastic shift of the peak of the $\sqrt{s}_{min}$ 
distribution to higher values, often by as much as a few TeV
\cite{Konar:2008ei,Papaefstathiou:2009hp,Papaefstathiou:2010ru}.
As a result, it appeared that unless effects from the underlying 
event could somehow be compensated for, the proposed measurement 
of the $\sqrt{s}_{min}$ peak would be of no practical value.

The main purpose of this paper is to propose two fresh
new approaches to dealing with the underlying event problem
which has plagued the $\sqrt{s}_{min}$ variable and
prevented its more widespread use in hadron collider 
physics applications. But before we discuss the two new ideas
put forth in this paper, we first briefly mention the two 
existing proposals in the literature on how to deal with 
the underlying event problem. 

First, it was recognized in Ref.~\cite{Konar:2008ei} that 
the contributions from the underlying event tend to be in 
the forward region, i.e. at large values of $|\eta|$.
Correspondingly, by choosing a suitable cut $|\eta|<\eta_{max}$,
designed to eliminate contributions from the very forward regions, 
one could in principle restore the proper behavior of 
the $\sqrt{s}_{min}$ distribution \cite{Konar:2008ei}. 
Unfortunately, there are no a priori guidelines 
on how to choose the appropriate
value of $\eta_{max}$, therefore this approach introduces an
uncontrollable systematic error and has not been pursued
further in the literature.

An alternative approach was proposed in 
Refs.~\cite{Papaefstathiou:2009hp,Papaefstathiou:2010ru},
which pointed out that the ISR effects on $\sqrt{s}_{min}$ 
are in principle calculable in QCD from first principles.
The calculations presented in Refs.~\cite{Papaefstathiou:2009hp,Papaefstathiou:2010ru}
could then be used to ``unfold'' the ISR effects and correct for the 
shift in the peak of the $\sqrt{s}_{min}$ distribution.
Unfortunately, in this analytical approach, the MPI effects 
would still be unaccounted for, and would have to be modeled 
and validated separately by some other means.
While such an approach may eventually bear fruit
at some point in the future, we shall not pursue it here.

We see that, for one reason or another, both of these
strategies appear unsatisfactory.
Therefore, here we shall pursue two different
approaches. We shall propose two new variants 
of the $\sqrt{s}_{min}$ variable, which we label
$\sqrt{s}_{min}^{(reco)}$ and $\sqrt{s}_{min}^{(sub)}$
and define in Secs.~\ref{sec:reco} and \ref{sec:sub}, 
correspondingly. We illustrate the properties of these 
two variables with several examples in 
Secs.~\ref{sec:top}-\ref{sec:gmsb}.
These examples will show that both 
$\sqrt{s}_{min}^{(reco)}$ and $\sqrt{s}_{min}^{(sub)}$
are unharmed by the effects from the underlying event,
thus resurrecting the original idea of Ref.~\cite{Konar:2008ei}
to use the peak in the $\sqrt{s}_{min}$ distribution as a first,
quick, model-independent estimate of the new physics mass scale.
In Section \ref{sec:comp} we compare the performance 
of $\sqrt{s}_{min}$ against some other inclusive variables which are 
commonly used in hadron collider physics for the purpose 
of estimating the new physics mass scale.
Section~\ref{sec:conclusions} is reserved for our 
main summary and conclusions.

\section{Definition of the RECO level variable $\sqrt{s}_{min}^{(reco)}$}
\label{sec:reco}

In the first approach, we shall {\em not} modify the original
definition of $\sqrt{s}_{min}$ and will continue to use the 
usual equation (\ref{smin_def}) (or its equivalent (\ref{smin_def_mvis})),
preserving the desired universal, global and inclusive 
character of the $\sqrt{s}_{min}$ variable. Then we shall
concentrate on the question, how should one calculate the 
observable quantities $E$, $\vec{P}$ and $\mpt$ entering 
the defining equations (\ref{smin_def}) and (\ref{smin_def_mvis}).

The previous $\sqrt{s}_{min}$ studies 
\cite{Konar:2008ei,Papaefstathiou:2009hp,Papaefstathiou:2010ru}
used calorimeter-based measurements of the 
total visible energy $E$ and momentum $\vec{P}$ as follows.
The total visible energy in the calorimeter $E_{(cal)}$ 
is simply a scalar sum over all calorimeter deposits
\beq
E_{(cal)} \equiv \sum_\alpha E_\alpha\ ,
\label{Ecal}
\eeq
where the index $\alpha$ labels the calorimeter towers,
and $E_\alpha$ is the energy deposit in the $\alpha$ tower.
As usual, since muons do not deposit significantly in 
the calorimeters, the measured $E_\alpha$ should first be corrected
for the energy of any muons which might be present in the event
and happen to pass through the corresponding tower $\alpha$. 
The three components of the total visible momentum $\vec{P}$
were also measured from the calorimeters as
\bea
P_{x(cal)} &=& \sum_\alpha E_\alpha \sin\theta_\alpha \cos\varphi_\alpha\ ,  \label{Pxcal}\\ [2mm]
P_{y(cal)} &=& \sum_\alpha E_\alpha \sin\theta_\alpha \sin\varphi_\alpha\ ,  \label{Pycal} \\ [2mm]
P_{z(cal)} &=& \sum_\alpha E_\alpha \cos\theta_\alpha\ ,  \label{Pzcal}
\eea
where $\theta_\alpha$ and $\varphi_\alpha$ are correspondingly 
the polar and azimuthal angular coordinates of the $\alpha$ calorimeter tower.
The missing transverse momentum can similarly be measured 
from the calorimeter as (see eq.~(\ref{PTcons}))
\beq
\not \!\! \vec{P}_{\,T(cal)} \equiv - \,\vec{P}_{T(cal)}.
\label{metcal}
\eeq
Using these calorimeter-based measurements (\ref{Ecal}-\ref{metcal}),
one can make the identification
\begin{eqnarray}
E &\equiv & E_{(cal)}        \, ,           \label{Ecalid} \\ [2mm]
\vec{P} &\equiv & \vec{P}_{(cal)}\, ,       \label{Pcalid} \\ [2mm]
\mptvec &\equiv & \not \!\! \vec{P}_{\,T(cal)} \label{METcalid} 
\end{eqnarray}
in the definition (\ref{smin_def})
and construct the corresponding ``calorimeter-based'' 
$\sqrt{s}_{min}$ variable as
\beq
\sqrt{s}_{min}^{(cal)}(\mmis) \equiv \sqrt{E_{(cal)}^2-P_{z(cal)}^2}+\sqrt{\mmis^2
+\not \!\! P_{T(cal)}^2}\ .
\label{smin_def_cal}
\eeq
This was precisely the quantity which was studied 
in \cite{Konar:2008ei,Papaefstathiou:2009hp,Papaefstathiou:2010ru}
and shown to exhibit extreme sensitivity to the physics of the 
underlying event.

Here we propose to evaluate the visible quantities $E$ and $\vec{P}$ 
at the RECO level, i.e. in terms of the reconstructed objects, namely
jets, muons, electrons and photons\footnote{This possibility was
briefly alluded to in \cite{Konar:2008ei}, but not pursued in any detail.}. 
To be precise, let there be
$N_{obj}$ reconstructed objects in the event, with 
energies $E_i$ and 3-momenta $\vec{P}_i$, $i=1,2,\ldots,N_{obj}$,
correspondingly. Then in place of (\ref{Ecalid}-\ref{METcalid}),
let us instead identify
\begin{eqnarray}
E &\equiv & E_{(reco)} \equiv \sum_{i=1}^{N_{obj}} E_i           \, ,        \label{Erecoid} \\ [2mm]
\vec{P} &\equiv & \vec{P}_{(reco)}  \equiv   \sum_{i=1}^{N_{obj}} \vec{P}_i \, ,   \label{Precoid} \\ [2mm]
\mptvec &\equiv & \not \!\! \vec{P}_{\,T(reco)}
                = - \vec{P}_{\,T(reco)}
\, , \label{METrecoid} 
\end{eqnarray}
and correspondingly define a ``RECO-level''
$\sqrt{s}_{min}$ variable as
\beq
\sqrt{s}_{min}^{(reco)}(\mmis) \equiv \sqrt{E_{(reco)}^2-P_{z(reco)}^2}+\sqrt{\mmis^2
+\not \!\! P_{T(reco)}^2}\ ,
\label{smin_def_reco}
\eeq
which can also be rewritten in analogy to (\ref{smin_def_mvis}) as
\beq
\sqrt{s}_{min}^{(reco)}(\mmis) \equiv \sqrt{M_{(reco)}^2+P_{T(reco)}^2}+\sqrt{\mmis^2
+\not \!\! P_{T(reco)}^2}\ ,
\label{smin_def_reco_mvis}
\eeq
where $\not \!\! P_{T(reco)}$ and $P_{T(reco)}$ are related as in eq.~(\ref{METrecoid}) and
the RECO-level total visible mass $M_{(reco)}$ is defined by
\beq
M^2_{(reco)} \equiv E_{(reco)}^2 - \vec{P}_{(reco)}^2\, .
\label{Mreco}
\eeq

What are the benefits from the new RECO-level $\sqrt{s}_{min}$ definitions 
(\ref{smin_def_reco},\ref{smin_def_reco_mvis}) in
comparison to the old calorimeter-based $\sqrt{s}_{min}$
definition in (\ref{smin_def_cal})?
In order to understand the basic idea, it is worth comparing 
the calorimeter-based missing transverse momentum $\mpt$
(which in the literature is commonly referred to as 
``missing transverse energy'' $\met$)
and the analogous RECO-level variable $\mht$, the ``missing $H_T$''.
The $\mhtvec$ vector is defined as the negative of 
the vector sum of the transverse momenta 
of all reconstructed objects in the event:
\beq
\mhtvec \equiv - \sum_{i=1}^{N_{obj}} \vec{P}_{Ti}\, .
\label{MHTvec}
\eeq
Then it is clear that in terms of our notation here, 
$\mht$ is nothing but $\not \!\! P_{T(reco)}$.

It is known that $\mht$ performs better than $\met$ \cite{GL}.
First, $\mht$ is less affected by a number of adverse
instrumental factors such as: electronic noise, 
faulty calorimeter cells, pile-up, etc. 
These effects tend to populate the calorimeter 
uniformly with unclustered energy, which will later
fail the basic quality cuts during object reconstruction.
In contrast, the {\em true} missing momentum
is dominated by clustered energy, which will be 
successfully captured during reconstruction. 
Another advantage of $\mht$ is that one can 
easily apply the known jet energy corrections 
to account for the nonlinear detector response.
For both of these reasons, CMS is now using
$\mht$ at both the trigger level and offline
\cite{GL}.

Now realize that $\sqrt{s}_{min}^{(cal)}$ is analogous
to the calorimeter-based $\met$, while our new variable
$\sqrt{s}_{min}^{(reco)}$ is analogous to the 
RECO-level $\mht$. Thus we may already expect that 
$\sqrt{s}_{min}^{(reco)}$ will inherit the advantages
of $\mht$ and will be better suited for determining the 
new physics mass scale than the calorimeter-based quantity
$\sqrt{s}_{min}^{(cal)}$. This expectation is confirmed 
in the explicit examples studied below in Secs.~\ref{sec:top}
and \ref{sec:susy}. Apart from the already mentioned 
instrumental issues, the most important advantage of 
$\sqrt{s}_{min}^{(reco)}$ from the physics point of view
is that it is much less sensitive to the effects from the 
underlying event, which had doomed its
calorimeter-based $\sqrt{s}_{min}^{(cal)}$ cousin.

Strictly speaking, the idea of $\sqrt{s}_{min}^{(reco)}$
does not solve the underlying event problem completely
and as a matter of principle. 
Every now and then the underlying event will still
produce a well-defined jet, which will have to be included 
in the calculation of $\sqrt{s}_{min}^{(reco)}$.
Because of this effect, we cannot any more guarantee that
$\sqrt{s}_{min}^{(reco)}$ provides a lower bound on the true
value $\sqrt{s}_{true}$ of the center-of-mass energy of the
hard scattering --- the additional jets formed out of ISR, pile-up, and so on,
will sometimes cause $\sqrt{s}_{min}^{(reco)}$ to exceed
$\sqrt{s}_{true}$. Nevertheless we find that this effect
modifies only the shape of the $\sqrt{s}_{min}^{(reco)}$
distribution, but leaves the location of its peak largely 
intact. To the extent that one is mostly interested in the 
peak location, $\sqrt{s}_{min}^{(reco)}$ should already be 
good enough for all practical purposes.

\section{Definition of the subsystem variable $\sqrt{s}_{min}^{(sub)}$}
\label{sec:sub}

In this section we propose an alternative modification
of the original $\sqrt{s}_{min}$ variable, which solves the 
underlying event problem completely and as a matter of 
principle. The downside of this approach is that it is not as general and 
universal as the one discussed in the previous section,
and can be applied only in cases where one can 
unambiguously identify a subsystem of the original event 
topology which is untouched by the underlying event.
The basic idea is schematically illustrated in Fig.~\ref{fig:subevent},
which is nothing but a slight rearrangement of Fig.~\ref{fig:metevent}
exhibiting a well defined subsystem (delineated by the black rectangle). 
\FIGURE[ht]{
{
\unitlength=1.5 pt
\SetScale{1.5}
\SetWidth{1.0}      
\normalsize    
{} \qquad\allowbreak
\begin{picture}(300,200)(0,0)
\SetColor{Gray}
\Line( 13,185)(125,185)
\Line( 50,150)(125,150)
\Line(110,130)(170,130)
\Line(110,100)(170,100)
\Text(175,100)[l]{\Black{$X_1$}}
\Text(175,130)[l]{\Black{$X_{n_{sub}}$}}
\Text(130,150)[l]{\Black{$X_{n_{sub}+1}$}}
\Text(130,185)[l]{\Black{$X_{n_{vis}}$}}
\DashLine(110, 90)(170, 90){2}
\DashLine(110, 65)(170, 65){2}
\Text(175, 90)[l]{$\chi_{n_{inv}}$}
\Text(175, 65)[l]{$\chi_{n_{\chi}+1}$}
\Line( 10,190)(55,120)
\Line( 10, 10)(55, 80)
\Text(20,155)[c]{\Black{$p(\bar{p})$}}
\Text(20, 45)[c]{\Black{$p(\bar{p})$}}
\DashLine(95,180)(95,155){1}
\DashLine(145, 85)(145, 70){1}
\DashLine(145,105)(145,125){1}
\CArc(10,167.5)(150,-9,9)
\LongArrow(167,167.5)(197,167.5)
\CArc(47,115)(150,-7,7)
\LongArrow(200,112.5)(230,112.5)
\SetColor{Red}
\DashLine(145, 35)(145, 50){1}
\CArc(55,   60)(144,-13,13)
\LongArrow(205,60)(235,60)
\SetWidth{1.2}      
\DashLine(110,55)(170,55){2}
\DashLine(110,30)(170,30){2}
\Text(175,55)[l]{\Red{$\chi_{n_\chi}$}}
\Text(175,30)[l]{\Red{$\chi_1$}}
\CBoxc(231,167.5)(55,20){Gray}{Yellow}
\Text(232,167.5)[c]{$E_{(up)},\vec{P}_{\,(up)}$}
\CBoxc(260,115)(50,20){Gray}{Yellow}
\Text(260,115)[c]{$E_{(sub)},\vec{P}_{(sub)}$}
\CBoxc(255, 60)(25,20){Red}{Yellow}
\Text(255,60)[c]{\Red{$\mptvec$}}
\SetColor{Blue}
\Line( 50,115)(110,115)
\Line( 50, 60)(110, 60)
\Line( 50, 45)(110, 45)
\DashLine(78,75)(78,110){1}
\Text(78, 51)[c]{\Blue{$P_1$}}
\Text(78, 66)[c]{\Blue{$P_2$}}
\Text(78,121)[c]{\Blue{$P_{n_p}$}}
\COval(50,100)(75,15)(0){Blue}{Green}
\COval(110, 80)(55,10)(0){Blue}{Green}
\SetColor{Black}
\Boxc(192, 80)(195,120)
\rText(110,80)[c][l]{$\sqrt{s}^{(sub)}$}
\end{picture}
}
\caption{A rearrangement of Fig.~\ref{fig:metevent}
into an event topology exhibiting a well defined subsystem
(delineated by the black rectangle) with total invariant mass
$\sqrt{s}^{(sub)}$. There are
$n_{sub}$ visible particles $X_i$, $i=1,2,\ldots, n_{sub}$, 
originating from within the subsystem, while the remaining 
$n_{vis}-n_{sub}$ visible particles $X_{n_{sub}+1},\ldots,X_{n_{vis}}$
are created upstream, outside the subsystem. 
The subsystem results from the production and 
decays of a certain number of parent particles $P_j$, $j=1,2,\ldots,n_p$,
(some of) which may decay semi-invisibly. All invisible particles $\chi_1,\ldots,\chi_{n_{inv}}$ 
are then assumed to originate from within the subsystem. 
}
\label{fig:subevent} 
}
The original $n_{vis}$ visible particle $X_i$
from Fig.~\ref{fig:metevent} have now been 
divided into two groups as follows:
\begin{enumerate}
\item There are $n_{sub}$ visible particles $X_1,\ldots,X_{n_{sub}}$
originating from within the subsystem. Their total energy and total 
momentum are denoted by $E_{(sub)}$ and $\vec{P}_{(sub)}$, correspondingly.
The subsystem particles are chosen so that to guarantee 
that they could not have come from the underlying event.
\item The remaining $n_{vis}-n_{sub}$ visible particles 
$X_{n_{sub}+1},\ldots,X_{n_{vis}}$ are created upstream
(outside the subsystem)
and have total energy $E_{(up)}$ and total momentum 
$\vec{P}_{(up)}$. The upstream particles may originate from
the underlying event or from decays of heavier particles
upstream -- this distinction is inconsequential at this point.
\end{enumerate}
We also assume that all invisible particles $\chi_1,\ldots,\chi_{n_{inv}}$ 
originate from within the subsystem, 
i.e. that no invisible particles are created upstream.
In effect, all we have done in Fig.~\ref{fig:subevent} is to
partition the original measured values of the total visible 
energy $E$ and 3-momentum $\vec{P}$
from Fig.~\ref{fig:metevent} into two separate components as
\begin{eqnarray}
E &=& E_{(up)} + E_{(sub)}\, ,  \label{Epartition} \\ [2mm]
\vec{P} &=& \vec{P}_{(up)} + \vec{P}_{(sub)}\, .  \label{Ppartition} 
\end{eqnarray}
Notice that now the missing transverse momentum is defined as
\beq
\mptvec \equiv -\vec{P}_{T(up)} - \vec{P}_{T(sub)}\, ,
\label{PTconssub}
\eeq
while the total visible invariant mass $M_{(sub)}$ of the subsystem is
given by
\beq
M_{(sub)}^2 = E_{(sub)}^2 - \vec{P}_{(sub)}^2\, .
\label{Mvissub}
\eeq

At this point the reader may be wondering what are the guiding 
principles for categorizing a given visible particle $X_i$ 
as a subsystem or an upstream particle. Since our goal is to
identify a subsystem which is shielded from the effects of the 
underlying event, the safest way to do the partition 
of the visible particles is to require that all QCD jets 
belong to the upstream particles, while the subsystem particles 
consist of objects which are unlikely to come from the underlying 
event, such as isolated electrons, photons and muons (and possibly
identified $\tau$-jets and, to a lesser extent, tagged $b$-jets).

With those preliminaries, we are now ready to ask the
usual $\sqrt{s}_{min}$ question: Given the measured values of
$E_{(up)}$, $E_{(sub)}$, $\vec{P}_{(up)}$ and $\vec{P}_{(sub)}$,
what is the minimum value $\sqrt{s}_{min}^{(sub)}$ of the 
{\em subsystem} Mandelstam invariant mass variable $\sqrt{s}^{(sub)}$, 
which is consistent with those measurements?
Proceeding as in \cite{Konar:2008ei}, once again we 
find a very simple universal answer, which, with the help of 
(\ref{PTconssub}) and (\ref{Mvissub}), can be equivalently 
written in several different ways as follows:
\begin{eqnarray}
\sqrt{s}_{min}^{(sub)}(\mmis) 
&=& 
\left\{ \left(\sqrt{E_{(sub)}^2-P_{z(sub)}^2}+\sqrt{\mmis^2+\mpt^2}\right)^2
-P_{T(up)}^2 \right\}^{\frac{1}{2}} \label{smin_def_sub} \\ [2mm]
&=& 
\left\{ \left(\sqrt{M_{(sub)}^2+P_{T(sub)}^2}+\sqrt{\mmis^2+\mpt^2}\right)^2
-P_{T(up)}^2 \right\}^{\frac{1}{2}} \label{smin_def_sub_mvis} \\ [2mm]
&=& 
\left\{ \left(\sqrt{M_{(sub)}^2+P_{T(sub)}^2}+\sqrt{\mmis^2+\mpt^2}\right)^2
-(\vec{P}_{T(sub)}+\mptvec)^2 \right\}^{\frac{1}{2}} \label{smin_def_sub_mvis2} \\ [2mm]
&=& ||p_{T(sub)} + \not\!\! p_T || \, , \label{smin_def_sub_pt}
\end{eqnarray}
where in the last line we have introduced the Lorentz 1+2 vectors
\begin{eqnarray}
p_{T(sub)}    &\equiv & \left( \sqrt{M_{(sub)}^2+P_{T(sub)}^2}\, ,\,  \vec{P}_{T(sub)} \right)\, ; \\ [2mm]
\not\!\! p_T &\equiv & \left( \sqrt{\mmis^2    +\mpt^2     }\, ,\, \mptvec \right)\, .
\end{eqnarray}
As usual, the length of a 1+2 vector is computed as
$||p|| = \sqrt{p\cdot p}=\sqrt{p_0^2-p_1^2-p_2^2}$.

Before we proceed to the examples of the next few sections, 
as a sanity check of the obtained result
it is useful to consider some limiting cases. 
First, by taking the upstream visible particles to be an empty set,
i.e. $\vec{P}_{T(up)}\to 0$, 
we recover the usual expression for $\sqrt{s}_{min}$
given in eqs.~(\ref{smin_def},\ref{smin_def_mvis}).
Next, consider a case with no invisible particles, i.e.
$\mmis=0$ and correspondingly, $\mptvec=0$.
In that case we obtain that $\sqrt{s}_{min}^{(sub)}=M_{(sub)}$,
which is of course the correct result.
Finally, suppose that there are no visible subsystem particles,
i.e. $E_{(sub)}=\vec{P}_{(sub)}=M_{(sub)}=0$.
In that case we obtain $\sqrt{s}_{min}^{(sub)}=\mmis$,
which is also the correct answer.

As we shall see, the subsystem concept of Fig.~\ref{fig:subevent}
will be most useful when the subsystem results from
the production and decays of a certain number $n_p$ of parent 
particles $P_j$ with masses $M_{P_j}$, $j=1,2,\ldots,n_p$,
correspondingly. Then the total combined mass of all parent particles 
is given by
\beq
M_p \equiv \sum_{j=1}^{n_p} M_{P_{j}}\, .
\label{Mp}
\eeq
By the conjecture of ref.~\cite{Konar:2008ei}, 
the location of the peak of the $\sqrt{s}_{min}^{(sub)}(\mmis)$
distribution will provide an approximate measurement of $M_p$ 
as a function of the unknown parameter $\mmis$. By construction,
the obtained relationship $M_p(\mmis)$ will then be completely
insensitive to the effects from the underlying event.

At this point it may seem that by excluding all QCD jets from the 
subsystem, we have significantly narrowed down the number of potential 
applications of the $\sqrt{s}_{min}^{(sub)}$ variable. Furthermore, 
we have apparently reintroduced a certain amount of model-dependence 
which the original $\sqrt{s}_{min}$ approach was trying so hard to avoid.
Those are in principle valid objections, which can be overcome by
using the $\sqrt{s}_{min}^{(reco)}$ variable introduced in the previous 
section. Nevertheless, we feel that the $\sqrt{s}_{min}^{(sub)}$
variable can prove to be useful in its own right, and in a wide variety of contexts.
To see this, note that the typical hadron collider signatures of 
the most popular new physics models (supersymmetry, extra dimensions, 
Little Higgs, etc.) are precisely of the form exhibited in 
Fig.~\ref{fig:subevent}. One typically considers production
of colored particles (squarks, gluinos, KK-quarks, etc.) whose
cross-sections dominate. In turn, these colored particles 
shed their color charge by emitting jets and decaying to
lighter, uncolored particles in an electroweak sector.
The decays of the latter often involve electromagnetic objects, 
which could be targeted for selection in the subsystem. 
The $\sqrt{s}_{min}^{(sub)}$ variable would then be the perfect 
tool for studying the mass scales in the electroweak sector
(in the context of supersymmetry, for example, the 
electroweak sector is composed of
the charginos, neutralinos and sleptons).

Before we move on to some specific examples illustrating these ideas, 
one last comment is in order. One may wonder whether the $\sqrt{s}_{min}^{(sub)}$
variable should be computed at the RECO-level or from the calorimeter.
Since the subsystem will usually be defined in terms of reconstructed objects, 
the more logical option is to calculate $\sqrt{s}_{min}^{(sub)}$
at the RECO-level and label it as $\sqrt{s}_{min}^{(sub,reco)}$.
However, to streamline our notation, 
in what follows we shall always omit the ``reco'' part 
of the superscript and will always implicitly assume that 
$\sqrt{s}_{min}^{(sub)}$ is computed at RECO-level.


\section{SM example: dilepton events from $t\bar{t}$ production}
\label{sec:top}

In this and the next two sections we illustrate the properties of the 
new variables $\sqrt{s}_{min}^{(reco)}$ and $\sqrt{s}_{min}^{(sub)}$
with some specific examples. In this section we discuss an example 
taken from the Standard Model, which is guaranteed to be available 
for early studies at the LHC. We consider dilepton events from $t\bar{t}$ 
pair production, where both $W$'s decay leptonically. In this event 
topology, there are two missing particles (two neutrinos).
Therefore, these events very closely resemble the typical SUSY-like 
events, in which there are two missing dark matter particles.
In the next two sections, we shall also consider some SUSY examples.
In all cases, we perform detailed event simulation, including 
the effects from the underlying event and detector resolution.

\subsection{Event simulation details}
\label{sec:simulation}

Events are generated with PYTHIA \cite{Sjostrand:2006za}
(using its default model of the underlying event)
at an LHC of 14 TeV, 
and then reconstructed with the PGS detector simulation package
\cite{PGS}. We have made certain modifications in the publicly 
available version of PGS to better match it to the CMS detector.
For example, we take the hadronic calorimeter resolution to be
\cite{Bayatian:2006zz}
\beq
\frac{\sigma}{E} = \frac{120\%}{\sqrt{E}}\, ,
\eeq
while the electromagnetic calorimeter resolution is
\cite{Bayatian:2006zz}
\beq
\left(\frac{\sigma}{E}\right)^2 
=  \left(\frac{S}{\sqrt{E}}\right)^2
+  \left(\frac{N}{E}\right)^2
+C^2\, ,
\eeq
where the energy $E$ is measured in GeV,
$S=3.63\%$ is the stochastic term, 
$N=0.124$ is the noise and $C=0.26\%$ is the constant term.
Muons are reconstructed within $|\eta|<2.4$, and
we use the muon global reconstruction efficiency 
quoted in \cite{Bayatian:2006zz}.
We use default $p_T$ cuts on the reconstructed objects as follows:
$3$ GeV for muons, $10$ GeV for electrons and photons, and 
$15$ GeV for jets. 

For the $t\bar{t}$ example presented in this section, we use 
the approximate next-to-next-to-leading order $t\bar{t}$ cross-section
of $\sigma_{t\bar{t}}=894\pm 4^{+73+12}_{-46-12}$ pb at a top mass of $m_t=175$ GeV
\cite{Kidonakis:2008mu}. For the SUSY examples in the next two 
sections we use leading order cross-sections.

Since our examples are meant for illustration purposes only, 
we do not include any backgrounds to the processes being considered,
nor do we require any specific triggers. 
A detailed study of the dilepton $t\bar{t}$ signature
including all those effects will appear elsewhere \cite{KKMP}.

\subsection{$\sqrt{s}_{min}^{(reco)}$ variable}

We first consider SUSY-like missing energy events
arising from $t\bar{t}$ production, where each
$W$-boson is forced to decay leptonically (to an electron 
or a muon). We do not impose any trigger or offline
requirements, and simply plot directly the output from 
PGS\footnote{Therefore, our plots in this subsection are normalized to 
a total number of events equal to $\sigma_{t\bar{t}}\times
BR(W\to e,\mu)^2$.}. We show various $\sqrt{s}$ 
quantities of interest in Fig.~\ref{fig:ttb}, setting $\mmis=0$, 
since in this case the missing particles are neutrinos and 
are massless.
\FIGURE[t]{
\centerline{
\epsfig{file=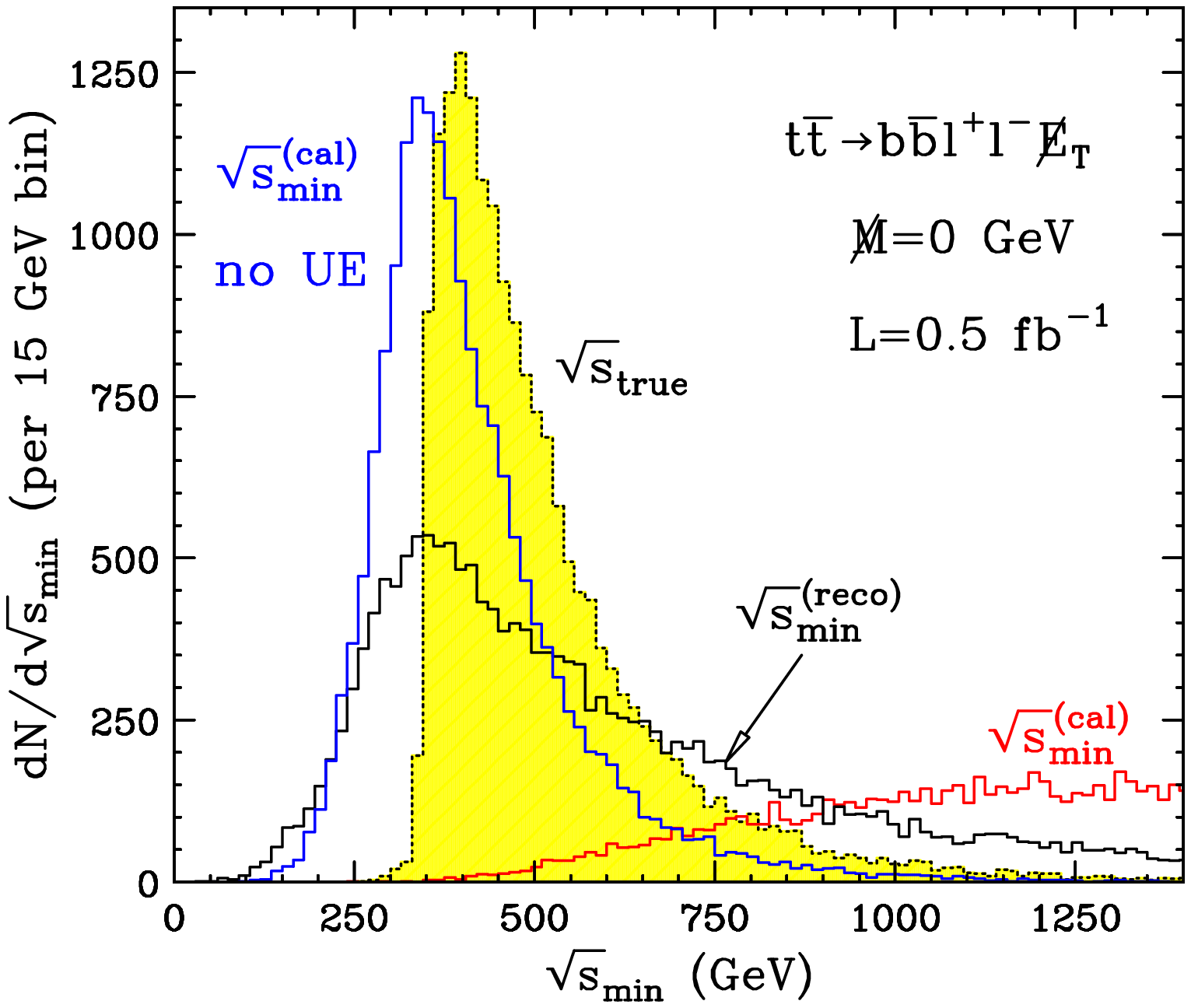,width=10cm}
}
\caption{Distributions of various $\sqrt{s}_{min}$ quantities
discussed in the text, for the dilepton
$t\bar{t}$ sample at the LHC with 14 TeV CM energy and 
0.5 ${\rm fb}^{-1}$ of data. The dotted (yellow-shaded) histogram 
gives the true $\sqrt{s}$ distribution of the $t\bar{t}$ pair.
The blue histogram is the distribution of the calorimeter-based
$\sqrt{s}_{min}^{(cal)}$ variable in the ideal case when all
effects from the underlying event are turned off.
The red histogram shows the corresponding result for 
$\sqrt{s}_{min}^{(cal)}$ in the presence of the underlying event.
The black histogram is the distribution of the 
$\sqrt{s}_{min}^{(reco)}$ variable introduced in Sec.~\ref{sec:reco}.
All $\sqrt{s}_{min}$ distributions are shown for $\mmis=0$.
}
\label{fig:ttb}
}
The dotted (yellow-shaded) histogram 
represents the true $\sqrt{s}$ distribution of the $t\bar{t}$ pair.
It quickly rises at the $t\bar{t}$ mass threshold
\beq
M_p\equiv 2m_t = 350\ {\rm GeV}
\eeq
and then eventually falls off at large $\sqrt{s}$
due to the parton density function suppression.
Because the top quarks are typically produced with some boost, the 
$\sqrt{s}_{true}$ distribution in Fig.~\ref{fig:ttb}
peaks a little bit above threshold:
\beq
\left(\sqrt{s}_{true}\right)_{peak} > M_p\, .
\eeq
It is clear that if one could directly measure the $\sqrt{s}_{true}$ 
distribution, or at least its onset, the $t\bar{t}$ mass scale
will be easily revealed. Unfortunately, 
the escaping neutrinos make such a measurement impossible,
unless one is willing to make additional model-dependent 
assumptions\footnote{For example, 
one can use the known values of the neutrino, $W$ and top masses 
to solve for the neutrino kinematics (up to discrete ambiguities). 
However, this method assumes that the full
mass spectrum is already known, and furthermore, uses the
knowledge of the top decay topology to perfectly solve the 
combinatorics problem discussed in the Introduction. As an example, 
consider a case where the lepton is produced first and the 
$b$-quark second, i.e. when the top first decays to a lepton and
a leptoquark, which in turn decays to a neutrino and a $b$-quark.
The kinematic method would then be using the wrong on-shell 
conditions. The advantage of the $\sqrt{s}_{min}$ approach 
is that it is fully inclusive and does not make any reference 
to the actual decay topology.}.

Fig.~\ref{fig:ttb} also shows two versions of the calorimeter-based
$\sqrt{s}_{min}^{(cal)}$ variable: the blue (red) histogram is 
obtained by switching off (on) the underlying event (ISR and MPI).
These curves reveal two very interesting phenomena. First,
without the UE, the peak of the $\sqrt{s}_{min}^{(cal)}$ 
distribution (blue histogram) is very close to the parent 
mass threshold \cite{Konar:2008ei}:
\beq
{\rm no \ UE}\ \Longrightarrow \ \left(\sqrt{s}_{min}^{(cal)}\right)_{peak} \approx M_p\, .
\label{noUEpeak}
\eeq
The main observation of Ref.~\cite{Konar:2008ei}
was that this correlation offers an alternative, 
fully inclusive and model-independent, method of 
estimating the mass scale $M_p$ of the parent particles, 
even when some of their decay products are invisible and not
seen in the detector. 

Unfortunately, the ``no UE'' limit of eq.~(\ref{noUEpeak}) is unphysical, and
the corresponding $\sqrt{s}_{min}^{(cal)}$ distribution 
(blue histogram in in Fig.~\ref{fig:ttb}) is unobservable.
What is worse, when one tries to measure the $\sqrt{s}_{min}^{(cal)}$
distribution in the presence of the UE (red histogram in  Fig.~\ref{fig:ttb}),
the resulting peak is very far from the physical threshold:
\beq
{\rm with \ UE}\ \Longrightarrow \ \left(\sqrt{s}_{min}^{(cal)}\right)_{peak} \gg M_p\, .
\label{withUEpeak}
\eeq
In the $t\bar{t}$ example of Fig.~\ref{fig:ttb}, the shift 
is on the order of 1 TeV! It appears therefore that in practice
the $\sqrt{s}_{min}^{(cal)}$ peak would be uncorrelated with any 
physical mass scale, and instead would be completely determined
by the (uninteresting) physics of the underlying event.
Once the nice model-independent correlation of eq.~(\ref{noUEpeak})
is destroyed by the UE, it becomes of only academic value 
\cite{Konar:2008ei,Papaefstathiou:2009hp,
Papaefstathiou:2010ru,Barr:2010zj,Brooijmans:2010tn}.

However, Fig.~\ref{fig:ttb} also suggests the solution 
to this difficult problem. If we look at the 
distribution of the $\sqrt{s}_{min}^{(reco)}$ 
variable (black solid histogram), we see that 
its peak has returned to the desired value:
\beq
\left(\sqrt{s}_{min}^{(reco)}\right)_{peak} \approx M_p\, ,
\label{recopeak}
\eeq
thus resurrecting the original proposal of Ref.~\cite{Konar:2008ei}.
In order to measure physical mass thresholds, one simply needs to 
investigate the distribution of the inclusive $\sqrt{s}_{min}^{(reco)}$ 
variable, which is calculated at RECO-level. 
Each peak in that distribution signals the opening of a 
new channel, and from (\ref{recopeak}) the location of the peak 
provides an immediate estimate of the total mass of all particles 
involved in the production. Of course, the $\sqrt{s}_{min}^{(reco)}$
distribution is now not as sharply peaked as the unphysical
``no UE'' case of $\sqrt{s}_{min}^{(cal)}$, but as long as
its peak is found in the right location, the method of 
Ref.~\cite{Konar:2008ei} becomes viable once again.

Our first main result is therefore nicely summarized in
Fig.~\ref{fig:ttb}, which shows a total of 4 distributions, 
3 of which are either 
unphysical (the blue histogram of $\sqrt{s}_{min}^{(cal)}$ 
{\em in the absence} of the UE), 
unobservable (the yellow-shaded histogram of $\sqrt{s}_{true}$), 
or useless (the red histogram of $\sqrt{s}_{min}^{(cal)}$ 
{\em in the presence} of the UE). 
The only distribution in Fig.~\ref{fig:ttb} which is physical, 
observable and useful at the same time, is the 
distribution of $\sqrt{s}_{min}^{(reco)}$ (solid black histogram).

\FIGURE[ht]{
\centerline{
\epsfig{file=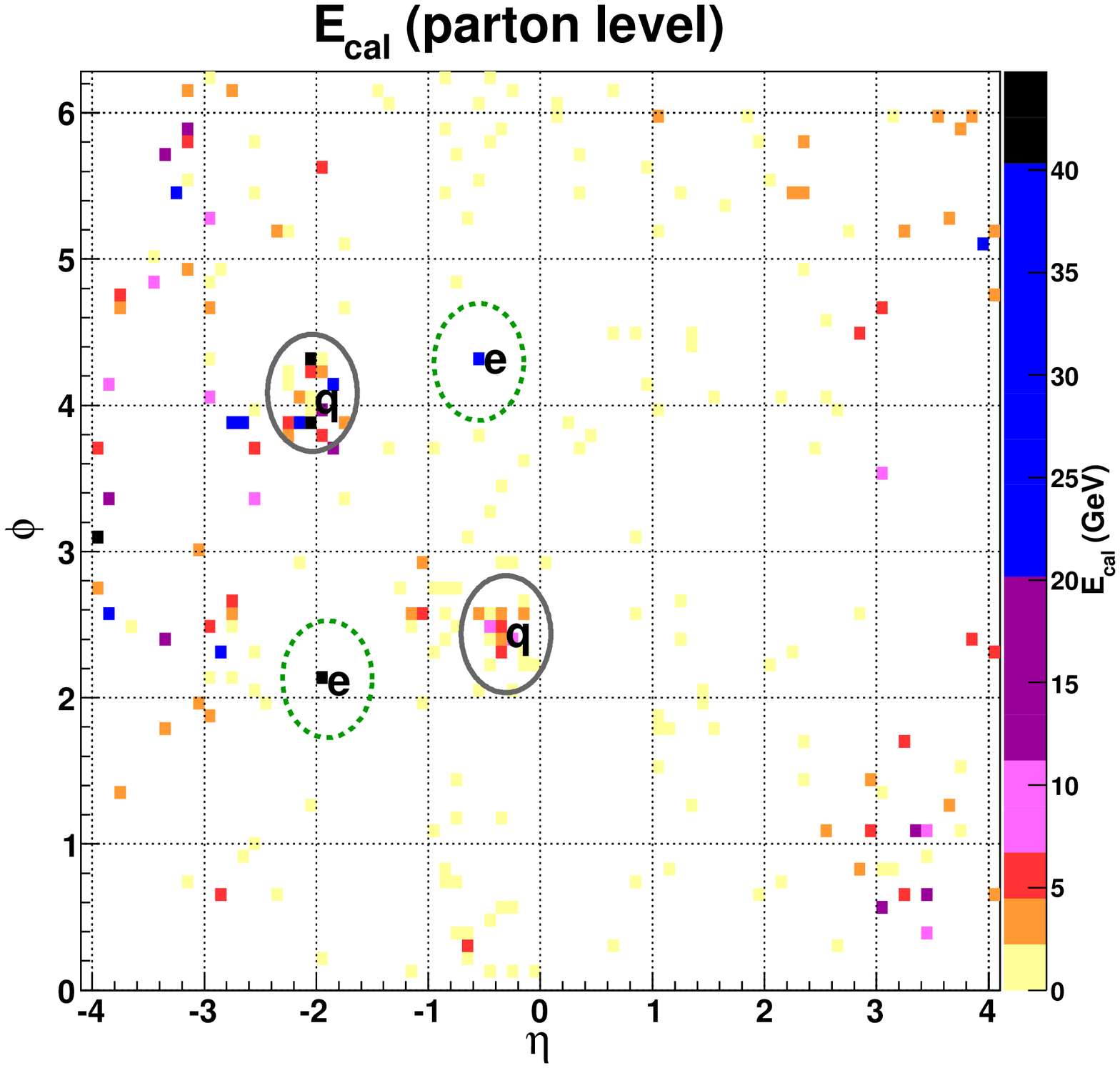,height=7.7cm}
\epsfig{file=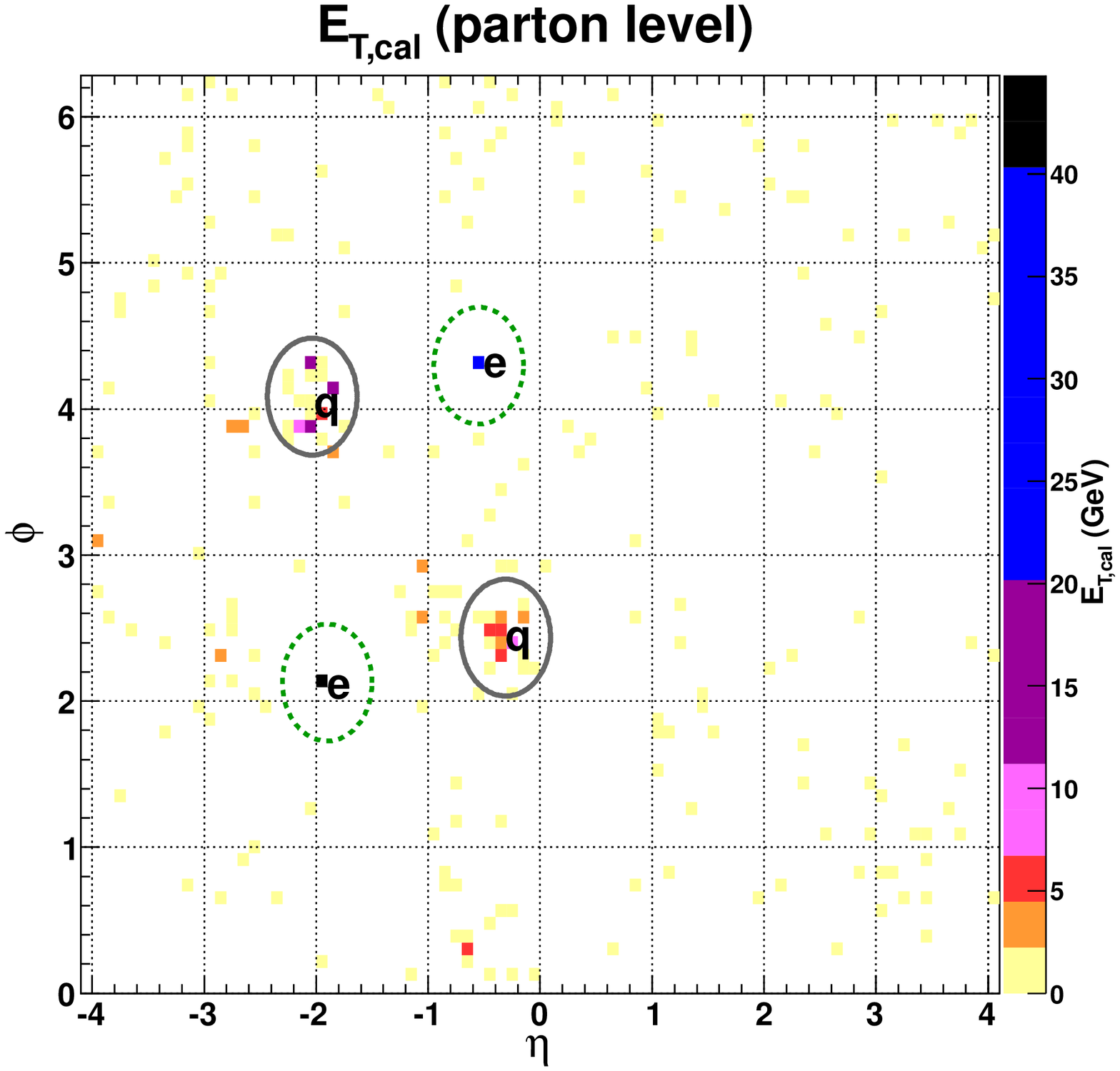,height=7.7cm}
}\\
\centerline{
\epsfig{file=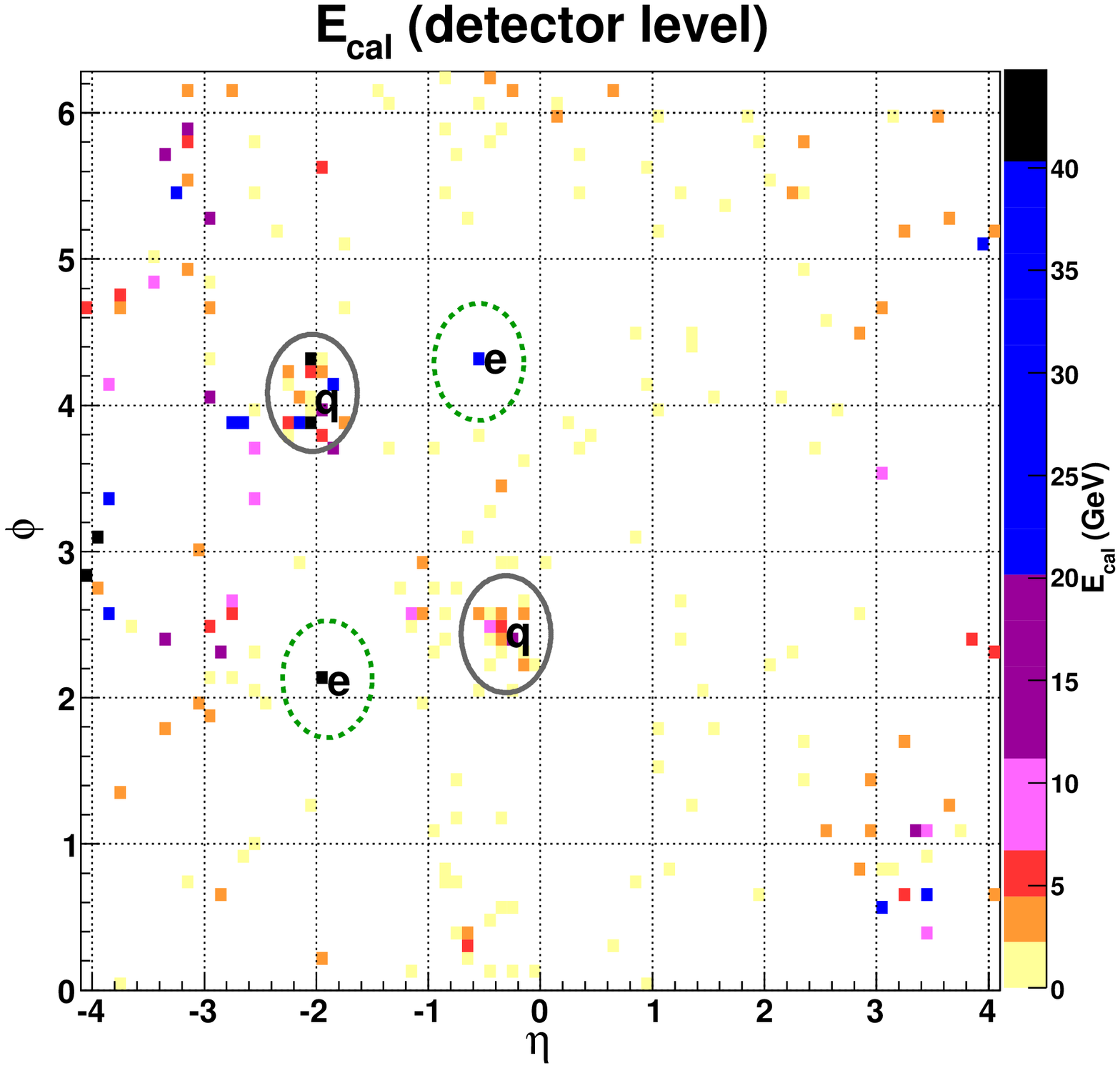,height=7.7cm}
\epsfig{file=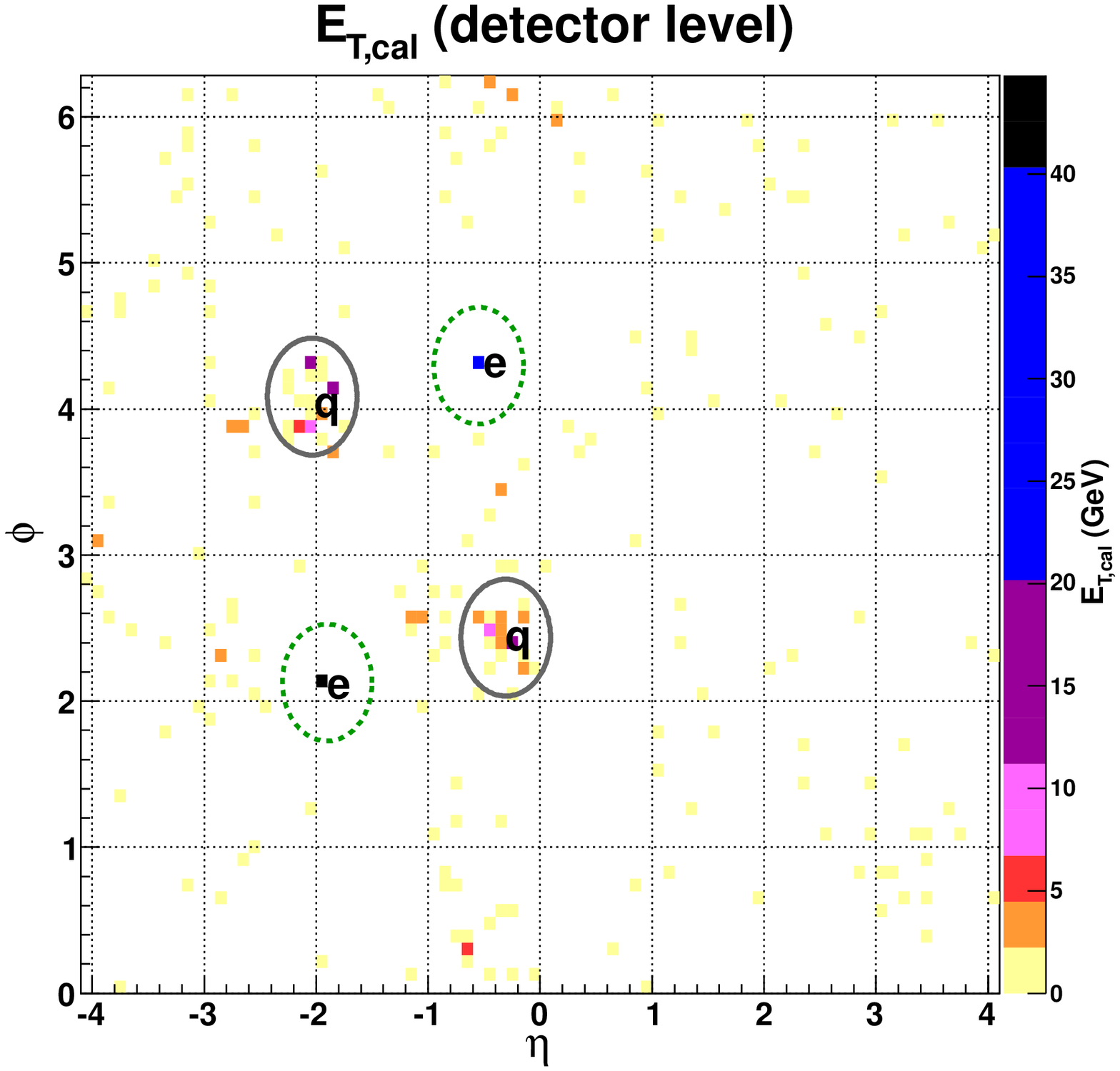,height=7.7cm}
}
\caption{PGS calorimeter map of the energy deposits, 
as a function of pseudorapidity $\eta$ and 
azimuthal angle $\phi$, for a dilepton
$t\bar{t}$ event with only two reconstructed jets.
At the parton level, this particular event has two $b$-quarks
and two electrons. The location of a $b$-quark (electron, muon)
is marked with the letter ``q'' (``e'', ``$\mu$'').
A grey circle delineates (the cone of) a 
reconstructed jet, while a green dotted circle
denotes a reconstructed lepton.
In the upper two plots the calorimeter is filled 
at the parton level directly from PYTHIA, while the lower 
two plots contain results after PGS simulation. 
The left plots show absolute energy deposits
$E_\alpha$, while in the right plots the energy 
in each tower is shown projected on the transverse plane
as $E_\alpha \cos\theta_\alpha$. 
}
\label{fig:ttb_cal1}
}

Before concluding this subsection, we explain the 
reason for the improved performance of the $\sqrt{s}_{min}^{(reco)}$
variable in comparison to the $\sqrt{s}_{min}^{(cal)}$ version.
As already anticipated in Sec.~\ref{sec:reco}, the basic idea
is that energy deposits which are due to hard particles 
originating from the hard scattering, tend to be clustered, 
while the energy deposits due to the UE tend to be more 
uniformly spread throughout the detector. In order to see this
pictorially, in Figs.~\ref{fig:ttb_cal1} and \ref{fig:ttb_cal2}
we show a series of calorimeter maps of the combined
ECAL+HCAL energy deposits as a function of the pseudorapidity 
$\eta$ and azimuthal angle $\phi$. Since the calorimeter in PGS is 
segmented in cells of $(\Delta\eta,\Delta\phi)=(0.1,0.1)$,
each calorimeter tower is represented by a square pixel, which
is color-coded according to the amount of energy present in
the tower. We have chosen the color scheme so that larger deposits
correspond to darker colors.

\FIGURE[ht]{
\centerline{
\epsfig{file=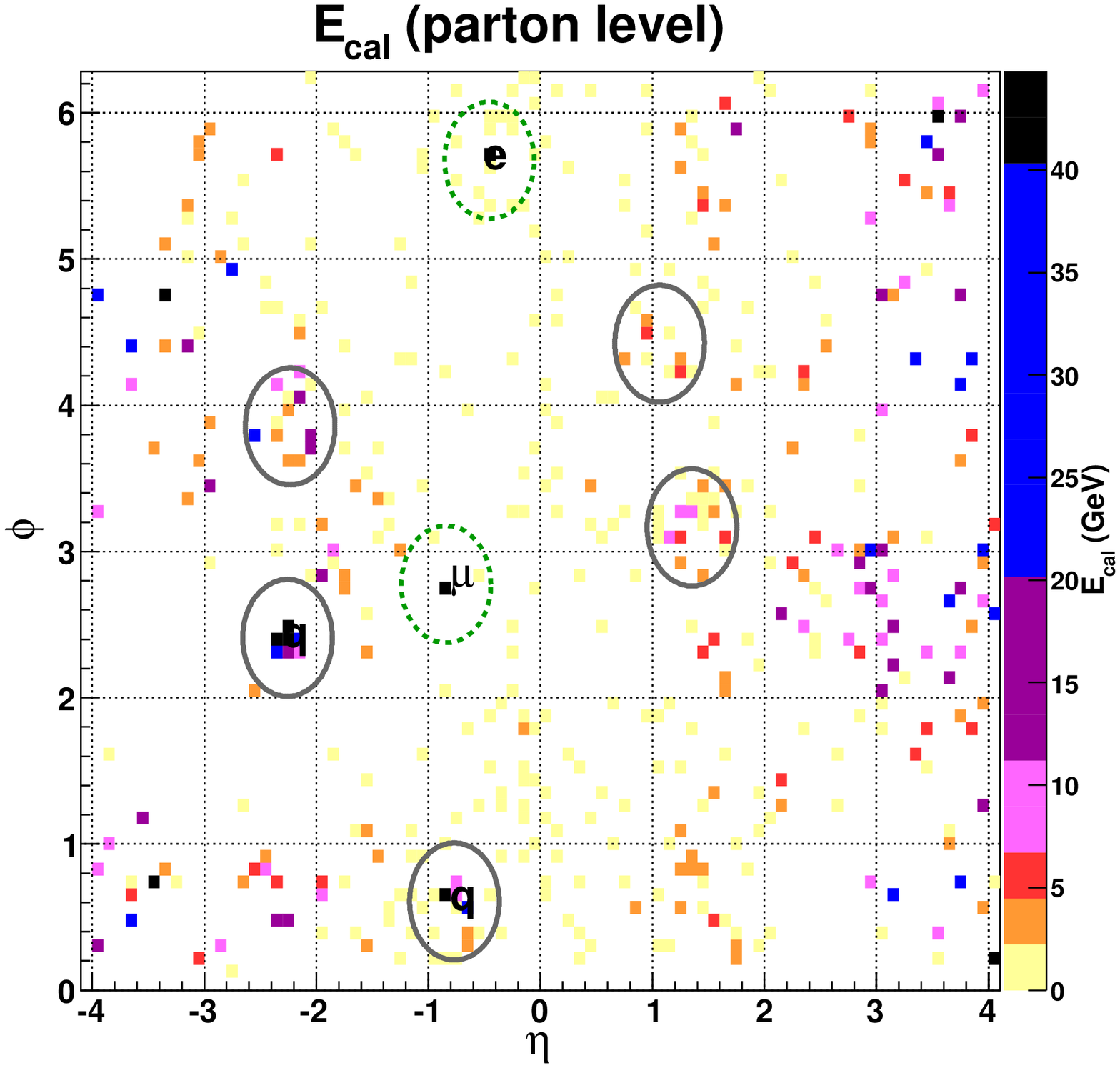,height=7.7cm}
\epsfig{file=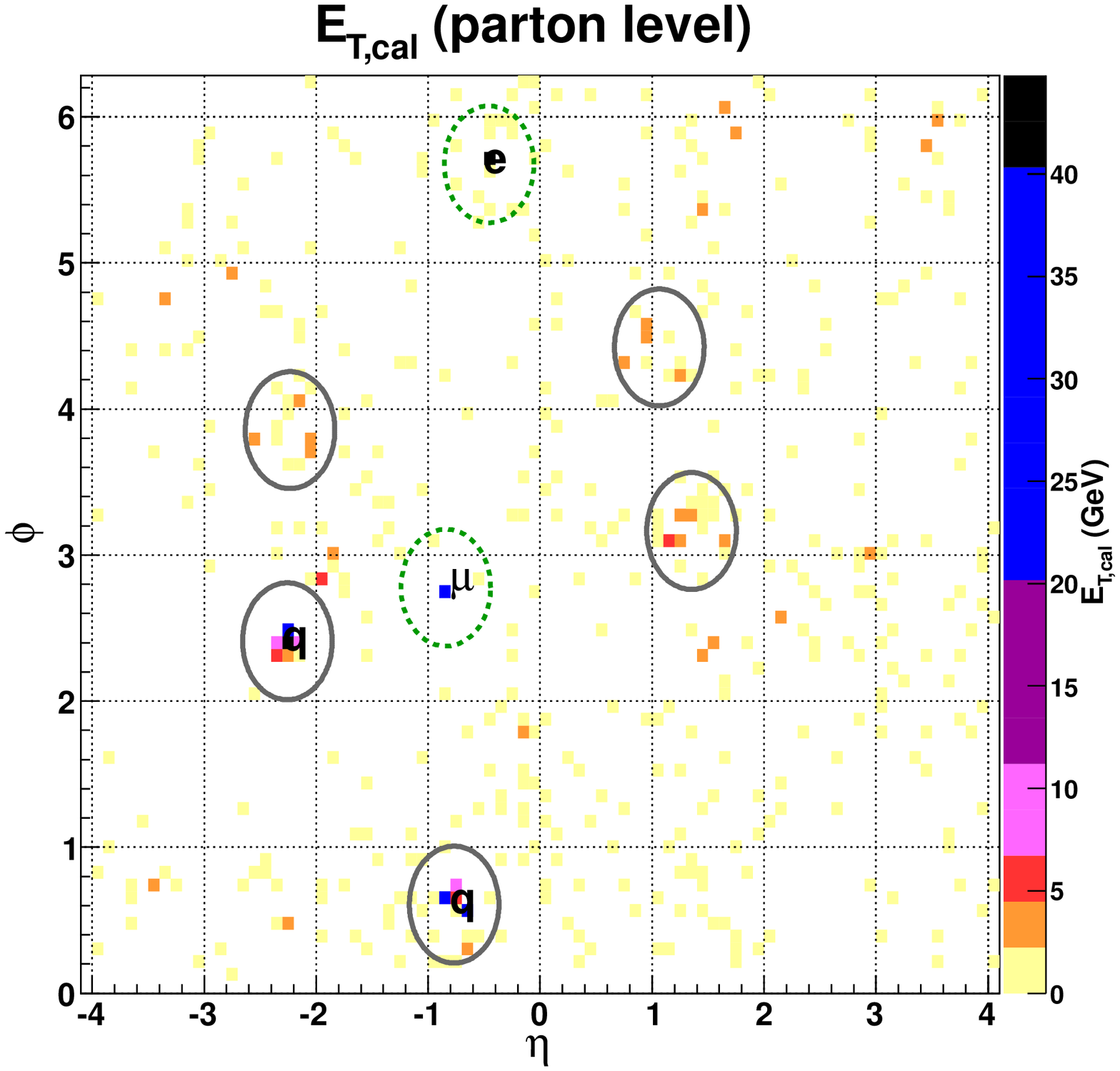,height=7.7cm}
}\\
\centerline{
\epsfig{file=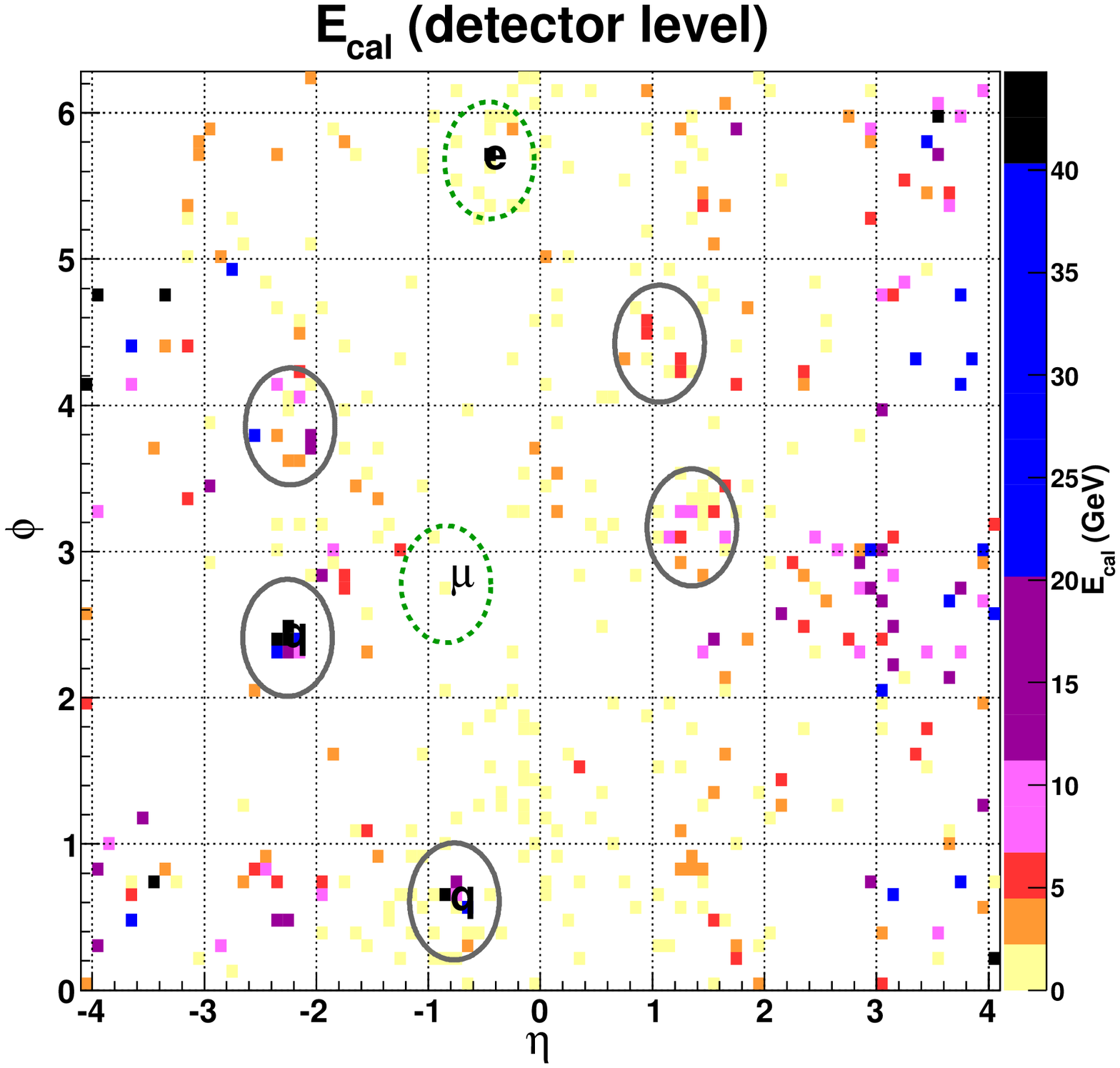,height=7.7cm}
\epsfig{file=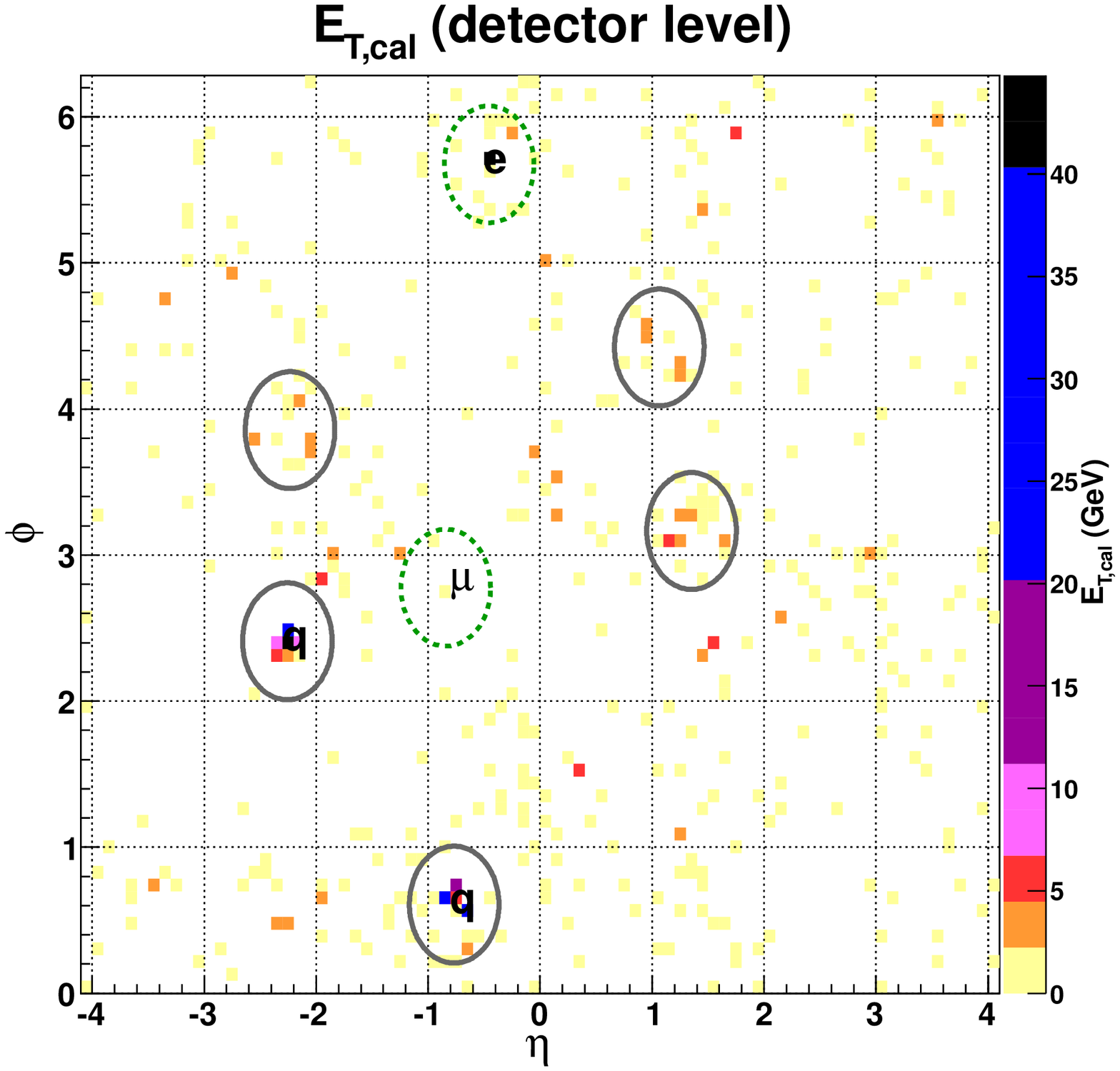,height=7.7cm}
}
\caption{The same as Fig.~\ref{fig:ttb_cal1}, but for 
an event with three additional reconstructed jets.
}
\label{fig:ttb_cal2}
}

Each calorimeter map figure below has four panels.
In the upper two panels the calorimeter is filled 
at the parton level directly from PYTHIA. This corresponds to a
perfect detector, where we ignore any smearing effects due to 
the finite energy resolution. The lower 
two plots in each figure show the corresponding 
results after PGS simulation. Thus by comparing the 
plots in the upper row to those in the bottom row, 
one can see the effect of the detector resolution. 
While the finite detector resolution does play some role,
we find that it is of no particular importance for 
understanding the reason behind the big swings in the 
$\sqrt{s}_{min}$ peaks observed in Fig.~\ref{fig:ttb}.

Let us instead concentrate on comparing the plots
in the left column versus those in the right column.
The left plots show the absolute energy deposit
$E_\alpha$ in the $\alpha$ calorimeter tower, 
while in the right plots this energy
is shown projected on the transverse plane
as $E_\alpha \cos\theta_\alpha$. 
The difference between the left and the right plots
is quite striking. The plots on the left 
exhibit lots of energy, which is deposited mostly 
in the forward calorimeter cells (at large $|\eta|$) \cite{Konar:2008ei}.
The plots on the right, on the other hand, show 
only a few clusters of energy, concentrated mostly 
in the central part of the detector. Those energy clusters
give rise to the objects (jets, electrons and photons)
which are reconstructed from the calorimeter. 
Furthermore, each energy cluster can be easily identified 
with a parton-level particle in the top decay chain. 
In order to exhibit this correlation, in Figs.~\ref{fig:ttb_cal1} 
and \ref{fig:ttb_cal2} we use the following notation for 
the parton-level particles: a $b$-quark (electron, muon)
is marked with the letter ``q'' (``e'', ``$\mu$'').
A grey circle delineates (the cone of) a 
reconstructed jet, while a green dotted circle
marks a reconstructed lepton (electron or muon). 
The lepton isolation requirement
implies that green circles should be void of large 
energy deposits off-center, 
and indeed we observe this to be the case.

In particular, Fig.~\ref{fig:ttb_cal1} shows a bare-bone
dilepton $t\bar{t}$ event with just two reconstructed  
jets and two reconstructed leptons (which happen to be both 
electrons). As seen in the figure, the two jets can be
easily traced back to the two $b$-quarks at the parton level,
and there are no additional reconstructed jets due to the UE activity.
Because the event is so clean and simple, one might 
expect to obtain a reasonable value for $\sqrt{s}_{min}$,
i.e. close to the $t\bar{t}$ threshold. However, this is not
the case, if we use the calorimeter-based measurement 
$\sqrt{s}_{min}^{(cal)}$. As seen in Table~\ref{table:data},
the measured value of $\sqrt{s}_{min}^{(cal)}$ is very far off ---
on the order of 1 TeV, even in the case of a perfect detector.
%
\TABULAR[ht]{|c||c|c||c|c|}{
\hline
Event type     & \multicolumn{2}{c||}{PYTHIA parton level}    
                                   & \multicolumn{2}{c|}{after PGS simulation}     \\
\cline{2-5}
               & $\sqrt{s}_{true}$  & $\sqrt{s}_{min}^{(cal)}$ 
                                   & $\sqrt{s}_{min}^{(cal)}$ 
                                   &  $\sqrt{s}_{min}^{(reco)}$  \\  \hline \hline
$t\bar{t}$ event in Fig.~\ref{fig:ttb_cal1}              
               & 427   &  1110 & 1179 & 363          \\  \hline
$t\bar{t}$ event in Fig.~\ref{fig:ttb_cal2}              
               & 638   &  2596 & 2761 & 736          \\  \hline
SUSY event in Fig.~\ref{fig:glu_cal}              
               & 1954  &  3539 & 3509 & 2085         \\  
\hline
}{\label{table:data} Selected $\sqrt{s}$ quantities (in GeV) for the 
events shown in Figs.~\ref{fig:ttb_cal1}, \ref{fig:ttb_cal2}  and \ref{fig:glu_cal}. 
The second column shows the true invariant mass $\sqrt{s}_{true}$ of the 
parent system: top quark pair in case of Figs.~\ref{fig:ttb_cal1}
and \ref{fig:ttb_cal2}, or gluino pair in case of Fig.~\ref{fig:glu_cal}.
The third column shows the value of the 
$\sqrt{s}_{min}^{(cal)}$ variable (\ref{smin_def_cal})
calculated at the parton level, without any PGS detector simulation,
but with the full detector acceptance cut of $|\eta|<4.1$.
The fourth column lists the value of $\sqrt{s}_{min}^{(cal)}$ 
obtained after PGS detector simulation, while the last column
shows the value of the $\sqrt{s}_{min}^{(reco)}$ variable
defined in (\ref{smin_def_reco}).
}
%
The reason for this discrepancy is now easy to understand from 
Fig.~\ref{fig:ttb_cal1}. Recall that $\sqrt{s}_{min}^{(cal)}$ is 
defined in terms of the {\em total} energy $E_{(cal)}$ in the 
calorimeter, which in turn is dominated by the large deposits in the 
forward region, which came from the underlying event.
More importantly, those contributions are more or less
equally spread over the forward and backward region of the detector, 
leading to cancellations in the calculation of the
corresponding longitudinal $P_{z(cal)}$ momentum component. 
As a result, the first term in (\ref{smin_def_cal})
becomes completely dominated by the UE contributions
\cite{Papaefstathiou:2009hp}.

Let us now see how the calculation of $\sqrt{s}_{min}^{(reco)}$
is affected by the UE. Since object reconstruction
is done with the help of minimum {\em transverse} cuts (for clustering
and object id), the relevant calorimeter plots are the
maps on the right side in Fig.~\ref{fig:ttb_cal1}.
We see that the large forward energy deposits which were 
causing the large shift in $\sqrt{s}_{min}^{(cal)}$
are not incorporated into any reconstructed objects, 
and thus do not contribute to the $\sqrt{s}_{min}^{(reco)}$
calculation at all. In effect, the RECO-level prescription
for calculating $\sqrt{s}_{min}$ is leaving out precisely 
the unwanted contributions from the UE, while keeping the
relevant contributions from the hard scattering.
As seen from Table~\ref{table:data}, the calculated 
value of $\sqrt{s}_{min}^{(reco)}$ for that event is 
363 GeV, which is indeed very close to the $t\bar{t}$ threshold.
It is also smaller than the true $\sqrt{s}$ 
value of $427$ GeV in that event, which is to be expected, 
since by design $\sqrt{s}_{min}\le \sqrt{s}$, and this event 
does not have any extra ISR jets to spoil this relation.

It is instructive to consider another, more complex
$t\bar{t}$ dilepton event, such as the one shown in 
Fig.~\ref{fig:ttb_cal2}. The corresponding calculated values 
for $\sqrt{s}_{min}^{(cal)}$ and $\sqrt{s}_{min}^{(reco)}$
are shown in the second row of Table~\ref{table:data}.
As seen in Fig.~\ref{fig:ttb_cal2}, this event has additional jets
and a lot more UE activity. As a result, the calculated value
of $\sqrt{s}_{min}^{(cal)}$ is shifted by almost 2 TeV from the
nominal $\sqrt{s}_{true}$ value. Nevertheless, the RECO-level prescription
nicely compensates for this effect, and the calculated
$\sqrt{s}_{min}^{(reco)}$ value is only $736$ GeV, which 
is within $100$ GeV of the nominal $\sqrt{s}_{true}=638$ GeV. 
Notice that in this example we end up with 
a situation where $\sqrt{s}_{min}^{(reco)}>\sqrt{s}_{true}$.
Fig.~\ref{fig:ttb} indicates that this happens quite often --- 
the tail of the $\sqrt{s}_{min}^{(reco)}$ distribution
is more populated than the (yellow-shaded) $\sqrt{s}_{true}$
distribution. This should be no cause for concern.  
First of all, we are only interested in the {\em peak} of the 
$\sqrt{s}_{min}^{(reco)}$ distribution, and we do 
not need to make any comparisons between
$\sqrt{s}_{min}^{(reco)}$ and $\sqrt{s}_{true}$.
Second, any such comparison would be meaningless, since
the value of $\sqrt{s}_{true}$ is a priori unknown, 
and unobservable.

\subsection{$\sqrt{s}_{min}^{(sub)}$ variable}

Before concluding this section, we shall use the $t\bar{t}$ 
example to also illustrate the idea of the subsystem $\sqrt{s}_{min}^{(sub)}$
variable developed in Sec.~\ref{sec:sub}. Dilepton $t\bar{t}$ 
events are a perfect testing ground for this idea, since
the $WW$ subsystem decays leptonically, without any jet activity.
We therefore define the subsystem as the two hard isolated 
leptons resulting from the decays of the $W$-bosons. Correspondingly,
we require two reconstructed leptons (electrons or muons) 
at the PGS level\footnote{The selection efficiency for the two
leptons is on the order of $60\%$, which explains the 
different normalization of the distributions in 
Figs.~\ref{fig:ttb} and \ref{fig:ttb_sub}.}, 
and plot the distribution of the 
leptonic subsystem $\sqrt{s}_{min}^{(sub)}$ variable in Fig.~\ref{fig:ttb_sub}.
\FIGURE[t]{
\centerline{
\epsfig{file=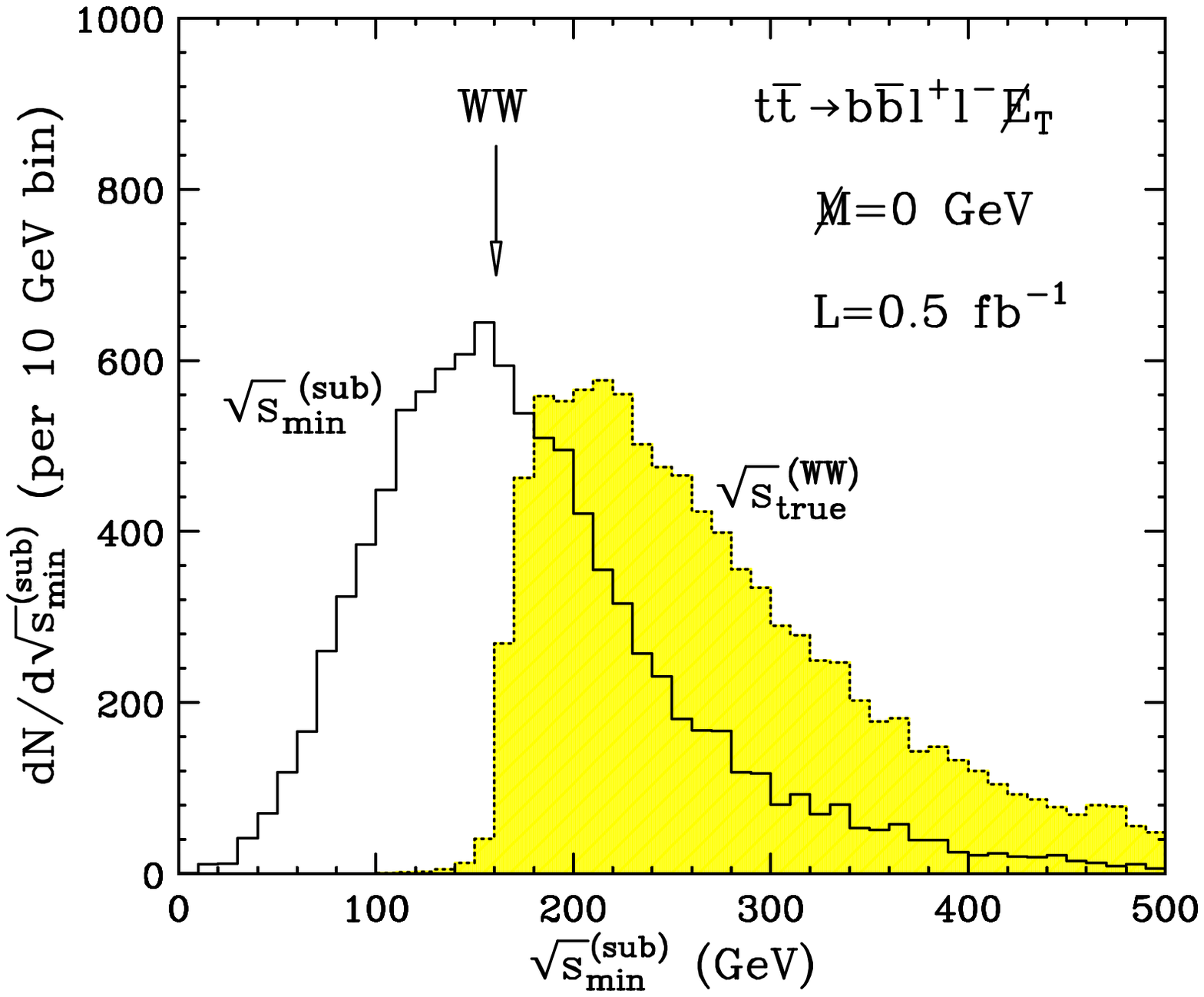,width=10cm}
}
\caption{The same as Fig.~\ref{fig:ttb}, but for the dilepton subsystem
in dilepton $t\bar{t}$ events with two reconstructed leptons
in PGS. The dotted (yellow-shaded) histogram 
gives the true $\sqrt{s}$ distribution of the $W^+W^-$ pair in those events.
The black histogram shows the distribution of the (leptonic) 
subsystem variable $\sqrt{s}_{min}^{(sub)}$ defined in Sec.~\ref{sec:sub}. 
In this case, the subsystem is defined by the two isolated leptons, 
while all jets are treated as upstream particles. The vertical arrow marks the 
$W^+W^-$ mass threshold.
}
\label{fig:ttb_sub}
}
As before, the dotted (yellow-shaded) histogram represents the
true $\sqrt{s}$ distribution of the $W^+W^-$ pair.
As expected, it quickly rises at the $WW$ threshold 
(denoted by the vertical arrow), then falls off at large $\sqrt{s}$. 
Since the $\sqrt{s}_{true}^{\,(WW)}$ distribution is unobservable, 
the best we can do is to study the corresponding 
$\sqrt{s}_{min}^{(sub)}$ distribution shown with the 
solid black histogram. In this subsystem example, 
all UE activity is lumped together with the upstream 
$b$-jets from the top quarks decays, and thus has no 
bearing on the properties of the leptonic $\sqrt{s}_{min}^{(sub)}$.
In particular, we find that the value of $\sqrt{s}_{min}^{(sub)}$ 
is always smaller than the true $\sqrt{s}_{true}^{\,(WW)}$.
More importantly, Fig.~\ref{fig:ttb_sub} demonstrates
that the peak in the $\sqrt{s}_{min}^{(sub)}$ 
distribution is found precisely at the mass threshold
of the particles (in this case the two $W$ bosons)
which initiated the subsystem. Therefore, in analogy to 
(\ref{recopeak}) we can also write
\beq
\left(\sqrt{s}_{min}^{(sub)}\right)_{peak} \approx M_p^{(sub)}\, ,
\label{subpeak}
\eeq
where $M_p^{(sub)}$ is the combined mass of all the parents initiating the subsystem.
Fig.~\ref{fig:ttb_sub} shows that in the $t\bar{t}$ example just considered,
this relation holds to a very high degree of accuracy.

This example should not leave the reader with the impression 
that hadronic jets are never allowed to be part of the subsystem.
On the contrary --- the subsystem may very well include 
reconstructed jets as well. The $t\bar{t}$ case considered here
in fact provides a perfect example to illustrate the idea.

Let us reconsider the $t\bar{t}$ dilepton sample, and redefine
the subsystem so that we now target the two {\em top quarks} 
as the parents initiating the subsystem. Correspondingly, 
in addition to the two leptons, let us allow the subsystem 
to include two jets, presumably coming from the two top quark decays. 
Unfortunately, in doing so, we must face a variant of 
the partitioning\footnote{By construction, 
the $\sqrt{s}_{min}$ and  $\sqrt{s}_{min}^{(sub)}$
variables never have to face the 
{\em ordering} combinatorial problem.}
combinatorial problem discussed in the introduction:
as seen in Fig.~\ref{fig:ttb_jetmulti},
the typical jet multiplicity in the events is relatively high,
and we must therefore specify the exact procedure how to select 
the two jets which would enter the subsystem.
\FIGURE[t]{
\centerline{
\epsfig{file=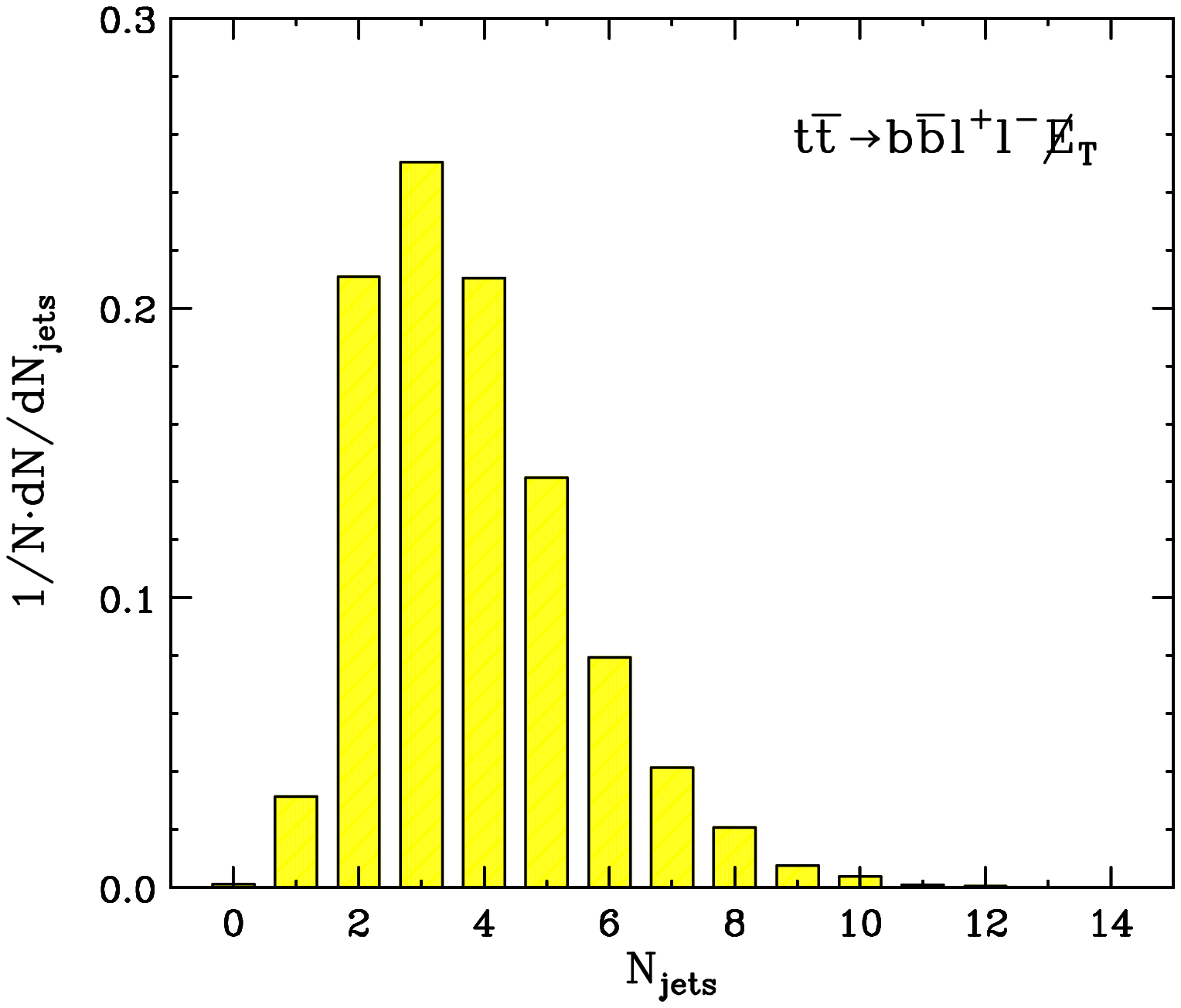,width=9cm}  
}
\caption{Unit-normalized distribution of jet multiplicity in 
dilepton $t\bar{t}$ events.}
\label{fig:ttb_jetmulti}
}
We shall consider three different approaches.
\begin{itemize}
\item {\em B-tagging.} We can use the fact that 
the jets from top quark decay are $b$-jets, 
while the jets from ISR are typically light flavor jets.
Therefore, by requiring exactly two $b$-tags, and 
including only the two $b$-tagged jets as part of the subsystem,
we can significantly increase the probability 
of selecting the correct jets. Of course, ISR will sometimes 
also contribute $b$-tagged jets from gluon splitting, but that 
happens rather rarely and the corresponding contribution can 
be suppressed by a further invariant mass cut on the two $b$-jets.
The resulting $\sqrt{s}_{min}^{(sub)}$ distribution for 
the subsystem of 2 leptons and 2 $b$-tagged jets 
is shown in Fig.~\ref{fig:ttb_sub_jets}
with the black histogram.
We see that, as expected, the 
distribution peaks at the $t\bar{t}$ threshold and this time
provides a measurement of the top quark mass:
\beq
\left(\sqrt{s}_{min}^{(sub)}\right)_{peak} \approx M_p^{(sub)} = 2 m_t = 350\ {\rm GeV}\, .
\eeq
The disadvantage of this method is the loss in statistics:
compare the normalization of the black histogram in Fig.~\ref{fig:ttb_sub_jets}
after applying the two $b$-tags,
to the dotted (yellow-shaded) distribution of the true 
$t\bar{t}$ distribution in the selected inclusive dilepton 
sample (without $b$-tags).
\item {\em Selection by jet $p_T$.} Here one can use the fact 
that the jets from top decays are on average harder
than the jets from ISR. Correspondingly, by choosing the
two highest $p_T$ jets (regardless of $b$-tagging), one also
increases the probability to select the correct jet pair.
The corresponding distribution is shown in Fig.~\ref{fig:ttb_sub_jets}
with the blue histogram, and is also seen to peak 
at the $t\bar{t}$ threshold.
An important advantage of this method is that one does not 
have to pay the price of reduced statistics due to the
two additional $b$-tags.
\item {\em No selection.} The most conservative approach would be 
to apply no selection criteria on the jets, and
include all reconstructed jets in the subsystem.
Then the subsystem $\sqrt{s}_{min}^{(sub)}$ variable
essentially reverts back to the RECO-level inclusive
variable $\sqrt{s}_{min}^{(reco)}$ already discussed in
the previous subsection. Not surprisingly, we find 
the peak of its distribution (red histogram in Fig.~\ref{fig:ttb_sub_jets})
near the $t\bar{t}$ threshold as well.
\end{itemize}

\FIGURE[t]{
\centerline{
\epsfig{file=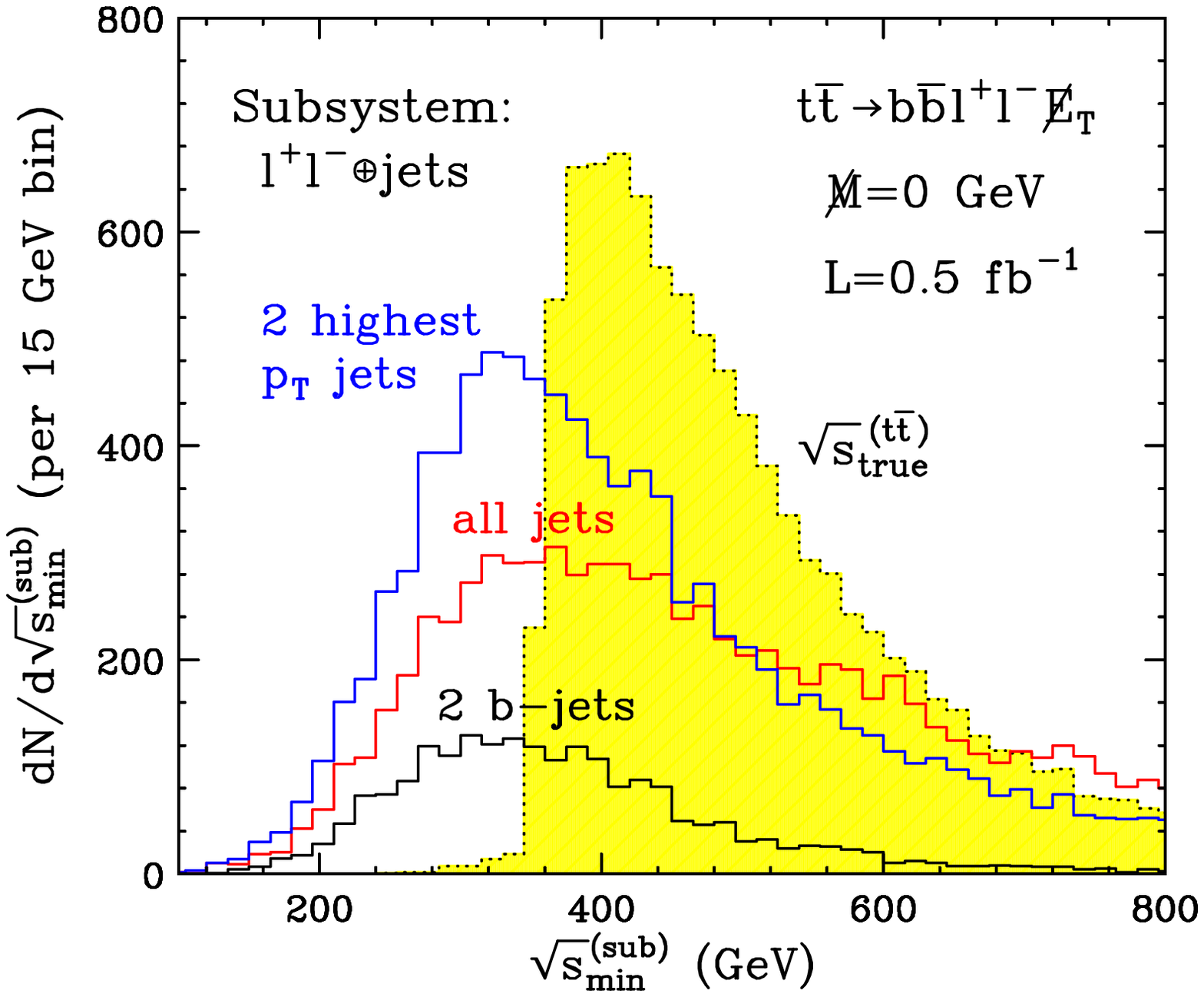,width=10cm}
}
\caption{The same as Fig.~\ref{fig:ttb_sub}, but in addition to the two leptons, 
the subsystem now also includes: exactly two $b$-tagged jets (black histogram);
the two highest $p_T$ jets (blue histogram); or all jets (red histogram). 
The dotted (yellow-shaded) histogram gives the true $\sqrt{s}$ distribution 
of the $t\bar{t}$ pair. 
}
\label{fig:ttb_sub_jets}
}

All three of these examples show that jets can also be usefully 
incorporated into the subsystem. The only question is whether one 
can find a reliable way of preferentially selecting jets which 
are more likely to originate from within the intended subsystem,
as opposed to from the outside. As we see in Fig.~\ref{fig:ttb_sub_jets},
in the $t\bar{t}$ case this is quite possible, although 
in general it may be difficult in other settings, like
the SUSY examples discussed in the next section.

\section{An exclusive SUSY example: multijet events from gluino production}
\label{sec:susy}

Since $\sqrt{s}_{min}$ is a fully inclusive variable, 
arguably its biggest advantage 
is that it can be applied to purely jetty events with 
large jet multiplicities, where no other method on 
the market would seem to work.
In order to simulate such a challenging case,
we consider gluino pair production in supersymmetry,
with each gluino forced to undergo a cascade decay
chain involving only QCD jets and nothing else.
For concreteness, we revisit the setup 
of Ref.~\cite{Konar:2008ei}, where two different possibilities 
for the gluino decays were considered:
\begin{itemize}
\item In one scenario, the gluino $\tilde g$ is forced to undergo 
a two-stage cascade decay to the LSP.
In the first stage, the gluino decays to the second-lightest 
neutralino $\tilde\chi^0_2$ and two quark jets:
$\tilde g \to q\bar{q}\tilde\chi^0_2$. In turn,
$\tilde\chi^0_2$ itself is then forced to decay 
via a 3-body decay to 2 quark jets and the LSP:
$\tilde \chi^0_2 \to q\bar{q}\tilde\chi^0_1$.
The resulting gluino signature is 4 jets plus missing energy:
\beq
\tilde g \to jj\tilde\chi^0_2\to jjjj\tilde\chi^0_1\ .
\label{4jet}
\eeq
Therefore, gluino pair production will nominally result
in 8 jet events. Of course, as shown in Fig.~\ref{fig:glu_jetmulti},
the actual number of reconstructed jets in such events is even higher, 
due to the effects of ISR, FSR and/or string fragmentation. 
\FIGURE[t]{
\centerline{
\epsfig{file=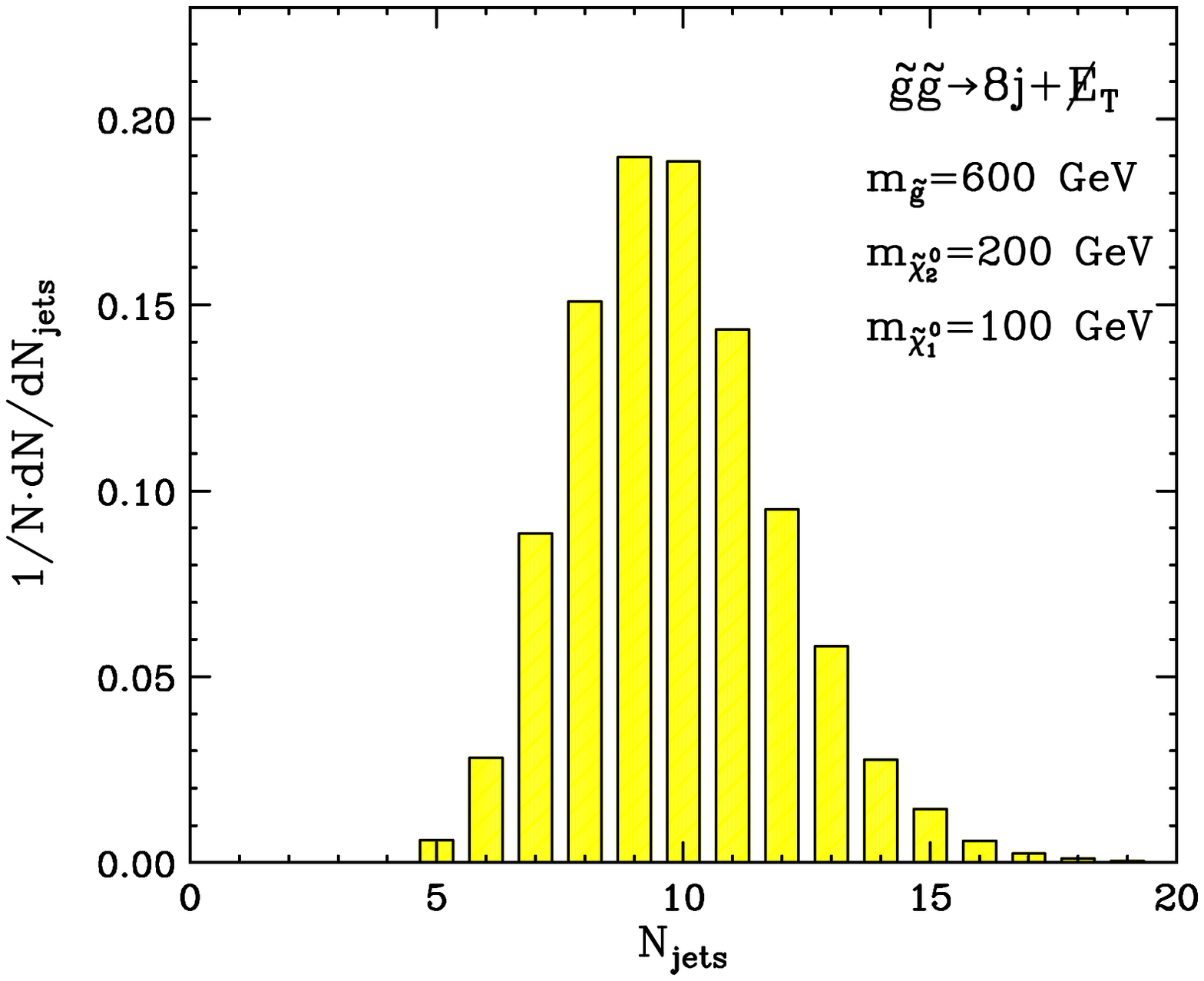,width=9cm}
}
\caption{Unit-normalized distribution of jet multiplicity in gluino pair 
production events, with each gluino decaying to four jets 
and a $\tilde\chi^0_1$ LSP as in (\ref{4jet}).}
\label{fig:glu_jetmulti}
}
As seen from the figure, each such event has on average
$\sim 10$ jets, presenting a formidable combinatorics problem.
We suspect that all\footnote{With the possible exception of 
the $M_{Tgen}$ method of Ref.~\cite{Lester:2007fq}, see Section \ref{sec:comp} 
below.} mass reconstruction methods on the market 
are doomed if they were to face such a scenario. It is therefore
of particular interest to see how well the $\sqrt{s}_{min}$ method
(which is advertized as universally applicable)
would fare under such dire circumstances. 
\item In the second scenario, the gluino decays 
directly to the LSP via a three-body decay
\beq
\tilde g \to jj\tilde\chi^0_1\ ,
\label{2jet}
\eeq
so that gluino pair-production events would nominally have 4 jets 
and missing energy. 
\end{itemize}
For concreteness, 
in each scenario we fix the mass spectrum as was done in 
\cite{Konar:2008ei}: we use the approximate gaugino  
unification relations to relate the gaugino and neutralino
masses as
\beq
m_{\tilde g} = 3 m_{\tilde\chi^0_2} = 6 m_{\tilde\chi^0_1} \ .
\label{gunif}
\eeq
We can then vary one of these masses, and choose the other two 
in accord with these relations.
Since we assume three-body decays in (\ref{2jet}) and (\ref{4jet}),
we do not need to specify the SUSY scalar mass parameters, which can be
taken to be very large. In addition, as implied by (\ref{gunif}),
we imagine that the lightest two neutralinos are gaugino-like, so that
we do not have to specify the higgsino mass parameter either, and it can 
be taken to be very large as well.

\FIGURE[t]{
\centerline{
\epsfig{file=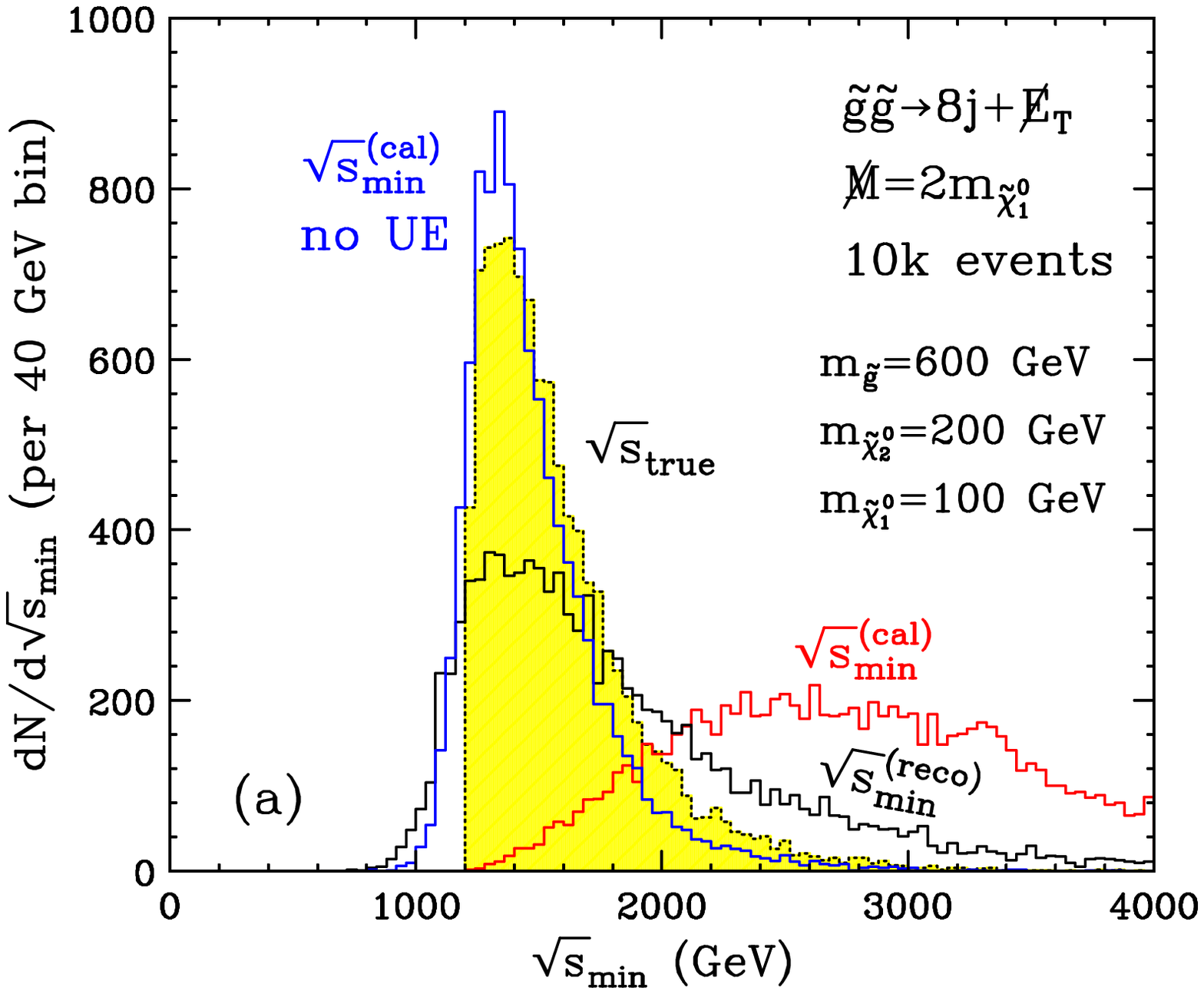,width=7.3cm}
\epsfig{file=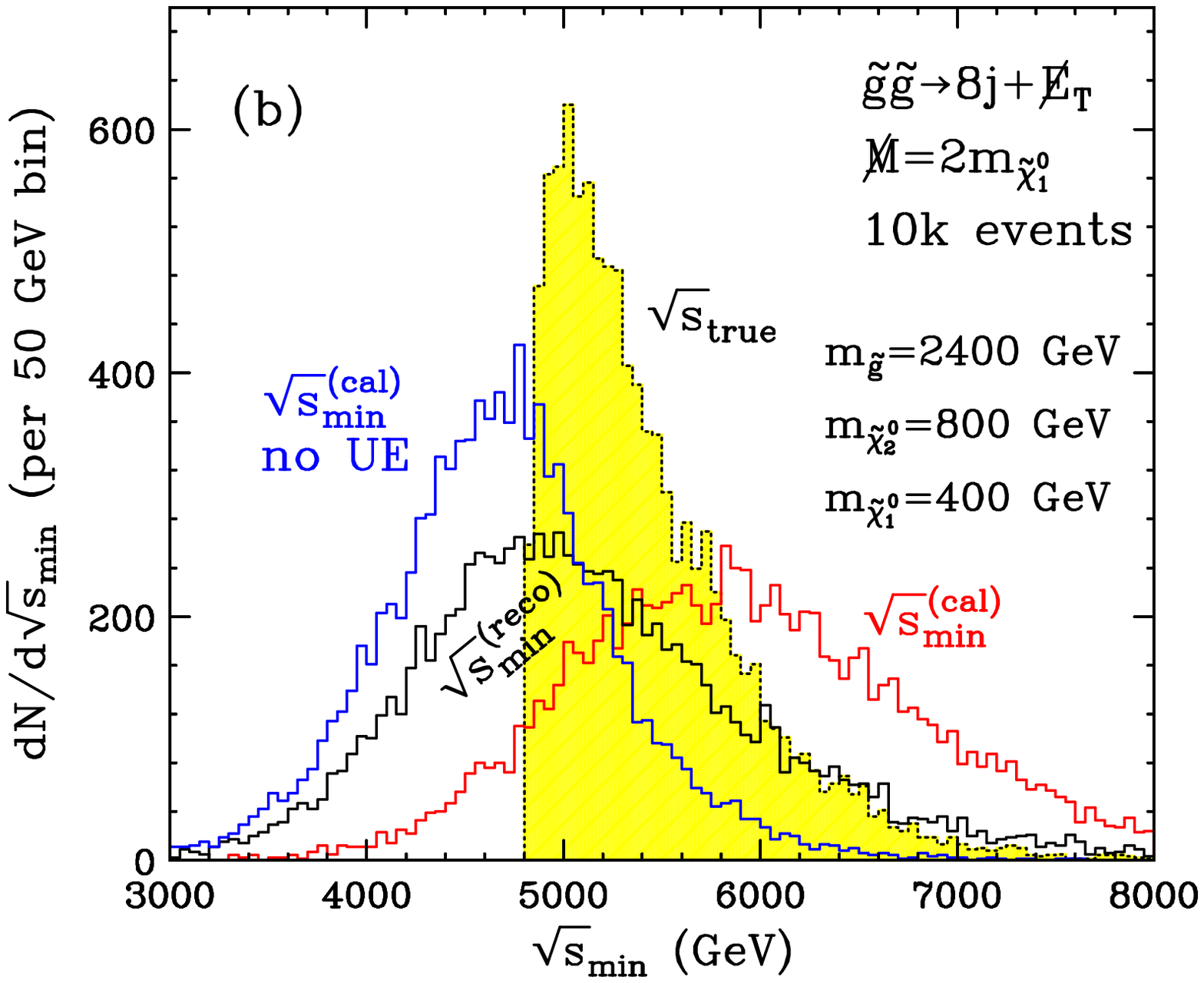,width=7.3cm}
}
\caption{The same as Fig.~\ref{fig:ttb}, but for 
a SUSY example of gluino pair production, with 
each gluino decaying to four jets 
and a $\tilde\chi^0_1$ LSP as indicated in (\ref{4jet}).
The mass spectrum is chosen as:
(a) $m_{\tilde g}=600$ GeV, $m_{\tilde\chi^0_2}=200$ GeV
and $m_{\tilde\chi^0_1}=100$ GeV; or
(b) $m_{\tilde g}=2400$ GeV, $m_{\tilde\chi^0_2}=800$ GeV
and $m_{\tilde\chi^0_1}=400$ GeV.
All three $\sqrt{s}_{min}$ distributions are plotted 
for the correct
value of the missing mass parameter, in this case
$\mmis=2m_{\tilde\chi^0_1}$.}
\label{fig:glu_4jet}
}

\FIGURE[t]{
\centerline{
\epsfig{file=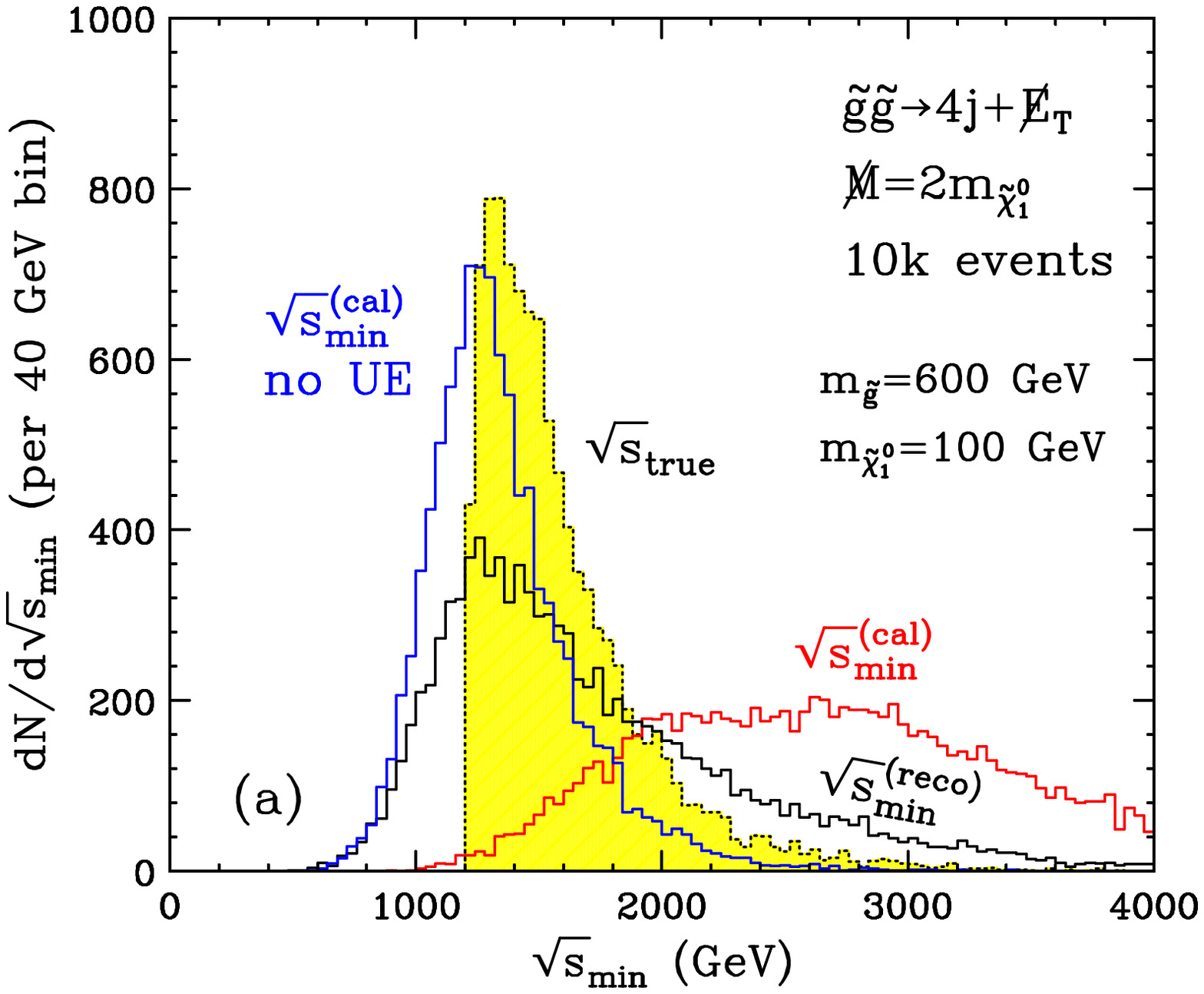,width=7.3cm}
\epsfig{file=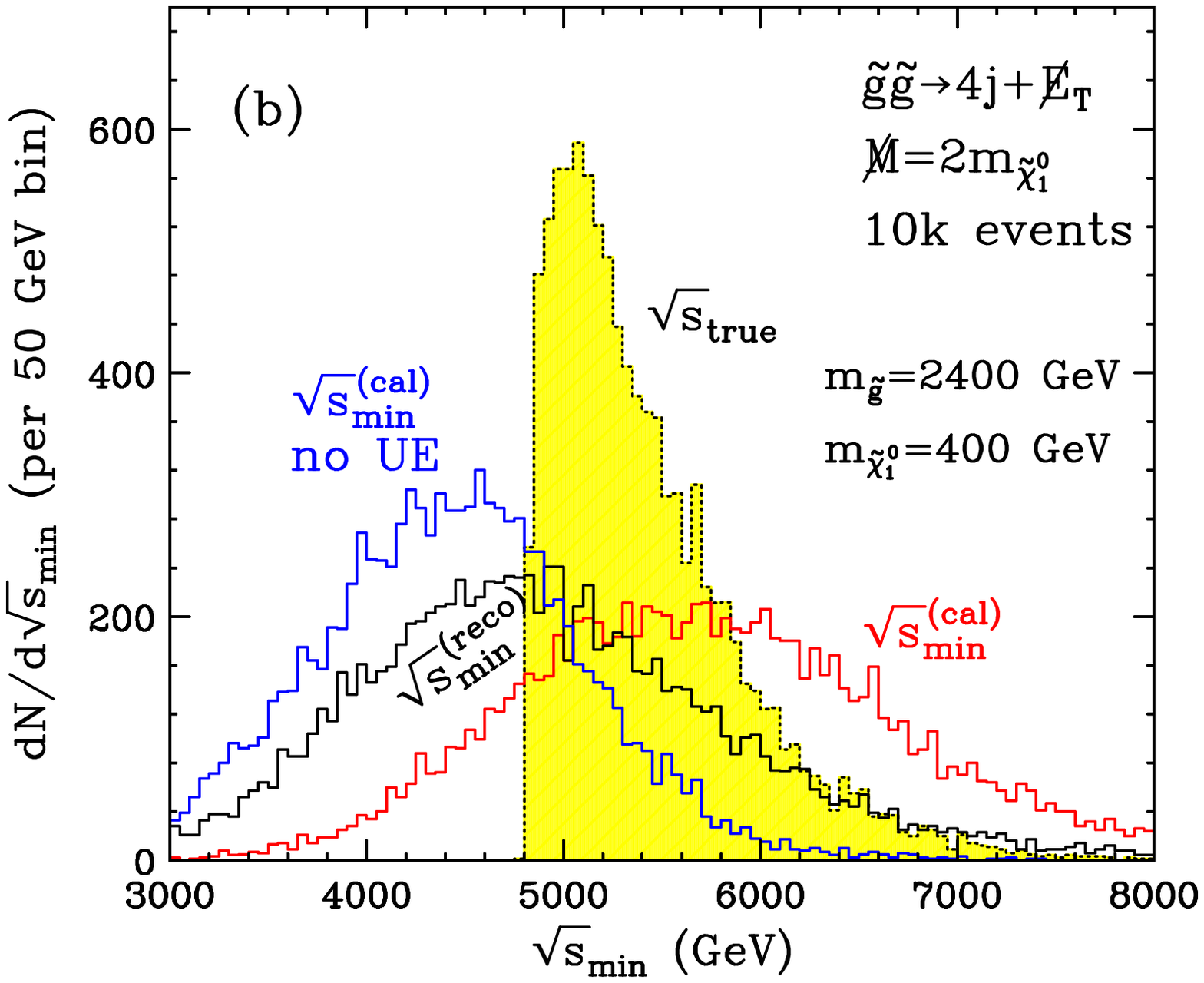,width=7.3cm}
}
\caption{The same as Fig.~\ref{fig:glu_4jet}, but for
the case of gluino decays to 2 jets and a $\tilde\chi^0_1$ 
LSP as in (\ref{2jet}).}
\label{fig:glu_2jet}
}

After these preliminaries, 
our results for these two scenarios are shown in 
Figs.~\ref{fig:glu_4jet} and \ref{fig:glu_2jet}, correspondingly.
In Fig.~\ref{fig:glu_4jet} (Fig.~\ref{fig:glu_2jet}) 
we consider the 8-jet signature arising from (\ref{4jet})
(the 4-jet signature arising from (\ref{2jet})).
In both figures, panels (a) correspond to a light mass spectrum
$m_{\tilde g}=600$ GeV, $m_{\tilde\chi^0_2}=200$ GeV
and $m_{\tilde\chi^0_1}=100$ GeV; while
panels (b) correspond to a heavy mass spectrum
$m_{\tilde g}=2400$ GeV, $m_{\tilde\chi^0_2}=800$ GeV
and $m_{\tilde\chi^0_1}=400$ GeV. Each plot shows the same 
four distributions as in Fig.~\ref{fig:ttb}.
The $\sqrt{s}_{min}$ 
distributions are all plotted for the correct
value of the missing mass parameter, namely
$\mmis=2m_{\tilde\chi^0_1}$.

\FIGURE[ht]{
\centerline{
\epsfig{file=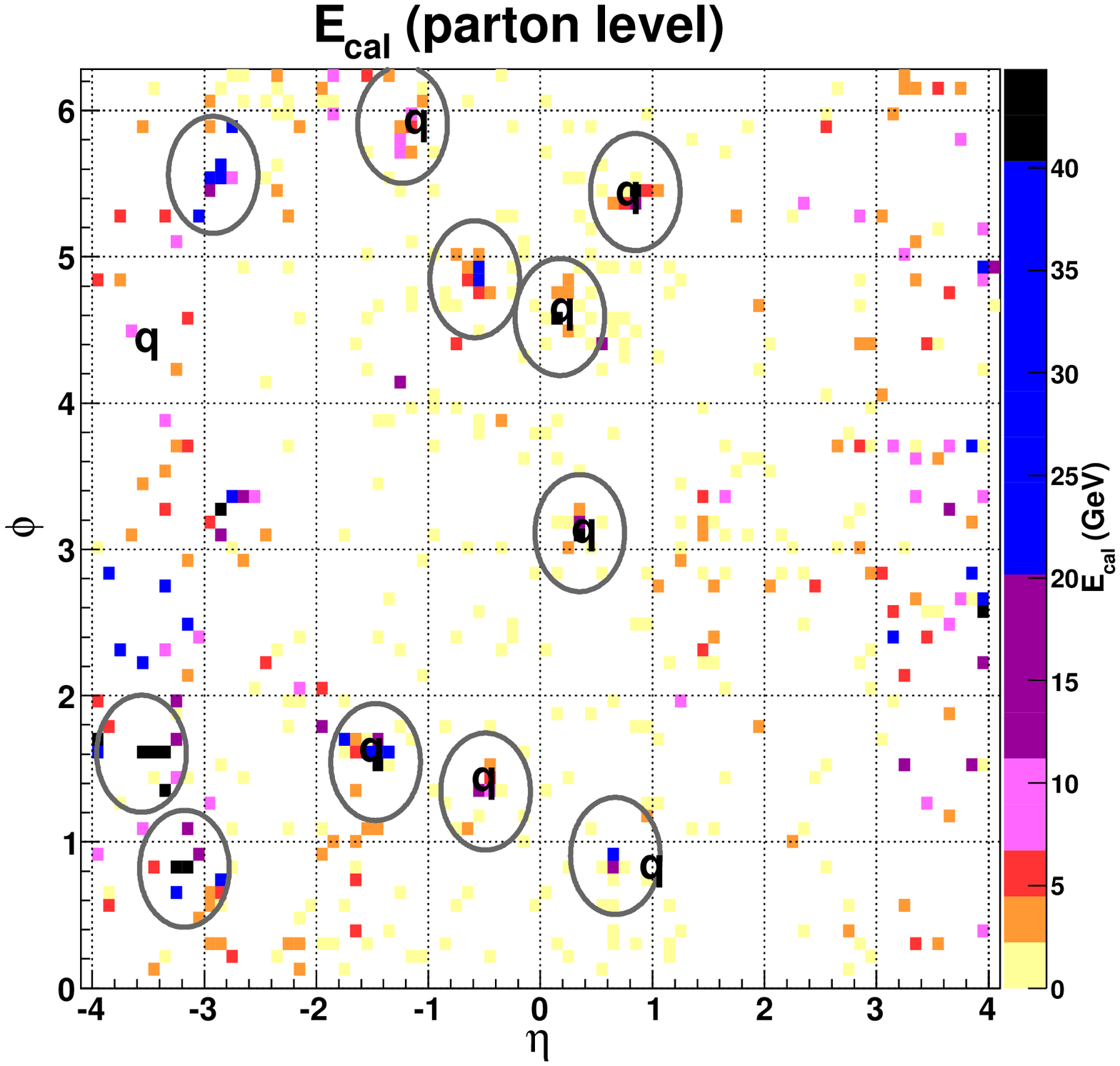,height=7.7cm}
\epsfig{file=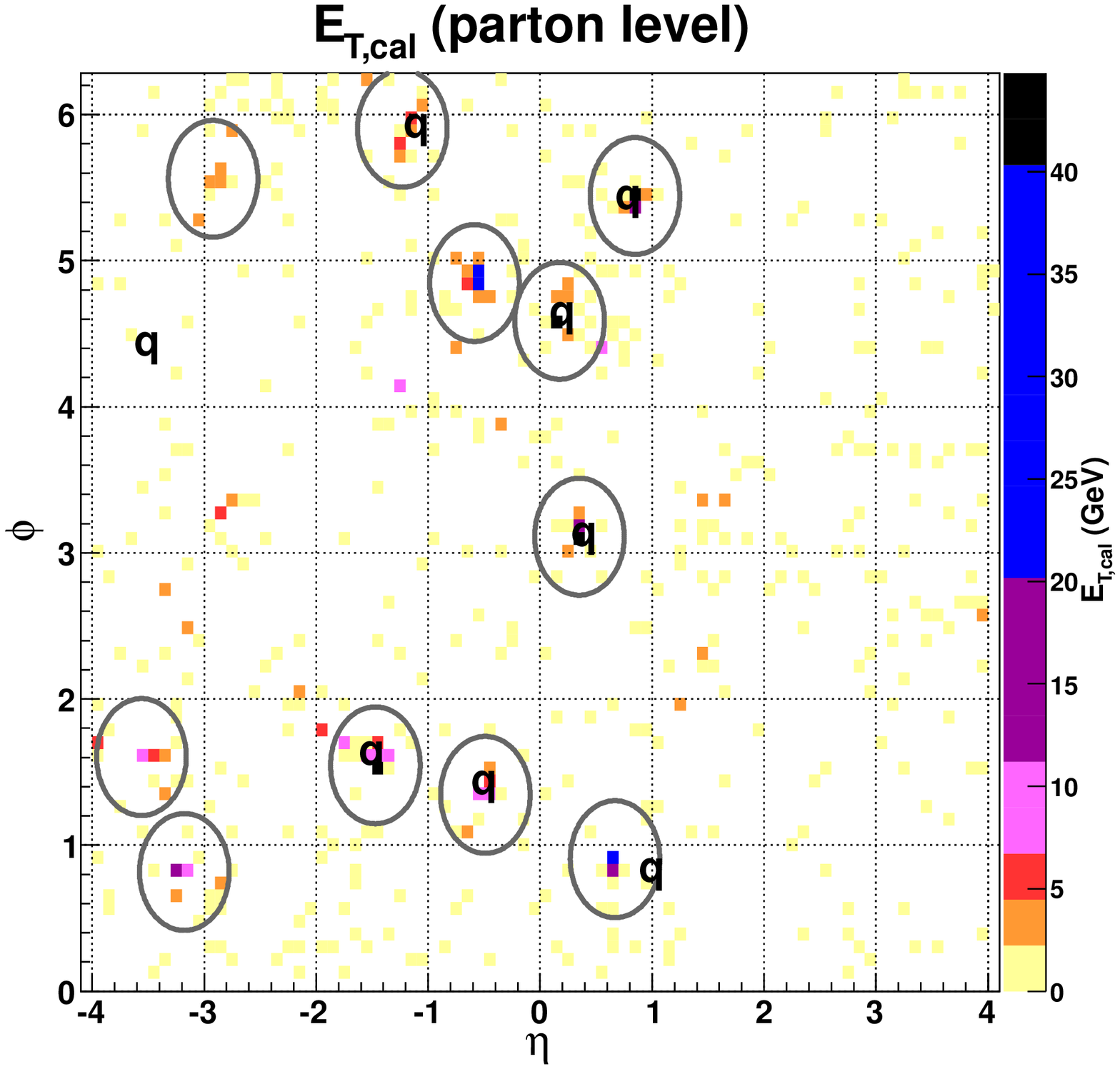,height=7.7cm}
}\\
\centerline{
\epsfig{file=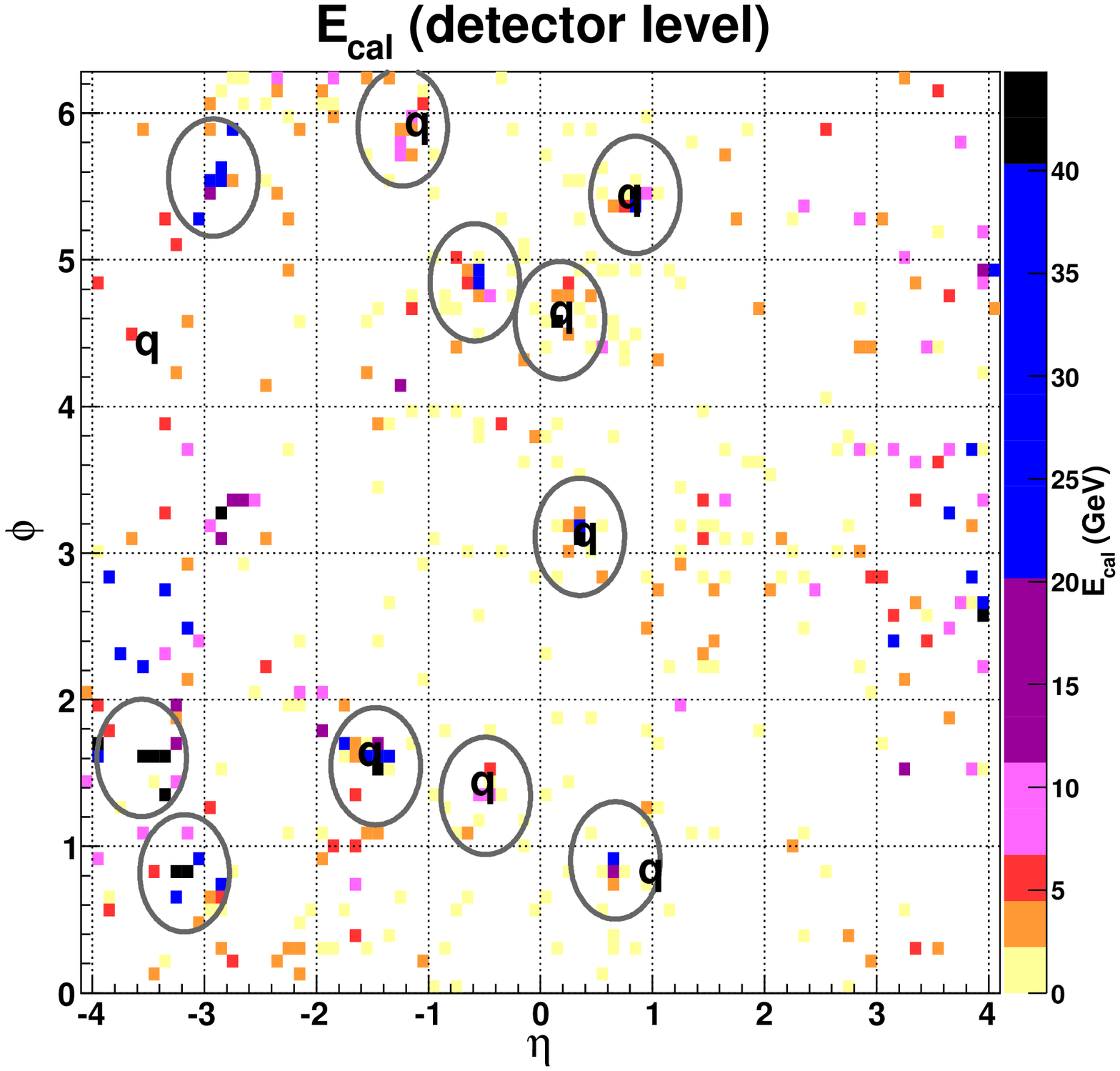,height=7.7cm}
\epsfig{file=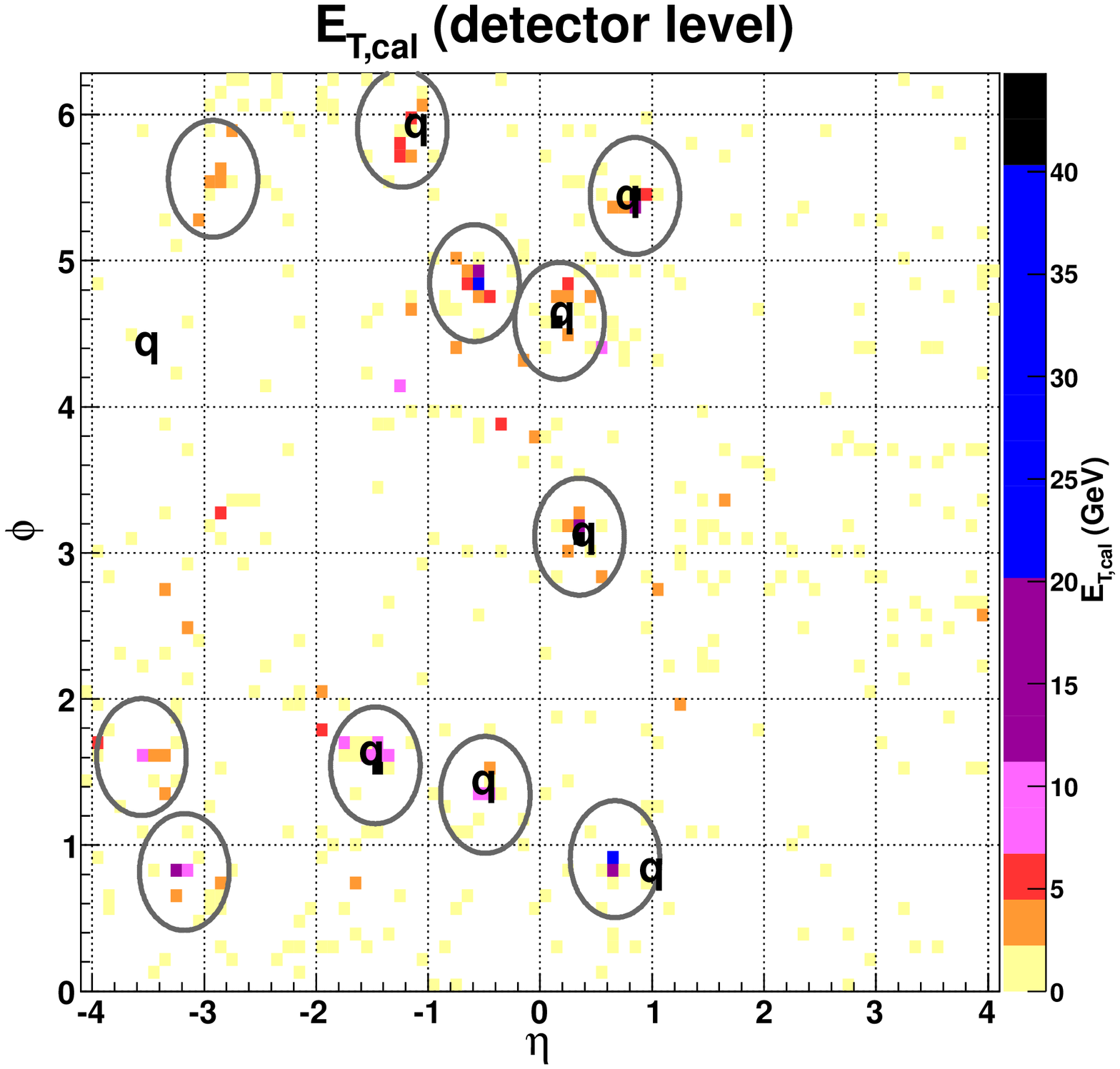,height=7.7cm}
}
\caption{The same as Fig.~\ref{fig:ttb_cal1}, but for 
a SUSY event of gluino pair production, with each gluino
forced to decay to 4 jets and the LSP as in (\ref{4jet}). The SUSY mass spectrum 
is as in Figs.~\ref{fig:glu_4jet}(a) and \ref{fig:glu_2jet}(a):
$m_{\tilde g}=600$ GeV, $m_{\tilde\chi^0_2}=200$ GeV
and $m_{\tilde\chi^0_1}=100$ GeV. As in Figs.~\ref{fig:ttb_cal1}
and \ref{fig:ttb_cal2}, the circles denote 
jets reconstructed in PGS, and here ``q'' marks 
the location of a quark from a gluino decay chain.
Therefore, a circle without a ``q'' inside
corresponds to a jet resulting from ISR or FSR, 
while a letter ``q'' without 
an accompanying circle represents a quark in the gluino 
decay chain which was not subsequently reconstructed as a jet.
}
\label{fig:glu_cal}
}

Overall, the results seen in Figs.~\ref{fig:glu_4jet} and \ref{fig:glu_2jet}
are not too different from what we already witnessed
in Fig.~\ref{fig:ttb} for the $t\bar{t}$ example.
The (unobservable) distribution $\sqrt{s}_{true}$
shown with the dotted yellow-shaded histogram
has a sharp turn-on at the physical mass threshold
$M_p=2m_{\tilde g}$. If the effects of the UE are ignored,
the position of this threshold is given rather well
by the peak of the $\sqrt{s}_{min}^{(cal)}$ distribution
(blue histogram). Unfortunately, the UE shifts the 
peak in $\sqrt{s}_{min}^{(cal)}$ by 1-2 TeV (red histogram). 
Fortunately, the distribution of the RECO-level variable
$\sqrt{s}_{min}^{(reco)}$ (black histogram) 
is stable against UE contamination, 
and its peak is still in the right place (near $M_p$).

Having already seen a similar behavior in the $t\bar{t}$ 
example of the previous section, these results may not 
seem very impressive, until one realizes just how 
complicated those events are. For illustration,
Fig.~\ref{fig:glu_cal} shows the previously discussed
calorimeter maps for one particular ``8 jet'' event.
This event happens to have 11 reconstructed jets, which is
consistent with the typical jet multiplicity 
seen in Fig.~\ref{fig:glu_jetmulti}.
The values of the $\sqrt{s}$ quantities of interest for this event
are listed in Table~\ref{table:data}. We see that
the RECO prescription for calculating $\sqrt{s}_{min}$
is able to compensate for a shift in $\sqrt{s}$ of more than 1.5 TeV!
A casual look at Fig.~\ref{fig:glu_cal} should be
enough to convince the reader just how daunting the 
task of mass reconstruction in such events is.
In this sense, the ease with which the 
$\sqrt{s}_{min}$ method reveals the gluino mass scale
in Figs.~\ref{fig:glu_4jet} and \ref{fig:glu_2jet}
is quite impressive.

\section{An inclusive SUSY example: GMSB study point GM1b}
\label{sec:gmsb}

In the Introduction we already mentioned that $\sqrt{s}_{min}$ is 
a fully inclusive variable. Here we would like to point out that
there are two different aspects of the inclusivity property of $\sqrt{s}_{min}$:
\begin{itemize}
\item {\em Object-wise inclusivity:}
$\sqrt{s}_{min}$ is inclusive with regards to the type of reconstructed objects.
The definition of $\sqrt{s}_{min}^{(reco)}$ does not distinguish between 
the different types of reconstructed objects (and $\sqrt{s}_{min}^{(cal)}$ 
makes no reference to any reconstructed objects at all). 
This makes $\sqrt{s}_{min}$ a very convenient variable to 
use in those cases where the newly produced particles have 
many possible decay modes, and restricting oneself to a single 
exclusive signature would cause loss in statistics.
For illustration, consider the gluino pair production
example from the previous section. Even though we are always 
producing the same type of parent particles (two gluinos), 
in general they can have several different decay modes, 
leading to a very diverse sample of events with 
varying number of jets and leptons.
Nevertheless, the $\sqrt{s}_{min}^{(reco)}$ distribution,
plotted over this whole signal sample, will still be able
to pinpoint the gluino mass scale, as explained in Sec.~\ref{sec:susy}. 
\item {\em Event-wise inclusivity:}
$\sqrt{s}_{min}$ is inclusive also with regards to the type of events,
i.e. the type of new particle production. For simplicity, 
in our previous examples we have been considering only 
one production mechanism at a time, but this is not really
necessary ---  $\sqrt{s}_{min}$ can also be applied in the 
case of several simultaneous production mechanisms. 
\end{itemize}

In order to illustrate the last point, in this section we 
shall consider the simultaneous production of the full
spectrum of SUSY particles at a particular benchmark point.
We chose the GM1b CMS study point \cite{CMSwiki}, 
which is nothing but a minimal gauge-mediated 
SUSY-breaking (GMSB) scenario on the SPS8 Snowmass slope
\cite{Allanach:2002nj}. The input parameters 
are $\Lambda$=80 TeV, $M_{mes}$=160 TeV, 
$N_{mes}$=1, $\tan\beta=15$ and $\mu>0$.
The physical mass spectrum is given in Table~\ref{table:mass}.
Point GM1b is characterized by a neutralino 
NLSP, which promptly decays (predominantly) 
to a photon and a gravitino. Therefore, 
a typical event has two hard photons
and missing energy, which provide good handles 
for suppressing the SM backgrounds. 

%
\TABULAR[t]{|c|c|c|c||c|c|c||c|c|c||c|}{
\hline
  $\tilde{u}_L$ & $\tilde{d}_L$ 
& $\tilde{u}_R$ & $\tilde{d}_R$ 
& $\tilde{\ell}_L$ & $\tilde{\nu}_{\ell}$ 
& $\tilde{\ell}_R$ & $\tilde{\chi}_2^\pm$ 
& $\tilde{\chi}_4^0$ & $\tilde{\chi}_3^0$ 
& $\tilde{g}$   \\ 
\hline
  908
& 911 
& 872
& 870 
& 289
& 278
& 145
& 371 
& 371
& 348 
& 690 \\
\hline
\hline
  $\tilde{t}_1$ & $\tilde{b}_1$ 
& $\tilde{t}_2$ & $\tilde{b}_2$ 
& $\tilde{\tau}_2$ & $\tilde{\nu}_{\tau}$ 
& $\tilde{\tau}_1$ & $\tilde{\chi}_1^\pm$ 
& $\tilde{\chi}_2^0$ & $\tilde{\chi}_1^0$ 
& $\tilde G$    \\ 
\hline
  806
& 863 
& 895
& 878 
& 290
& 277
& 138
& 206
& 206 
& 106
& 0 \\
\hline
}
{\label{table:mass} Masses (in GeV) of the
SUSY particles at the GM1b study point.
Here $\tilde u$ and $\tilde d$ 
($\tilde \ell$ and $\tilde \nu_\ell$) 
stand for either of the first two generations
squarks (sleptons).
}
%

We now consider inclusive production of all
SUSY subprocesses and plot the $\sqrt{s}_{min}$ 
distributions of interest in Fig.~\ref{fig:gmsb}.
As usual, the dotted yellow-shaded histogram is the true $\sqrt{s}$
distribution of the parent pair of SUSY particles
produced at the top of each decay chain. 
Since we do not fix the production subprocess, the
identity of the parent particles varies from 
event to event. 
Naturally, the most common parent particles are the ones 
with the highest production cross-sections.
For point GM1b, at a 14 TeV LHC, strong SUSY production dominates,
and is $87\%$ of the total cross-section.
A few of the dominant subprocesses and their cross-sections 
are listed in Table~\ref{table:xsec}.

\FIGURE[t]{
\centerline{
\epsfig{file=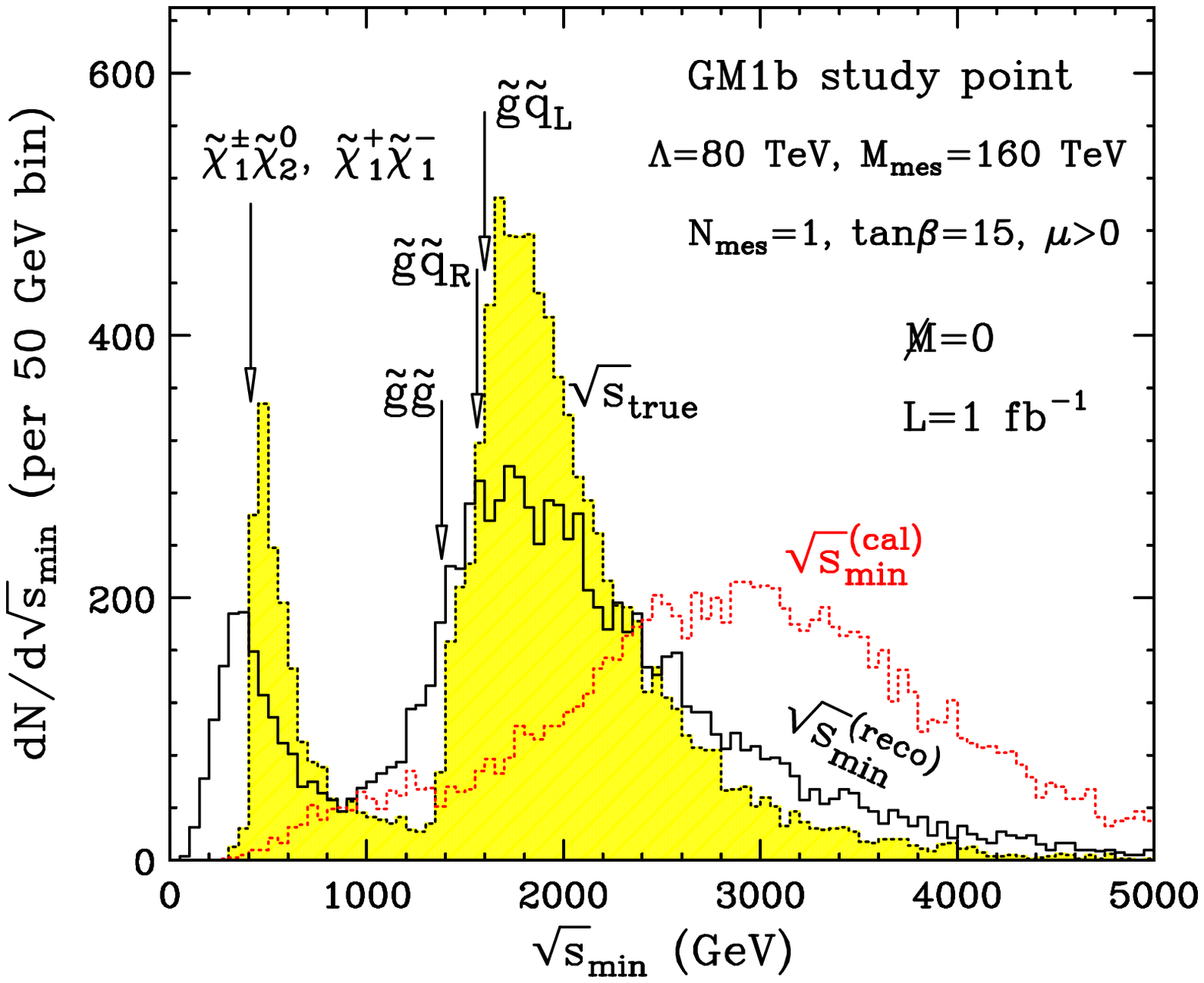,width=10cm}
}
\caption{Distribution of the $\sqrt{s}_{min}^{(cal)}$ (dotted red)
and $\sqrt{s}_{min}^{(reco)}$ (solid black) variables 
in inclusive SUSY production for the GMSB GM1a benchmark study point with
parameters $\Lambda=80$ TeV, $M_{mes}=160$ TeV, $N_{mes}=1$,
$\tan\beta=15$ and $\mu>0$.
The dotted yellow-shaded histogram is the true $\sqrt{s}$
distribution of the parent pair of SUSY particles
produced at the top of each decay chain
(the identity of the parent particles varies from event to event).
The $\sqrt{s}_{min}$ distributions are shown for $\mmis=0$ and 
are normalized to $1\ {\rm fb}^{-1}$ of data. The vertical arrows mark
the mass thresholds for a few dominant SUSY pair-production processes.
}
\label{fig:gmsb}
}

The true $\sqrt{s}$ distribution in Fig.~\ref{fig:gmsb}
exhibits an interesting 
double-peak structure, which is easy to understand as follows.
As we have seen in the exclusive examples from
Secs.~\ref{sec:top} and \ref{sec:susy}, at hadron colliders
the particles tend to be produced with $\sqrt{s}$ 
close to their mass threshold. As seen in Table~\ref{table:mass},
the particle spectrum of the GM1b point can be broadly divided 
(according to mass) into two groups of superpartners: electroweak sector 
(the lightest chargino $\tilde\chi^\pm_1$, 
second-to-lightest neutralino $\tilde\chi^0_2$ and sleptons)
with a mass scale on the order of $200$ GeV and a strong sector
(squarks and gluino) with masses of order $700-900$ GeV.
The first peak in the true $\sqrt{s}$ distribution 
(near $\sqrt{s}\sim 500$ GeV) arises from the pair production 
of two particles from the electroweak sector, while the second,
broader peak in the range of $\sqrt{s}\sim 1500-2300$ GeV
is due to the pair production of two colored 
superpartners\footnote{The attentive reader may also notice 
two barely visible bumps (near $950$ GeV and $1150$ GeV) 
reflecting the associated production of one colored 
and one uncolored particle:
$\tilde g \tilde\chi^\pm_1,\tilde g \tilde\chi^0_2$ and 
$\tilde q \tilde\chi^\pm_1,\tilde q \tilde\chi^0_2$, 
correspondingly.}.
Each one of those peaks is made up of several contributions 
from different individual subprocesses, but because their
mass thresholds\footnote{A few individual mass thresholds 
are indicated by vertical arrows in Fig.~\ref{fig:gmsb}.} are so close, 
in the figure they cannot be individually resolved, 
and appear as a single bump.

%
\TABULAR[t]{|c||c|c|c|c|c|c|c|c|c|}{
\hline
Process & $\tilde{\chi}_1^\pm \tilde{\chi}_2^0$ 
        & $\tilde{\chi}_1^+ \tilde{\chi}_1^-$ 
        & $\tilde{g} \tilde{g}  $ 
        & $\tilde{g} \tilde{q}_R$
        & $\tilde{g} \tilde{q}_L$  
        & $\tilde{q}_R \tilde{q}_R$  
        & $\tilde{q}_L \tilde{q}_R$  
        & $\tilde{q}_L \tilde{q}_L$  
\\ \hline
$\sigma$ (pb)
        & 0.83
        & 0.43
        & 2.03
        & 2.17
        & 1.90
        & 0.36
        & 0.50
        & 0.28
\\ \hline
$M_p$ (GeV)  
        &  412
        &  412
        & 1380
        & $\sim 1560$
        & $\sim 1600$
        & $\sim 1740$
        & $\sim 1780$
        & $\sim 1820$
\\
\hline
}{\label{table:xsec} Cross-sections (in pb) 
and parent mass thresholds (in GeV) for the dominant
production processes at the GM1b study point.
The listed squark cross-sections are summed over the 
light squark flavors and conjugate states. 
The total SUSY cross-section at point GM1b is 9.4 pb. 
}
%

If one could somehow directly observe the true $\sqrt{s}$ SUSY 
distribution (the dotted yellow-shaded histogram in Fig.~\ref{fig:gmsb}),
this would lead to some very interesting conclusions.
First, from the presence of two separate peaks one would 
know immediately that there are two widely separated 
scales in the problem. Second, the normalization of each peak
would indicate the relative size of the total inclusive 
cross-sections (in this example, of the particles in the 
electroweak sector versus those in the strong sector). 
Finally, the broadness of each peak is indicative of the 
total number of contributing subprocesses, as well as the 
typical mass splittings of the particles within each sector.
It may appear surprising that one is able to draw so many 
conclusions from a single distribution of an inclusive variable,
but this just comes to show the importance of $\sqrt{s}$ 
as one of the fundamental collider physics variables.

Unfortunately, because of the missing energy due to the
escaping invisible particles, the true $\sqrt{s}$ distribution
cannot be observed, and the best one can do to approximate it is to 
look at the distributions of our inclusive $\sqrt{s}_{min}$ variables
discussed in Sec.~\ref{sec:reco}:
the calorimeter-based $\sqrt{s}_{min}^{(cal)}$ 
variable (dotted red histogram in Fig.~\ref{fig:gmsb})
and the RECO-level $\sqrt{s}_{min}^{(reco)}$ variable 
(solid black histogram in Fig.~\ref{fig:gmsb}).
In the figure, both of those are plotted for $\mmis=0$.

First let us concentrate on the calorimeter-based version 
$\sqrt{s}_{min}^{(cal)}$ (dotted red histogram).
We can immediately see the detrimental effects of the UE:
first, the electroweak production peak has been almost completely
smeared out, while the strong production peak has been shifted 
upwards by more than a TeV! This behavior is not too surprising,
since the same effect was already encountered in our previous 
examples in Secs.~\ref{sec:top} and \ref{sec:susy}.
Fortunately, we now also know the solution to this problem:
one needs to consider the RECO-level variable
$\sqrt{s}_{min}^{(reco)}$ instead, which tracks 
the true $\sqrt{s}$ distribution much better.
We can see evidence of this in Fig.~\ref{fig:gmsb} as well:
notice how the (black) $\sqrt{s}_{min}^{(reco)}$ histogram exhibits the 
same features as the (yellow-shaded) true $\sqrt{s}$ distribution.
In particular, $\sqrt{s}_{min}^{(reco)}$ does show two separate peaks 
(indicating that SUSY production takes place at two different mass scales), 
the peaks are in their proper locations (relative to the missing 
mass scale $\mmis$), and have the correct 
relative width, hinting at the size of the mass splittings in 
each sector. We thus conclude that all of the interesting physics
conclusions that one would be able to reach from looking at the
true $\sqrt{s}$ distributions, can still be made based on 
the inclusive distribution of our RECO-level  
$\sqrt{s}_{min}^{(reco)}$ variable.

Before concluding this section, we shall take the opportunity to
use the GM1b example to also illustrate the $\sqrt{s}_{min}^{(sub)}$
variable proposed in Sec.~\ref{sec:sub}. As already mentioned,
the GM1b study point corresponds to a GMSB scenario with a promptly decaying
Bino-like $\tilde\chi^0_1$ NLSP. Most events therefore contain 
two hard photons from the two $\tilde\chi^0_1$ decays to gravitinos.
Then it is quite natural to define the exclusive subsystem
in Fig.~\ref{fig:subevent} in terms of these two photons.
\FIGURE[t]{
\centerline{
\epsfig{file=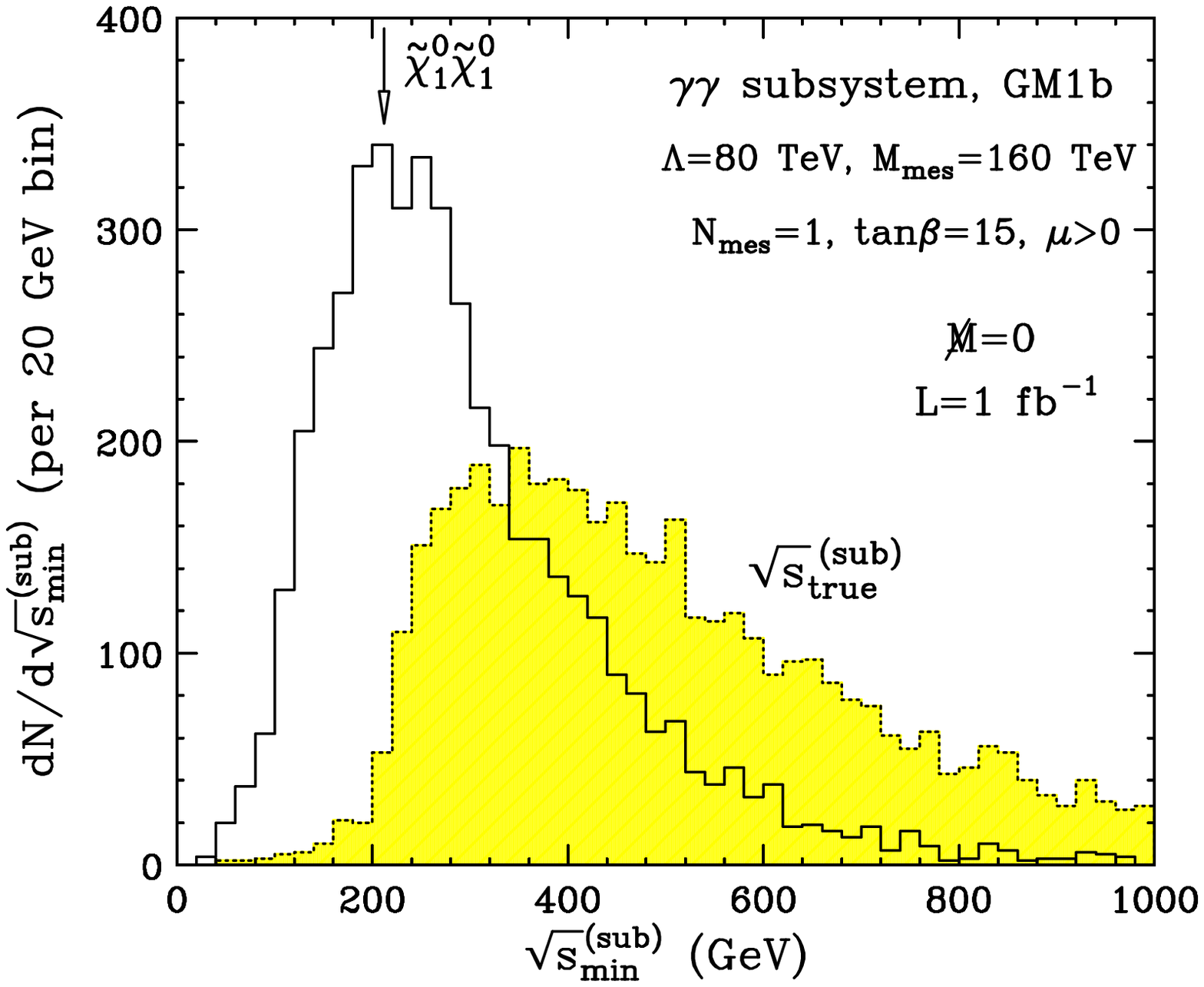,width=10cm}
}
\caption{The same as Fig.~\ref{fig:ttb_sub}, but for the 
GMSB SUSY example considered in Fig.~\ref{fig:gmsb}.
Here the subsystem is defined in terms of the two hard photons 
resulting from the two $\tilde\chi^0_1\to \tilde G+\gamma$
decays. The vertical arrow marks the onset for inclusive 
$\tilde\chi^0_1\tilde\chi^0_1$ production.}
\label{fig:gmsb2}
}
The corresponding $\sqrt{s}_{min}^{(sub)}$ distribution is 
shown in Fig.~\ref{fig:gmsb2} with the black solid histogram.
For completeness, in the figure we also show the 
true $\sqrt{s}$ distribution of the $\tilde\chi^0_1$ pair
(dotted yellow-shaded histogram). The vertical arrow marks the 
location of the $\tilde\chi^0_1\tilde\chi^0_1$ mass threshold.
We notice that the peak of the $\sqrt{s}_{min}^{(sub)}$ distribution 
nicely reveals the location of the neutralino mass threshold, 
and from there the neutralino mass itself.
We see that the method of $\sqrt{s}_{min}^{(sub)}$
provides a very simple way of measuring the 
NLSP mass in such GMSB scenarios (for an alternative
approach based on $M_{T2}$, see \cite{Hamaguchi:2008hy}).

\section{Comparison to other inclusive collider variables}
\label{sec:comp}

Having discussed the newly proposed variables 
$\sqrt{s}_{min}^{(reco)}$ and $\sqrt{s}_{min}^{(sub)}$
in various settings in Secs.~\ref{sec:top}-\ref{sec:gmsb},
we shall now compare them to some other global inclusive
variables which have been discussed in the literature in
relation to determining a mass scale of the new physics.
For simplicity here we shall concentrate only on the most 
model-independent variables, which do not suffer from the
topological and combinatorial ambiguities mentioned in the 
Introduction. 

At the moment, there are only a handful of such variables.
Depending on the treatment of the unknown masses of the 
invisible particles, they can be classified into one of the 
following two categories:
\begin{itemize}
\item {\em Variables which do not depend on an unknown invisible 
mass parameter.}
The most popular members of this class are the ``missing $H_T$'' 
variable
\beq
\mht \equiv \left| - \sum_{i=1}^{N_{obj}} \vec{P}_{Ti}\right|\, ,
\label{MHT}
\eeq
which is simply the magnitude of the $\mhtvec$ vector
from eq.~(\ref{MHTvec}),
and the scalar $H_T$ variable
\beq
H_T \equiv\ \mht + \sum_{i=1}^{N_{obj}} P_{Ti}\, .
\label{HT}
\eeq
Here we follow the notation from Sec.~\ref{sec:reco}, where
$\vec{P}_{Ti}$ is the measured transverse momentum of the 
$i$-th reconstructed object in the event $(i=1,2,\ldots,N_{obj})$.
The main advantage of $\mht$ and $H_T$ is their simplicity: 
both are very general, and are defined purely in terms of 
observed quantities, without any unknown mass parameters. 
The downside of $\mht$ and $H_T$ 
is that they cannot be directly correlated 
with any physical mass scale in a model-independent 
way\footnote{Some early studies of $H_T$-like variables 
found interesting linear correlations between the peak 
in the $H_T$ distribution and a suitably defined SUSY mass scale
in the context of specific SUSY models, 
e.g.~minimal supergravity (MSUGRA) \cite{Hinchliffe:1996iu,Tovey:2000wk,Hisano:2003qu},
minimal GMSB \cite{Tovey:2000wk}, or mixed moduli-mediation
\cite{Kitano:2006gv}. However, any such
correlations do not survive further scrutiny in more generic
SUSY scenarios, see e.g.~\cite{Bityukov:2002fs}.}.
\item {\em Variables which exhibit dependence on one or more
invisible mass parameters.} As two representatives
from this class we shall consider $M_{Tgen}$ from Ref.~\cite{Lester:2007fq} 
and $\sqrt{s}_{min}^{(reco)}$ from Sec.~\ref{sec:reco} here. 
We shall not repeat the technical definition of $M_{Tgen}$, 
and instead refer the uninitiated 
reader to the original paper \cite{Lester:2007fq}.
Suffice it to say that the method of $M_{Tgen}$ starts out by 
assuming exactly two decay chains in each event.
The arising combinatorial problem is then solved by brute force ---
by considering all possible partitions of the event into two sides, 
computing $M_{T2}$ for each such partition, and taking the 
minimum value of $M_{T2}$ found in the process. 
Both $M_{Tgen}$ and $\sqrt{s}_{min}^{(reco)}$ introduce
a priori unknown parameters related to the mass scale of the
missing particles produced in the event. In the case of
$\sqrt{s}_{min}^{(reco)}$, this is simply the single parameter
$\mmis$, measuring the {\em total} invisible mass (in the sense of 
a scalar sum as defined in eq.~(\ref{minv})). The $M_{Tgen}$ variable, on the
other hand, must in principle introduce two separate missing mass 
parameters $\mmis_1$ and $\mmis_2$ (one for each side of the event). 
However, the existing applications of $M_{Tgen}$ in the literature
have typically made the assumption that $\mmis_1=\mmis_2$, although
this is not really necessary and one could just as easily work in terms of two 
separate inputs $\mmis_1$ and $\mmis_2$ \cite{Barr:2009jv,Konar:2009qr}.
The inconvenience of having to deal with unknown mass parameters 
in the case of $M_{Tgen}$ and $\sqrt{s}_{min}^{(reco)}$ is greatly 
compensated by the luxury of being able to relate certain features 
of their distributions to a fundamental physical mass scale in a robust, 
model-independent way. In particular, the upper {\em endpoint} 
$M_{Tgen}^{(max)}$ of the $M_{Tgen}$ distribution gives the 
larger of the two parent masses $\max\{M_{P_1},M_{P_2}\}$
\cite{Barr:2009wu}.
Therefore, if the two parent masses are the same, i.e. $M_{P_1}=M_{P_2}$,
then the parent mass threshold $M_p=M_{P_1}+M_{P_2}$ is simply given by
\begin{equation}
M_p = 2 M_{Tgen}^{(max)}.
\label{MTgenmax}
\end{equation} 
On the other hand, as we have 
already seen in Secs.~\ref{sec:top}-\ref{sec:gmsb},
the {\em peak} of the $\sqrt{s}_{min}^{(reco)}$ is 
similarly correlated with the parent mass threshold,
see eq.~(\ref{recopeak}).
\end{itemize}
In principle, all four\footnote{We caution the reader that
$H_T$ is often defined in a more narrow sense than eq.~(\ref{HT}). 
For example, sometimes the $\mht$ term is omitted,  
sometimes the sum in eq.~(\ref{HT}) is limited to
the reconstructed jets only; or to the four highest $p_T$ jets only; 
or to all jets, but starting from the second-highest $p_T$ one.} 
of these variables are inclusive both object-wise and event-wise.
It is therefore of interest to compare them with respect to:
\begin{enumerate}
\item The degree of correlation with the new physics mass scale $M_p$.
\item Stability of this correlation against the detrimental effects of the UE.
\end{enumerate}
\FIGURE[t]{
\centerline{
\epsfig{file=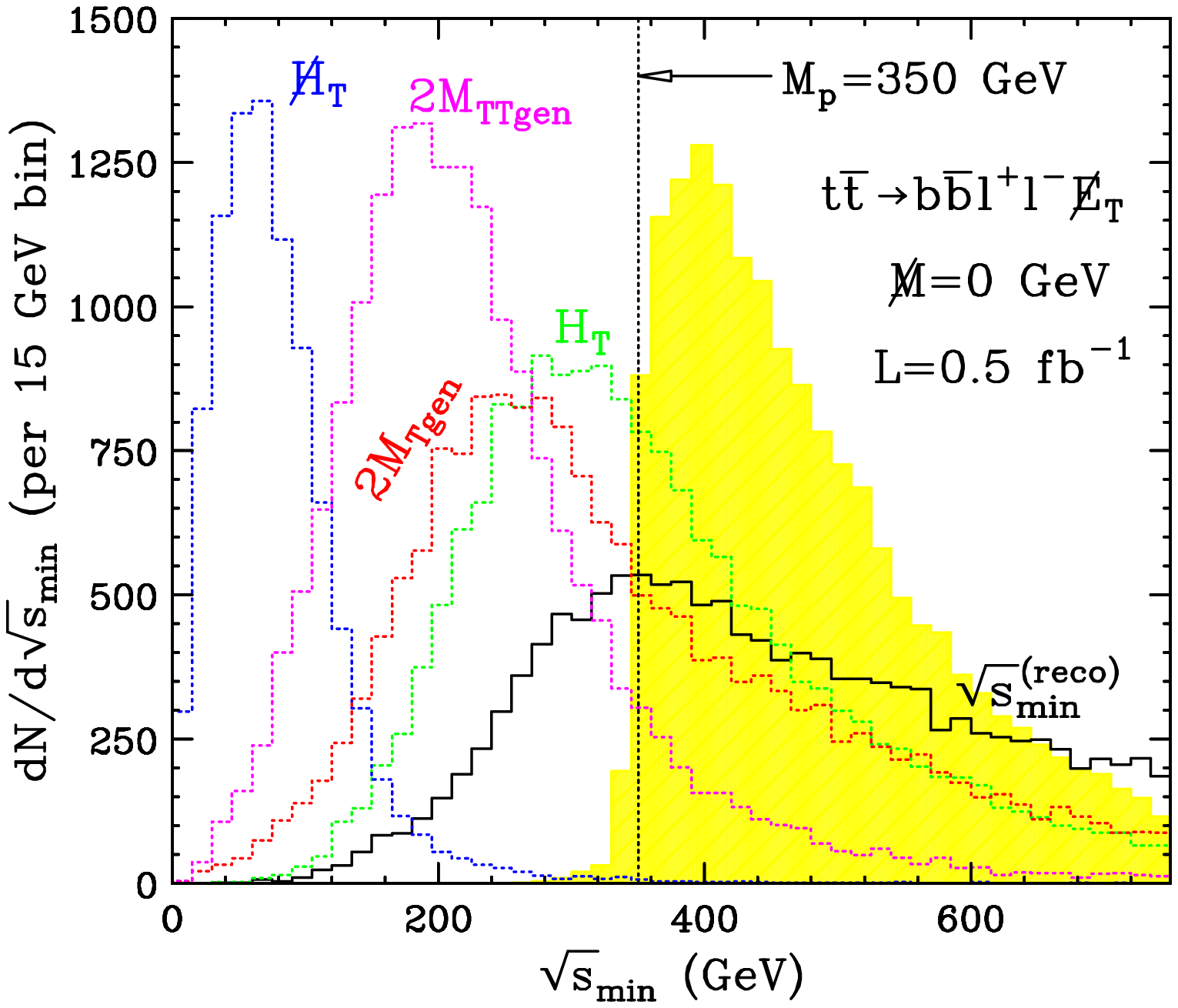,width=10cm}
}
\caption{The same as Fig.~\ref{fig:ttb}, but now in addition 
to the true $\sqrt{s}$ (yellow shaded) and
$\sqrt{s}_{min}^{(reco)}$ (black) distribution,
we also plot the distributions of
$2M_{Tgen}$ (red dots), $2M_{TTgen}$ (magenta dots), 
$H_T$ (green dots) and $\mht$ (blue dots),
all calculated at the RECO-level.
All results include the full simulation of the underlying event. 
For plotting convenience, the $\mht$ distribution
is shown scaled down by a factor of 2. The vertical 
dotted line marks the $t\bar{t}$ mass threshold $M_p=2m_t=350$ GeV.
}
\label{fig:comp_top}
}
Figs.~\ref{fig:comp_top}, \ref{fig:comp_glu4jet}
and \ref{fig:comp_glu2jet}
allow for such comparisons. 

In Fig.~\ref{fig:comp_top} we first revisit the case of the dilepton
$t\bar{t}$ sample discussed in Sec.~\ref{sec:top}.
In addition to the true $\sqrt{s}$ (yellow shaded) and
$\sqrt{s}_{min}^{(reco)}$ (black) distribution
already appearing in Fig.~\ref{fig:ttb}, we now 
also plot the distributions of $2M_{Tgen}$ (red dots), 
$H_T$ (green dots) and $\mht$ (blue dots), 
all calculated at the RECO-level.
For completeness, in Fig.~\ref{fig:comp_top} we also show
a variant of $M_{Tgen}$, called $M_{TTgen}$ (magenta dots),
where all visible particle momenta are first projected on the
transverse plane, before computing $M_{Tgen}$ in the usual 
way \cite{Lester:2007fq}\footnote{We caution the reader that 
the definition of $M_{TTgen}$ cannot be found in the 
published version of Ref.~\cite{Lester:2007fq}
--- the $M_{TTgen}$ discussion was added in a
recent replacement on the archive, which
appeared more than two years after the original publication.}.
All results include the full simulation of the underlying event. 
For plotting convenience, the $\mht$ distribution
is shown scaled down by a factor of 2. 

\FIGURE[t]{
\centerline{
\epsfig{file=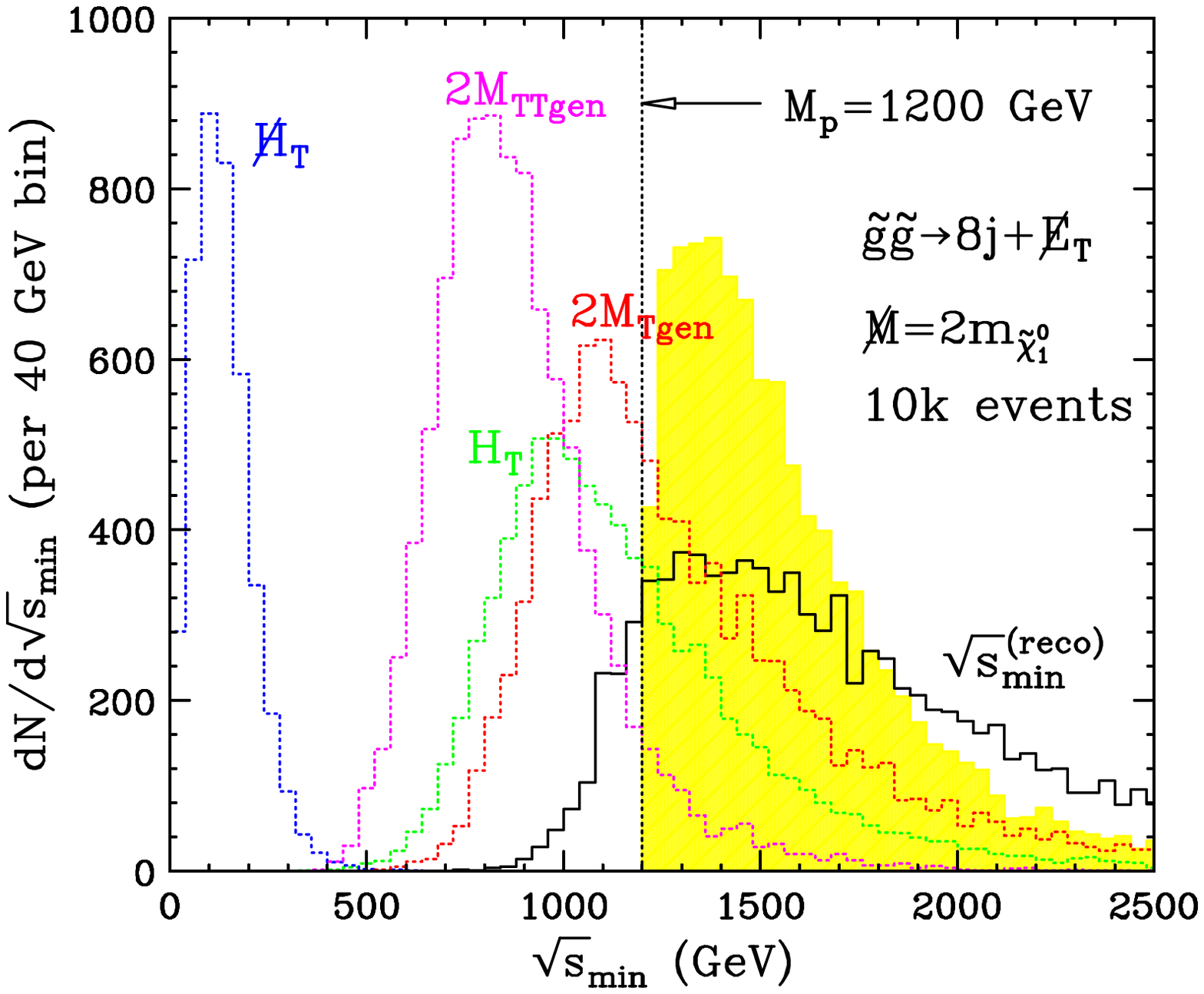,width=10cm}
}
\caption{The same as Fig.~\ref{fig:comp_top}, but 
for the gluino pair production example from Sec.~\ref{sec:susy},
with each gluino decaying to 4 jets as in (\ref{4jet}).
We use the light SUSY mass spectrum from Fig.~\ref{fig:glu_4jet}(a).
The vertical dotted line now shows the $\tilde g \tilde g$
mass threshold $M_p=2m_{\tilde g}=1200$ GeV.
}
\label{fig:comp_glu4jet}
}

Based on the results from Fig.~\ref{fig:comp_top}, we can now
address the question, which inclusive distribution shows the 
best correlation with the parent mass scale (in this case the 
parent mass scale is the $t\bar{t}$ mass threshold 
$M_p=2m_t=350$ GeV marked by the vertical dotted line
in Fig.~\ref{fig:comp_top}). Let us begin with the two variables,
$\mht$ and $H_T$, which do not depend on any unknown mass 
parameters. Fig.~\ref{fig:comp_top} 
reveals that the $\mht$ distribution
peaks very far from threshold, and therefore 
does not reveal much information about the 
new physics mass scale. Consequently, any attempt at extracting
new physics parameters out of the missing energy 
distribution alone, must make some additional 
model-dependent assumptions \cite{Hubisz:2008gg}.
On the other hand, the $H_T$ distribution appears to correlate better 
with $M_p$, since its peak is relatively close to 
the $t\bar{t}$ threshold. However, this relationship is
purely empirical, and it is difficult to know
what is the associated systematic error. 

Moving on to the variables which carry a dependence
on a missing mass parameter, $\sqrt{s}_{min}^{(reco)}$,
$2M_{Tgen}$ and $2M_{TTgen}$, we see that all three are 
affected to some extent by the presence of the UE.
In particular, the distributions of $2M_{Tgen}$ and $2M_{TTgen}$
are now smeared and extend significantly beyond their
expected endpoint (\ref{MTgenmax}). Not surprisingly, the
UE has a larger impact on $2M_{Tgen}$ than on $2M_{TTgen}$.
In either case, there is no obvious endpoint.
Nevertheless, one could in principle try to extract
an endpoint through a straight-line fit, for example,
but it is clear that the obtained value will be wrong by
a certain amount (depending on the chosen region for fitting and 
on the associated backgrounds). All these difficulties
with $2M_{Tgen}$ and $2M_{TTgen}$ are simply a reflection
of the challenge of measuring a mass scale from an endpoint
as in (\ref{MTgenmax}), instead of from a peak as in
(\ref{recopeak}). By comparison,
the determination of the new physics 
mass scale from the $\sqrt{s}_{min}^{(reco)}$ distribution
is much more robust. As shown in Fig.~\ref{fig:comp_top},
the $\sqrt{s}_{min}^{(reco)}$ peak is barely affected by 
the UE, and is still found precisely in the right location.

\FIGURE[t]{
\centerline{
\epsfig{file=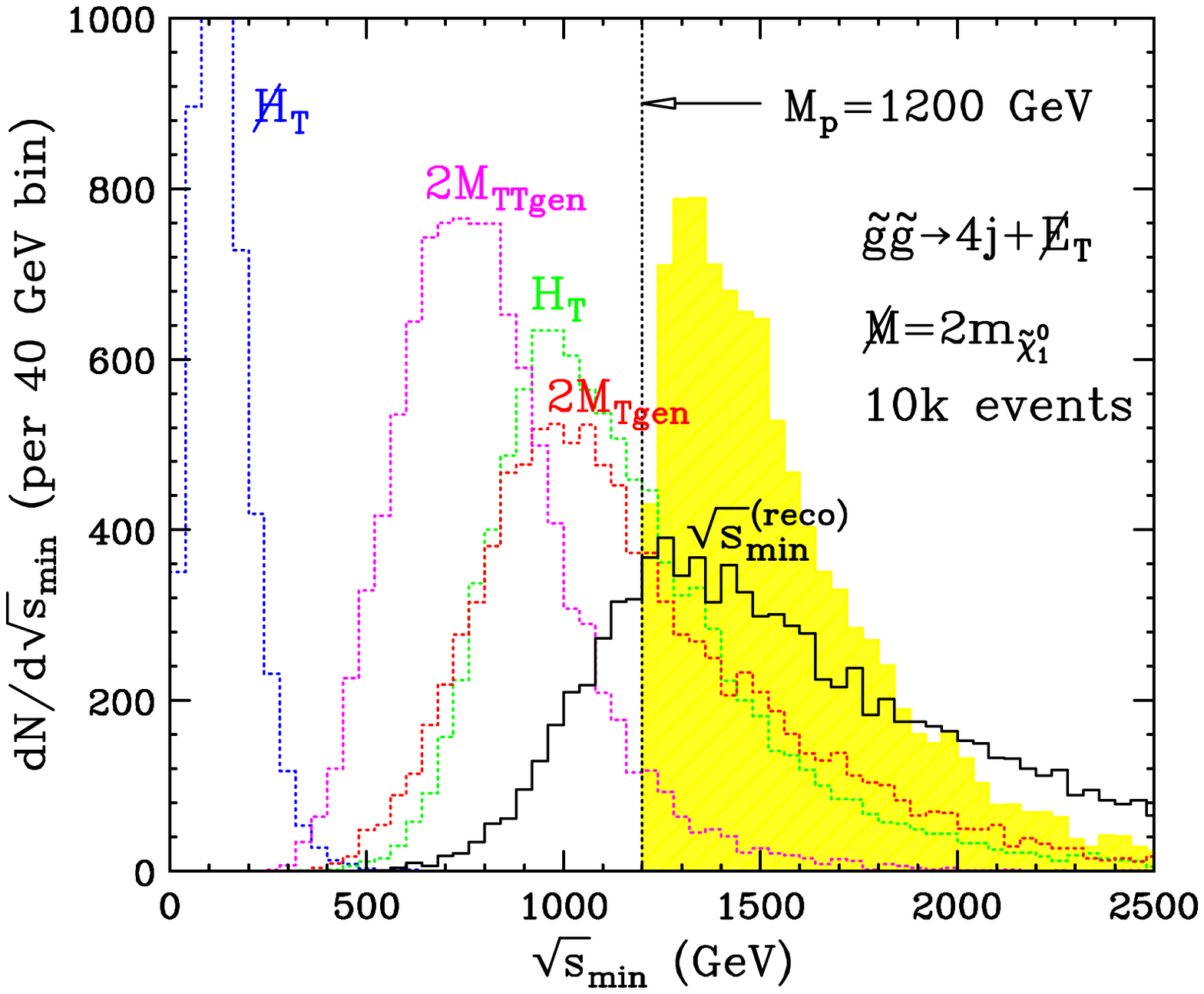,width=10cm}
}
\caption{The same as Fig.~\ref{fig:comp_glu4jet}, but 
with each gluino decaying to 2 jets as in (\ref{2jet}).
Compare to Fig.~\ref{fig:glu_2jet}(a).}
\label{fig:comp_glu2jet}
}

All of the above discussion can be directly 
applied to the SUSY examples considered in Sec.~\ref{sec:susy}
as well. As an illustration, Figs.~\ref{fig:comp_glu4jet}
and \ref{fig:comp_glu2jet} revisit two of the
gluino examples from Section~\ref{sec:susy}.
In both figures, we consider gluino pair-production with
a light SUSY spectrum ($m_{\tilde\chi^0_1}=100$ GeV,
$m_{\tilde\chi^0_2}=200$ GeV and $m_{\tilde g}=600$ GeV). 
Then in Fig.~\ref{fig:comp_glu4jet} each gluino decays
to 4 jets as in eq.~(\ref{4jet}), while in 
Fig.~\ref{fig:comp_glu2jet} each gluino decays
to 2 jets as in eq.~(\ref{2jet}). (Thus
Fig.~\ref{fig:comp_glu4jet} is the analogue of
Fig.~\ref{fig:glu_4jet}(a), while 
Fig.~\ref{fig:comp_glu2jet} is the analogue of
Fig.~\ref{fig:glu_2jet}(a).)

The conclusions from Figs.~\ref{fig:comp_glu4jet} 
and \ref{fig:comp_glu2jet} are very similar. Both
figures confirm that $\mht$ is not very helpful in 
determining the gluino mass scale $M_p=2m_{\tilde g}=1200$ GeV
(indicated by the vertical dotted line).
The $H_T$ distribution, on the other hand,
has a nice well-defined peak, but the location 
of the $H_T$ peak always underestimates the gluino mass 
scale (by about $250$ GeV in each case).  
Figs.~\ref{fig:comp_glu4jet} and \ref{fig:comp_glu2jet}
also confirm the effect already seen in Fig.~\ref{fig:comp_top}:
that the underlying event causes the 
$2M_{Tgen}$ and $2M_{TTgen}$ distributions to 
extend well beyond their upper kinematic endpoint,
thus violating (\ref{MTgenmax}) and making the 
corresponding extraction of $M_p$ rather problematic.
In fact, just by looking at Figs.~\ref{fig:comp_glu4jet} 
and \ref{fig:comp_glu2jet}, one might be tempted to 
deduce that, if anything, it is the {\em peak}
in $2M_{Tgen}$ that perhaps might indicate the value of the
new physics mass scale and not the $2M_{Tgen}$ endpoint.
Finally, the $\sqrt{s}_{min}^{(reco)}$ distribution
also feels to some extent the effects from the UE, but
always has its peak in the near vicinity of $M_p$.
Therefore, among the five inclusive variables
under consideration here, $\sqrt{s}_{min}^{(reco)}$
appears to provide the best estimate of the new physics
mass scale. 
The correlation (\ref{recopeak}) advertized in this paper
is seen to hold very well in Fig.~\ref{fig:comp_glu2jet}
and reasonably well in Fig.~\ref{fig:comp_glu4jet}.

\section{Summary and conclusions}
\label{sec:conclusions}

Since the original proposal of the $\sqrt{s}_{min}$ variable
in Ref.~\cite{Konar:2008ei}, its practicability has been 
called into question in light of the effects from the underlying
event, in particular initial state radiation and multiple 
parton interactions. In this paper we proposed two variations
of the $\sqrt{s}_{min}$ variable which are intended to avoid
this problems. 
\begin{enumerate}
\item {\em RECO-level $\sqrt{s}_{min}^{\,(reco)}$.} 
The first variant, the RECO-level variable
$\sqrt{s}_{min}^{(reco)}$ introduced in Sec.~\ref{sec:reco},
is basically a modification of the
{\em prescription for computing} the original $\sqrt{s}_{min}$ 
variable: instead of using (muon-corrected) calorimeter deposits,
as was done in \cite{Konar:2008ei,Papaefstathiou:2009hp},
one could instead calculate $\sqrt{s}_{min}$ with the help of the
reconstructed objects (jets and isolated photons, electrons 
and muons). Our examples in Sections \ref{sec:top}, \ref{sec:susy} and
\ref{sec:gmsb} showed that this procedure tends to automatically 
subtract out the bulk of the UE contributions, rendering the 
$\sqrt{s}_{min}^{(reco)}$ variable safe. 
\item {\em Subsystem $\sqrt{s}_{min}^{(sub)}$.} Our second suggestion,
discussed in Sec.~\ref{sec:sub}, was to apply $\sqrt{s}_{min}$
to a {\em subsystem} of the observed event, which is suitably 
defined so that it does not include the contributions from
the underlying event. The easiest way to do this is to veto
jets from entering the definition of the subsystem. In this case, 
the subsystem variable $\sqrt{s}_{min}^{(sub)}$ is completely
unaffected by the underlying event. However,
depending on the particular scenario, in principle one could also
allow (certain kinds of) jets to enter the subsystem. As long
as there is an efficient method (through cuts) of selecting 
jets which (most likely) did not originate from the UE, this
should work as well, as demonstrated in Fig.~\ref{fig:ttb_sub}
with our $t\bar{t}$ example.
\end{enumerate}

Being simply variants of the original $\sqrt{s}_{min}$ variable,
both $\sqrt{s}_{min}^{\,(reco)}$ and $\sqrt{s}_{min}^{\,(sub)}$
automatically inherit the many nice properties of $\sqrt{s}_{min}$:
\begin{itemize}
\item Both $\sqrt{s}_{min}^{\,(reco)}$ and $\sqrt{s}_{min}^{\,(sub)}$
have a clear physical meaning: the minimum CM energy in the 
(sub)system, which is required in order to explain the 
observed signal in the detector.
\item Both $\sqrt{s}_{min}^{\,(reco)}$ and $\sqrt{s}_{min}^{\,(sub)}$
are defined in a manifestly 1+3 Lorentz invariant way. As a 
consequence, their definitions utilize the available information
about the longitudinal momentum components of the particles 
observed in the detector.
\item Both $\sqrt{s}_{min}^{\,(reco)}$ and $\sqrt{s}_{min}^{\,(sub)}$
can be computed by simple analytical formulas, 
eqs.~(\ref{smin_def_reco},\ref{smin_def_reco_mvis})
and (\ref{smin_def_sub}-\ref{smin_def_sub_pt}), correspondingly.
\item $\sqrt{s}_{min}^{\,(reco)}$ (and to some extent $\sqrt{s}_{min}^{\,(sub)}$)
is a general, global, and inclusive variable, which can be 
applied to {\em any} type of events, regardless of the 
event topology, number or type of reconstructed objects, 
number or type of missing particles, etc. For example, 
all of the arbitrariness associated with the number and type of 
missing particles is encoded by a {\em single} parameter $\mmis$.
\item The most important property of both 
$\sqrt{s}_{min}^{\,(reco)}$ and $\sqrt{s}_{min}^{\,(sub)}$ is that they
exhibit a {\em peak} in their distributions, which {\em directly correlates}
with the mass scale $M_p$ of the parent particles.
In this regard we remind the reader that, compared to a kinematic endpoint,
a peak is a feature which is much easier to observe and subsequently measure
precisely over the SM backgrounds. This point was specifically
illustrated in Sec.~\ref{sec:comp}, where we contrasted 
the observability of the peak in the $\sqrt{s}_{min}^{\,(reco)}$
distribution to the observability of the endpoints of the
$2M_{Tgen}$ and $2M_{TTgen}$ distributions.
\end{itemize}

At the same time, compared to the original calorimeter-based 
$\sqrt{s}_{min}$ variable considered in 
Ref.~\cite{Konar:2008ei}, the new variables  
$\sqrt{s}_{min}^{\,(reco)}$ and $\sqrt{s}_{min}^{\,(sub)}$
proposed here have one crucial advantage:
they have very little sensitivity to the 
effects from the underlying event (ISR and MPI). 
As a result, the measurement of the corresponding mass scale
from the peak in the distribution of $\sqrt{s}_{min}^{\,(reco)}$ 
or $\sqrt{s}_{min}^{\,(sub)}$ is robust and 
physically meaningful.

In conclusion, we have shown that the variables 
$\sqrt{s}_{min}^{\,(reco)}$ and $\sqrt{s}_{min}^{\,(sub)}$
have certain important advantages, and we 
feel that the experimental collaborations at the Tevatron and the LHC
can only benefit from including them among their arsenal of observables.

\bigskip

\acknowledgments
We thank A.~Barr, C.~Lester, F.~Moortgat, L.~Pape and B.~Webber 
for stimulating discussions and correspondence.
This work is supported in part by a US Department of Energy 
grant DE-FG02-97ER41029. 
SLAC is operated by Stanford University for the US Department of Energy 
under contract DE-AC02-76SF00515.


\listoftables           
\listoffigures          


\end{document}